\documentstyle{article}

\newlength{\dinwidth}
\newlength{\dinmargin}
\makeatletter

\new@fontshape{msb}{m}{n}{%
   <5>msbm5%
   <6>msbm6%
   <7>msbm7%
   <8>msbm8%
   <9>msbm9%
   <10>msbm10%
   <11>msbm10%
   <12>msbm10 at10.95pt%
   <14>msbm10 at14.4pt%
   <17>msbm10 at17.28pt%
   <20>msbm10 at20.74pt%
   <25>msbm10 at24.88pt}{}
\extra@def{msb}{}{}
\newmathalphabet*\bbl{msb}{m}{n}

\new@fontshape{euf}{m}{n}{%
   <5>eufm5%
   <6>eufm6%
   <7>eufm7%
   <8>eufm8%
   <9>eufm9%
   <10>eufm10%
   <11>eufm10 at10.95pt%
   <12>eufm10 at12pt%
   <14>eufm14%
   <17>eufm14 at17.28pt%
   <20>eufm14 at20.74pt%
   <25>eufm14 at24.88pt}{}
\extra@def{euf}{\hyphenchar#1\m@ne
       \@tempdima\fontdimen2#1%
       \fontdimen3#1.4\@tempdima
       \fontdimen4#1.3\@tempdima}{}%
\newmathalphabet*\got{euf}{m}{n}

\newif\if@fewtab\@fewtabtrue

\def\draftdate{\number\day.\number\month.\number\year\ \ \ \hourmin }
{\count255=\time\divide\count255 by 60
\xdef\hourmin{\number\count255}
\multiply\count255 by-60\advance\count255 by\time
\xdef\hourmin{\hourmin:\ifnum\count255<10 0\fi\the\count255}}
\def\ps@draft{\let\@mkboth\@gobbletwo
    \def\@oddhead{}
    \def\@oddfoot
       {\hbox to 7 cm{$\scriptstyle\bf Draft\ version:\ \draftdate$
       \hfil}\hskip -7cm\hfil\rm\thepage \hfil}
    \def\@evenhead{}\let\@evenfoot\@oddfoot}

\def\label#1{\ifnum\draftcontrol=1
 \global\def\draftnote{\scriptsize\tt #1}\fi
 \@bsphack\if@filesw {\let\thepage\relax
   \def\protect{\noexpand\noexpand\noexpand}%
\xdef\@gtempa{\write\@auxout{\string
      \newlabel{#1}{{\@currentlabel}{\thepage}}}}}\@gtempa
   \if@nobreak \ifvmode\nobreak\fi\fi\fi
  \@esphack}

\def\@eqnnum{\hbox to 3cm{\phantom{\rm(\theequation)} \draftnote
                         \hfil}\hskip -3cm {\rm(\theequation)}}

\def\eqnarray{\def\draftnote{{}}\global\@fewtabtrue
\stepcounter{equation}\let\@currentlabel=\theequation
\global\@eqnswtrue
\global\@eqcnt\z@\tabskip\@centering\let\\=\@eqncr
$$\halign to \displaywidth\bgroup\@eqnsel\hskip\@centering\@eqcnt\z@
  $\displaystyle\tabskip\z@{##}$&\global\@eqcnt\@ne
  \hskip 1\arraycolsep \hfil${##}$\hfil
  &\global\@eqcnt\tw@ \hskip 1\arraycolsep
$\displaystyle\tabskip\z@{##}$
\hfil  \tabskip\@centering&\global\@eqcnt\thr@@\llap{##}\tabskip\z@
\cr}

\def\endeqnarray{\@@eqncr\egroup
      \global\advance\c@equation\m@ne$$\global\@ignoretrue}

\def\@@eqncr{\let\@tempa\relax
    \ifcase\@eqcnt \def\@tempa{& & &}\or \def\@tempa{& &}
      \or \def\@tempa{&}
      \or\def\@tempa{}
\fi\@tempa
\if@eqnsw
\if@fewtab\@eqnnum\fi
\stepcounter{equation}\fi\global
\@eqnswtrue\global\@eqcnt\z@\global\@fewtabtrue\cr}

\@addtoreset{equation}{section}

\def\ct#1{\ifnum\draftcontrol=1{\rm [#1]}\else{\cite{#1}}\fi}
\def\ctz#1#2{\ifnum\draftcontrol=1{\rm [#1,#2]}\else{\cite[#1]{#2}}\fi}
\def\draftcite#1{\ifnum\draftcontrol=1#1\else{}\fi}

\def\@lbibitem[#1]#2{\item{}\hskip -3cm \hbox to 2cm
{\hfil$\scriptstyle\draftcite{#2}$}\hskip
1cm[\@biblabel{#1}]\if@filesw
     {\def\protect##1{\string ##1\space}\immediate
      \write\@auxout{\string\bibcite{#2}{#1}}}\fi\ignorespaces}

\def\@bibitem#1{\item\hskip -3cm \hbox to 2cm
{\hfil \scriptsize\tt\draftcite{#1}}\hskip 1cm
\if@filesw \immediate\write\@auxout
       {\string\bibcite{#1}{\the\value{\@listctr}}}\fi\ignorespaces}

\def\cases#1{\left\{\,\vcenter{\normalbaselines\m@th
    \ialign{$\displaystyle{##}\hfil$&\quad##\hfil\crcr#1\crcr}}\right.}

\makeatother

\let\accentc=\c
\let\accentv=\v

\def\C{{\bbl C}}
\def\R{{\bbl R}}
\def\Q{{\bbl Q}}
\def\Z{{\bbl Z}}
\def\N{{\bbl N}}

\newtheorem{theo}{Theorem}

\def\lb#1{\label{#1}}
\def\Ref#1{(\ref{#1})}
\def\theequation{{\thesection.\arabic{equation}}}
\def\[{\begin{eqnarray}}
\def\nn{\nonumber}
\def\non{\nonumber \\ }
\def\]{\end{eqnarray}}
\def\leqn#1#2{$$ \displaylines{\hspace{#1em}{#2}\hfill\cr} $$}
\newlength{\breite}
\def\LL#1&#2&#3\RR{#1 & #2 & #3
                   \settowidth{\breite}{$\displaystyle #1#2#3$}
                   \addtolength{\breite}{-.973\hsize}
                   \hspace*{-\breite}\phantom{X}}
\def\HH{\hspace*{-2pt}}

\def\een{\end{enumerate}}
\def\ben{\begin{enumerate}}
\def\IF#1{\quad\mbox{if #1}}

\def\FOR#1{\quad\mbox{for #1}}
\def\noi{\noindent}
\def\a{\alpha}
\def\b{\beta}
\def\c{\gamma}
\def\d{\delta}
\def\e{\epsilon}
\def\f{\varphi}
\def\m{\mu}
\def\n{\nu}
\def\p{\psi}
\def\r{\rho}
\def\s{\sigma}
\def\x{\xi}
\def\y{\eta}
\def\ba{\mbox{\boldmath$\alpha$}}
\def\bd{\mbox{\boldmath$\delta$}}
\def\be{\mbox{\boldmath$\eta$}}
\def\bL{\mbox{\boldmath$\Lambda$}}
\def\br{\mbox{\boldmath$\rho$}}

\def\bw{\mbox{\boldmath$\omega$}}
\def\bx{\mbox{\boldmath$\xi$}}
\def\bz{\mbox{\boldmath$\zeta$}}
\def\D{\Delta}
\def\L{\Lambda}
\def\P{\Psi}
\def\cl{{\ell}}
\def\cD{{\cal D}}
\def\cF{{\cal F}}
\def\cH{{\cal H}}
\def\cV{{\cal V}}
\def\cP{{\cal P}}
\def\cY{{\cal Y}}
\def\Fz#1{\cF_{(#1)}}
\def\Pz#1{\cP_{(#1)}}
\def\gg{{\got g}}
\def\gh{{\got h}}
\def\gn{{\got n}}
\def\sS{{\rm  S}}
\def\sL{{\got L}}
\def\ew{{\got w}}
\def\eC{{\cal C}}
\def\eG{{\cal G}}
\def\eL{{\cal L}}
\def\eW{{\got W}}
\def\va{{\bf a}}
\def\val{{\ba}}
\def\vb{{\bf b}}
\def\vd{{\bd}}
\def\ve{{\bf e}}
\def\vk{{\bf k}}
\def\vl{{\bf l}}
\def\vL{{\bL}}
\def\vo{{\bf 0}}
\def\vp{{\bf p}}
\def\vq{{\bf q}}
\def\vQ{{\bf Q}}
\def\vr{{\bf r}}
\def\vri#1{{{\bf r}_{#1}}}
\def\vkx{{\bf k}^\times(z)}
\def\vs{{\bf s}}
\def\vt{{\bf t}}
\def\vu{{\bf u}}
\def\vx{{\bf x}}
\def\vy{{\bf y}}
\def\vro{{\br}}
\def\vxi{{\bx}}
\def\8{{\bx_8}}
\def\vxa{{\bx_\a}}
\def\vxb{{\bx_\b}}

\def\vet{{\be}}
\def\vze{{\bz}}
\def\End{{\rm End}\,}
\def\Vir{{\rm Vir}}
\def\Aut{{\rm Aut}\,}
\def\img{{\rm im}\,}
\def\mod{\,{\rm mod}\,}
\def\mult{{\rm mult}}
\def\ad#1#2{({\rm ad}\,#1)^{#2}}
\def\exp#1{{\rm e}^{#1}}
\def\frc#1#2{{\textstyle \frac{#1}{#2}}}
\def\2{\frc12}

\def\9{{E_9}}
\def\0{E_{10}}
\def\dz{\frac{d}{dz}}
\def\|{\,|\,}
\def\.{\cdot}
\def\X{\!\cdot\!}
\def\XO{\otimes}
\def\res{{\rm Res}}
\def\Res#1#2{{\rm Res}_{#1}\left[#2\right]}
\def\Dm#1#2#3{z_#1^{-1}\d\left(\frac{z_#2-z_#3}{z_#1}\right)}
\def\Dp#1#2#3{z_#1^{-1}\d\left(\frac{z_#2+z_#3}{z_#1}\right)}
\def\Di#1#2#3{-z_#1^{-1}\d\left(\frac{-z_#2+z_#3}{z_#1}\right)}
\def\Lp#1{{\rm L}_{(#1)}}
\def\Lm#1{{\rm L}_{(-#1)}}
\def\Ln{{\rm L}_{(n)}}
\def\1{{\bf 1}}
\def\vertex{$(\cF,\cV,\1,\bw)$}
\def\Vp{\cV(\p,z)}
\def\Vf{\cV(\f,z)}
\def\VP#1{\cV(\p,z_{#1})}
\def\VF#1{\cV(\f,z_{#1})}
\def\VV#1{\cV(\cV(\p,#1)\f,z_2)}
\def\<{\langle}
\def\>{\rangle}
\def\FLie{\mbox{$\cF\big/\Lm1\cF$\,}}
\def\PLie{\mbox{$\Pz1\big/\Lm1\Pz0$\,}}
\def\Pr{\P_{\vr}}
\def\Ps{\P_{\vs}}
\def\amm{\a^\m_m}
\def\ann{\a^\n_n}
\def\:{\mbox{\bf:}}
\def\ord{\mbox{\large\bf:}}
\def\Ord{_\times^\times}
\def\II{I\hspace{-.2em}I_{9,1}}
\def\III{I\hspace{-.2em}I_{25,1}}
\def\ggA{\gg(A)}
\def\ggF{\gg_\cF}
\def\ggL{\gg_\L}
\def\ggI{\gg_{\II}}
\def\gfake{\gg_{\III}}
\def\ghA{\gh(A)}
\begin{document}
\def\draft{\pagestyle{draft}\thispagestyle{draft}
\global\def\draftcontrol{1}}
\global\def\draftcontrol{0}

\arraycolsep3pt

\thispagestyle{empty}
\renewcommand{\thefootnote}{\fnsymbol{footnote}}
\begin{flushright} hep-th/9406175 \\
                   DESY 94-106    \end{flushright}
\vspace*{2cm}
\begin{center}
{\LARGE \sc On $\0$ and the DDF Construction%
            \footnote[1]{submitted to Comm. Math. Phys.}}\\
 \vspace*{1cm}
       {\sl Reinhold W. Gebert\footnote[2]{Supported by
        Konrad-Adenauer-Stiftung e.V.} and Hermann Nicolai}\\
 \vspace*{6mm}
     IInd Institute for Theoretical Physics, University of Hamburg\\
     Luruper Chaussee 149, D-22761 Hamburg, Germany\\
 \vspace*{6mm}
\ifnum\draftcontrol=1{\LARGE \bf Draft version: \draftdate \\}
                     \else{June 24, 1994 \\}\fi

\vspace*{1cm}
\begin{minipage}{11cm}\footnotesize
An attempt is made to understand the root spaces of Kac Moody
algebras of hyperbolic type, and in particular $\0$,
in terms of a DDF construction appropriate to a subcritical
compactified bosonic string. While the level-one root spaces can be
completely characterized in terms of transversal DDF states
(the level-zero elements just span the affine subalgebra),
longitudinal DDF states are shown to appear beyond level one.
In contrast to previous treatments of such algebras, we find
it necessary to make use of a rational extension of the self-dual root
lattice as an auxiliary device, and to admit non-summable operators
(in the sense of the vertex algebra formalism). We demonstrate the
utility of the method by completely analyzing a non-trivial
level-two root space, obtaining an explicit and comparatively
simple representation for it. We also emphasize the occurrence
of several Virasoro algebras, whose interrelation is expected
to be crucial for a better understanding of the complete
structure of the Kac Moody algebra.
\end{minipage}
\end{center}
\renewcommand{\thefootnote}{\arabic{footnote}}
\setcounter{footnote}{0}
\newpage
\section{Introduction}
Affine Kac Moody algebras \ct{Kac90, GodOli86}, which first
appeared in physics in the guise of (two-dimensional) current algebras,
have come to play an increasingly important role in string theory
and conformal field theory as well as other branches of mathematical
physics.
By contrast, Kac Moody algebras based on indefinite Cartan matrices
have not yet found applications in physics. In view of the
scarcity of results about such algebras, it is remarkable that
they have nevertheless been suggested as natural candidates for
the still elusive fundamental symmetry of string theory (and hence
of nature). Being vastly larger than affine Kac Moody algebras,
Kac Moody algebras of indefinite type might certainly be
``sufficiently big'' for a unified and background independent
formulation of string (field) theory, but an even more compelling
argument supporting such speculations
is the intimate link that exists between Kac Moody algebras
and the vertex operator construction of string theory (this connection
has been known for a long time \ct{BarHal71}). More specifically,
it has been established that the generators
making up a Kac Moody algebra of finite or affine type can be
explicitly realized in terms of tachyon and photon emission
vertex operators of a compactified open bosonic string
\ct{FreKac80, GodOli85}. On the basis of these results, it has been
conjectured that generalized Kac Moody algebras of indefinite
type might not only furnish new symmetries
of string theory, but might themselves be understood in terms
of string vertex operators associated with
the higher excited (massive) states of a compactified
bosonic string \ct{GodOli85, Fren85}. Despite its great appeal,
however, this idea has not led to a truly satisfactory understanding
of these Kac Moody algebras until now.

Disregarding possible physical applications in string theory,
very little is known about indefinite Kac Moody algebras beyond
their mere existence and the remarkable result that the Weyl-Kac
character formula continues to hold for them \ct{Mood79, Kac90}.
The basic problem here is the proliferation
of timelike roots (having {\it negative} (length)$^2$) and the
concomitant exponential growth in the dimension of the corresponding
root spaces. For a limited number of cases, and in particular
for roots of level two at most\footnote{The notion
of ``level of a root'' is defined in section 4.1.},
one knows explicit multiplicity
formulas counting the dimension of the root spaces \ct{KaMoWa88},
but the complete root multiplicities are not known
for a single Kac Moody algebra of indefinite type
(root multiplicities can be determined in principle from the
Peterson recursion formula \ct{KMPS90}, but this formula quickly
becomes too unwieldy for practical use). Unfortunately, the
available results have not shed much light on the structure
of the corresponding root spaces, and, in contrast to affine
Kac Moody algebras, a manageable representation of the root space
elements has not been found so far. In an interesting recent
development (more concerned with understanding the monster group than
with applications in physics), complete and explicit multiplicity
formulas were derived for the so-called fake monster Lie algebra
based on the 26-dimensional Lorentzian even self-dual lattice
$I\hspace{-.2em}I_{25,1}$ \ct{Borc90}; this algebra is, however,
not a conventional Kac Moody algebra in that it has imaginary
simple roots beside the usual simple roots \ct{Borc88}
(the extra simple roots correspond to new
Lie algebra elements that cannot be generated by multiple
commutators of the conventional Chevalley generators and must
therefore be adjoined ``by hand''). These results rely
heavily on special properties of 26 dimensions such as
the no-ghost theorem, and so far no other example has
been fully worked out\footnote{As an amusing aside, we note that
the very notion of what the monster Lie algebra should be has
undergone several metamorphoses since it was first proposed in
\ct{BoCoQuSl84}.}.

In this paper, an attempt is made to understand Kac Moody algebras
of hyperbolic type, and in particular the maximally extended
hyperbolic algebra $\0$, from a more ``physical'' (i.e. pedestrian)
point of view and to examine the known results as well as
the difficulties from what we believe to be a novel perspective.
We here (somewhat immodestly) concentrate on $\0$ not only
because, in our opinion, it is the most interesting, containing
$E_8$ and $\9$ as subalgebras, and because the basic problems
are not simplified in any substantial way by considering lower
rank hyperbolic algebras instead. Rather, we shall need to make use
of two very special features of $\0$ not shared by other algebras of
this type, namely the self-duality its root lattice, the unique
Lorentzian lattice $\II$, and secondly the property,
crucial for our construction, that the fundamental Weyl chamber
of $\0$ contains precisely one null direction (i.e. touches
the light-cone in root space only once), in contrast to
generic Kac Moody algebras of indefinite type,
whose Weyl chambers contain several linearly independent
null vectors, and also in contrast to strictly hyperbolic
algebras, whose Weyl chambers lie entirely within the light-cone and
therefore contain no null vectors at all.

Our key observation is that, beyond level one, there
appear {\it longitudinal} string states and vertex operators, whose
significance in this context has so far not been recognized.
A central role in our analysis is played by the DDF construction,
which provides the most direct and explicit solution of the physical
state constraints in string theory \ct{DeDiFu72}. The physical states,
which by definition are annihilated by the Virasoro constraints, are
simply obtained in this scheme by acting on a tachyonic groundstate
with the DDF operators, which commute with all
Virasoro generators and form a spectrum generating algebra.
For their definition, one must choose a special Lorentz frame,
in terms of which one can distinguish transversal and longitudinal
DDF operators. As is well known, the longitudinal states (or
operators) have zero norm and hence decouple in 26 dimensions
by the no-ghost theorem \ct{GodTho72, Brow72, Scher75}.
Above 26 dimensions, no consistent string theories and hence no
consistent Kac Moody algebras are expected to exist,
as there are always negative norm states. Below 26 dimensions,
on the other hand, the longitudinal DDF operators
create extra positive norm states (also referred to as Liouville
states), and thus modify the spectrum in an essential way.
It is therefore hardly surprising that longitudinal states should
also play a role in the construction of Lorentzian and hyperbolic
Kac Moody algebras of ``subcritical'' rank. Our results thus
suggest a connection between these algebras and Liouville
(or subcritical string) theory, the precise nature of which remains
to be elucidated, however.

The adaptation of the DDF construction to the present context
involves a discretization of the string vertex operator
formalism. As is well known \ct{GodOli85}, the allowed momenta
of the string excitations must be elements of the weight lattice
of the corresponding (affine or indefinite) Kac Moody algebra.
A curious feature of our construction, not encountered in previous
studies, is that we are here forced to make use of a rational extension
of the self-dual root lattice as an auxiliary device in order to
understand the root spaces associated with higher level roots
in terms of the DDF construction.
To be sure, the intermediate states belonging to momenta
not on the root lattice do not correspond to elements of the
Kac Moody algebra, but they are nonetheless indispensable for the
construction of a complete basis for any given root space, which
we here obtain in terms of transversal and longitudinal DDF states.
The problem of characterizing the root spaces is thereby reduced
to finding the ``missing states'', which cannot be reached
through multiple commutators of Chevalley generators.
To illustrate our method we will completely analyze one non-trivial
example of a level-two root space in terms of the DDF decomposition.
The simplicity of the explicit root space representation (cf.\
\Ref{E10-lev2}) obtained in this way is especially gratifying in view
of the lengthy intermediate expressions arising in the calculation;
to appreciate the relative efficiency of our method readers need only
contemplate the problem of classifying the states
in terms of 75-fold multiple commutators of the Chevalley
generators for this example.
Notwithstanding eventual refinements which may become necessary
at a later stage, we thus believe that our procedure provides
a workable method to probe higher level root spaces, yielding
a manageable representation of the level-two root space elements
for the first time.

Of course, we are aware that vertex operators have been utilized in
previous work on indefinite Kac Moody algebras \ct{GodOli85, Fren85}.
However, at least to the best of our knowledge,
neither the explicit representation of the level-one elements
in terms of transversal DDF operators
(the level-zero elements just span the affine subalgebra), nor the
emergence of longitudinal states and
vertex operators at level two and beyond have been exhibited
in previous treatments. Our results are couched in the language
of vertex algebras \ct{Borc86, FLM88, Gebe93} (see also \ct{LiZu94}
for a treatment in the BV formalism), which is entirely
equivalent to the formulation of \ct{GodOli85, Godd86}, but
slightly more convenient for our purposes because of its economy of
notation. An important technical point is that the longitudinal
vertex operators cannot be associated with definite states, as their
action cannot be defined on all of Fock space. Put differently,
they do not correspond to summable operators in the sense of
\ct{FLM88}; in this respect, vertex algebras
encompassing longitudinal states transcend
the definition given in \ct{Borc86, FLM88}.

A remarkable property of the vertex operator realization
of hyperbolic Kac Moody algebras,
which has not received due attention so far, is the occurrence
of several Virasoro algebras in the construction.
The first is the usual one with central charge $c=d$ (where $d$ is
the rank of the algebra, or, equivalently, the dimension of the
root lattice), by means of which physical states
can be distinguished from unphysical ones. Furthermore, there
are two sets of longitudinal DDF operators, called
$A_m^-$ and $\sL_m$ in this paper, which generate Virasoro
algebras of central charge $c=26-d$ and (independently of the
rank) $c=24$, respectively; the first choice
is the standard one used in proofs of the no-ghost theorem
(see e.g.\ \ct{GSW88}). These two sets are related
through a GKO construction \ct{GoKeOl85}, since they just differ
by a Sugawara type contribution built out of the transversal
DDF operators (see e.g. \ct{GodOli86} for an explanation of the
Sugawara construction). Since all DDF operators depend on the tachyon
momentum, there are actually infinitely many ``longitudinal
Virasoro algebras'', related to one another by Weyl rotations
on the (Lorentzian) root lattice (the Virasoro generators defining
the physical states are, of course, invariant under the Weyl group).
Our final results clearly suggest the structure of product
representations of the affine subalgebra and a Virasoro algebra
(already known to exist from the general representation theory
of Kac Moody algebras \ct{Kac90}). The action of the corresponding
Sugawara generators of level $\ell >1$ on the states remains to be
fully worked out, however. We believe that a proper
understanding of these Virasoro algebras is one of
the keys to unlocking the secrets of indefinite Kac Moody algebras.

Since this paper brings together various different concepts and
ideas, not all of which may be universally familiar, we have
tried to make it self-contained as far as possible.
In section 2, the formalism
of vertex algebras is briefly reviewed (for more details, the reader
should consult \ct{FLM88, Gebe93}). Section 3 is devoted to a
rather detailed exposition of the discrete DDF construction in
the framework of vertex (operator) algebras with particular
emphasis on the longitudinal states and operators. In section 4,
we apply this formalism to a study of the maximal hyperbolic Kac Moody
algebra $\0$, first recalling some of the few results known about it.
Sections 3 and 4 contain the bulk of our new results.
Explicit formulas for transversal and longitudinal DDF states as
well as a number of Lie algebra commutators giving level-two
root space elements are collected in appendices.

\section{General setup}
\subsection{Vertex algebras}
We shall provide a short primer to formal calculus and vertex algebras
following closely \ct{Gebe93}. In \ct{FLM88} the subject is treated
thoroughly.

In contrast to conformal field theory (see e.g.\ \ct{BePoZa84} or
\ct{Gins89}), in the vertex algebra approach we use {\it formal}
variables $z, z_0, z_1, z_2,$ etc.. The objects we will work with are
formal power series. For a vector space $W$, we set
\[   W[\![z,z^{-1}]\!]&=&\bigg\{\sum_{n\in\Z}w_nz^n\|w_n\in W\bigg\},\\
     W[\![z]\!]&=&\bigg\{\sum_{n\in\N}w_nz^n\|w_n\in W\bigg\}, \\
     W[z,z^{-1}]&=&\bigg\{\sum_{n\in\Z}w_nz^n\|w_n\in W
                    \mbox{, almost all }w_n=0\bigg\}
                    \qquad\mbox{(Laurent polynomials)}, \\
     W[z]&=&\bigg\{\sum_{n\in\N}w_nz^n\|w_n\in W
                    \mbox{, almost all }w_n=0\bigg\}
                    \qquad\mbox{(polynomials)}, \]
where ``almost all'' means ``all but finitely many.''
These sets are $\C$-vector spaces under obvious pointwise
operations. We can generalize above spaces in a straightforward way
to the case of several commuting formal variables, e.g.
$W[\![z_1,z_2^{-1}]\!]=\{\sum_{m,n\in\N}w_{mn}z_1^mz_2^{-n}\|w_{mn}
\in W\}$.

Since we will often multiply formal series or add up an infinite number
of series, it is necessary to introduce the notion of algebraic
summability. Let $(x_i)_{i\in I}$ be a family in $\End W$, the vector
space of endomorphisms of $W$ ($I$ is an index set). We say
that $(x_i)_{i\in I}$ is {\bf summable} if for every $w\in W$, $x_iw=0$
for all but a finite number of $i\in I$. Then the operator $\sum_{i\in I
}x_i$ is well-defined. In general an algebraic limit or a product of
formal series is defined if and only if the coefficient of {\it every}
monomial in the formal variables in the formal expression is summable.

If we define
\[ \d(z)=\sum_{n\in\Z}z^n\quad\in\C[\![z,z^{-1}]\!], \]
then, formally, this is the Laurent expansion of the classical
$\d$-function at $z=1$. Indeed, $\d(z)$ enjoys the following
fundamental properties: \\
Let $w(z)\in W[z,z^{-1}]\ ,\ a\in\C^\times$. Then
\[ w(z)\d(az)=w(a^{-1})\d(az). \lb{del1} \]
Let $X(z_1,z_2)\in(\End W)[\![z_1,z_1^{-1},z_2,z_2^{-1}]\!]$ be such
that $\lim_{z_1\to z_2}X(z_1,z_2)$ exists (algebraically) and let
$a\in\C^\times$. Then
\[ X(z_1,z_2)\d\left(a\frac{z_1}{z_2}\right)
             &=& X(a^{-1}z_2,z_2)\d\left(a\frac{z_1}{z_2}\right) \non
             &=& X(z_1,az_1)\d\left(a\frac{z_1}{z_2}\right).\lb{del2} \]
Note that $w(z)$ must be a Laurent polynomial to ensure existence of the
product with the $\d$-series. For explicit calculations it is useful
to keep in mind that the substitutions in the arguments
correspond formally to $az=1$ and $az_1/z_2=1$, respectively.

Now we want to introduce the tools for formal calculus which
correspond to contour integrals and residues for complex variables.
We will need quotients of formal power series which will often
be expressed by analytic functions of $z$ and $z^{-1}$, respectively.
They are understood as formal Taylor or Laurent expansions. E.g., for
$a\in\C$ we have
\[ (1+z)^a &=& \sum_{n\in\N}{a \choose n}z^n\in\C[\![z]\!], \\
   (1+z^{-1})^a&=&\sum_{n\in\N}{a \choose n}z^{-n}\in\C[\![z^{-1}]\!].\]
In the following we will always (though sometimes not explicitly stated)
refer to the {\bf binomial convention} which says that {\it all binomial
expressions are to be expanded in nonnegative integral powers of the
second variable}. This is the only point in explicit calculations at
which one must not be too sloppy. For example, for $a\in\C$ the
following expressions are in general not the same:
\[ \left(\frac{z_1-z_2}{z_0}\right)^a &=&
      \sum_{n\in\N}{a \choose n}(-1)^nz_0^{-a}z_1^{a-n}z_2^n, \\
      \left(\frac{-z_2+z_1}{z_0}\right)^a &=&
      \sum_{n\in\N}{a \choose n}(-1)^{a-n}z_0^{-a}z_1^nz_2^{a-n}. \]
With the binomial convention we can rewrite the generating function for
the derivatives of the $\d$-series as
\[ \Dm012=z_0^{-1}\exp{-z_2\frac\partial{\partial z_1}}
               \d\left(\frac{z_1}{z_0}\right).\lb{del3} \]
As an important exercise one may prove subsequent identities which will
be extremely useful for vertex operator calculus:
\[ \Dm012=\Dp102,\lb{del4} \]
\[ \Dm012\Di021=\Dm210, \lb{del5} \] \noi
where all binomial expressions are expanded in nonnegative integral
powers of the second variable.

We shall use the following residue notation. For a formal series
\[ w(z)=\sum_{n\in\Z}w_nz^n\in W[\![z,z^{-1}]\!], \]
we write \[ \Res{z}{w(z)}=w_{-1}, \]
so that we may think of $\Res{z}{\ldots}$ as the operation
$\oint_0\frac{dz}{2\pi i}[\ldots]$ in complex analysis. Indeed,
formal residue enjoys some properties of contour integration.
For $w(z)\in W[\![z,z^{-1}]\!]$ as above and $n\in\Z$ we find that
\[ w_n=\Res{z}{z^{-n-1}w(z)}; \lb{res0} \]
for $v(z),w(z)\in W[\![z,z^{-1}]\!]$ integration by parts reads as
\[ \Res{z}{v(z)\dz w(z)}=-\Res{z}{w(z)\dz v(z)}. \lb{res1} \]

We have already used exponentials of derivatives in \Ref{del3} to
obtain a formula for the higher derivatives of $\d(z)$.
However, one might also expect $\exp{z_0\dz}$ to act somehow as a
one-parameter group of automorphisms (parametrized by $z_0$).
This turns out to be also true in formal calculus.
Let $w(z)=\sum_{m\in\Z}w_mz^m\in W[\![z,z~{-1}]\!],\
     y\in z_0\C[\![z_0]\!]$ and
write $D_n=-z^{n+1}\dz,\ n\in\N$. Then we have
\ben
\item (Translation)
   \[ \exp{-yD_{-1}}w(z)\equiv\exp{y\dz}w(z)=w(z+y), \lb{trans} \]
\item (Scaling)
\[ \left(\exp y\right)^{-D_0}w(z)\equiv\exp{yz\dz}w(z)=
                                         w(\exp yz), \lb{scale} \]
\item (Projective change)
   \[ \exp{yD_n}w(z)=w\left((z^{-n}+ny)^{-1/n}\right)
                                       \FOR{}n\neq0, \lb{proj} \]
\een
with binomial convention.

We shall give a definition of vertex algebra \ct{FrHuLe93} using the
notation of \ct{Godd89b} which we believe is more accessible to
physicists.

A {\bf vertex algebra} is a $\Z$-graded vector space,
\[ \cF=\bigoplus_{n\in\Z}\Fz{n}, \]
equipped with a linear map $\cV:\cF\to(\End\cF)[\![z,z^{-1}]\!]$, which
assigns to each state $\p\in\cF$ a {\bf vertex operator} $\Vp$, and
the vertex operators satisfy the following axioms:
\ben
\item {\bf (Regularity)} If $\p,\f\in\cF$ then
\[ \Res{z}{z^n\Vp\f}=0 \FOR{ $n$ sufficiently large}, \lb{vert1} \]
and $n$ depending on $\p$ and $\f$.
\item {\bf (Vacuum)} There is a preferred state $\1\in\cF$, called the
vacuum, satisfying
\[ \cV(\1,z)={\rm id}_\cF. \lb{vert2} \]
\item {\bf (Injectivity)} There is a one-to-one correspondence between
states and vertex operators,
\[ \Vp=0\quad\iff\quad\p=0. \lb{vert3} \]
\item {\bf (Conformal vector)}
There is a preferred state $\bw\in\cF$, called the
conformal vector, such that its vertex operator
\[ \cV(\bw,z)=\sum_{n\in\Z}\Ln z^{-n-2}, \lb{vert4} \]  \ben
\item gives the {\bf Virasoro algebra} with some central charge
      $c\in\R$,
\[ [\Lp{m},\Ln]=(m-n)\Lp{m+n}+\frac{c}{12}(m^3-m)\delta_{m+n,0};
                                                      \lb{vert4a} \]
\item provides a {\bf translation generator}, $\Lm1$,
\[ \cV(\Lm1\p,z)=\dz\Vp \FOR{ every }\p\in\cF; \lb{vert4b} \]
\item gives the grading of $\cF$ via the eigenvalues of $\Lp0$,
\[ \Lp0\p=n\p\equiv\D_\p\p\FOR{ every }\p\in\Fz{n},n\in\Z,
                                                      \lb{vert4c} \]
      the eigenvalue $\D_\p$ is called the {\bf (conformal) weight    }
      of $\p$.
\een
\item {\bf (Jacobi identity)} For every $\p,\f\in\cF$,
\leqn4{\Dm012\VP{1}\VF{2}\Di021\VF{2}\VP{1}}
\[       =\Dm210\VV{z_0}, \lb{vert5} \]
where binomial expressions have to be expanded in nonnegative
integral powers of the second variable.
\een
\noindent We denote the vertex algebra just defined by \vertex.

We may think of $\cF$ as the space of finite occupation number states
in a Fock space so that $\cF$ is a dense subspace of the Hilbert space
$\cH$ of states. The regularity axiom states that, given $\p,\f\in\cF$
, there is always a high enough power $z^n$ such that $z^n\Vp\f$ is (at
``$z=0$'') a regular formal series. In other words, the regularity axiom
ensures that any $\Vp\f$ contains only a {\it finite} number of
singular (at ``$z=0$'') expressions. In terms of creation and
annihilation operators it reflects the fact that any finite occupation
number state $\f$ is killed by a finite but large enough number of
annihilation operators contained in (the normal ordered expression)
$\p_n$. We also mention that in physical applications the vertex
operator of the conformal vector corresponds to the stress--energy
tensor of the field theory.

A {\bf vertex operator algebra} is a vertex algebra with the additional
assumptions that \ben
\item the spectrum of $\Lp0$ is bounded below,
\item the eigenspaces $\Fz{n}$ of $\Lp0$ are finite-dimensional. \een
The first condition is an immediate consequence of a physical postulate.
As we will see $\Lp0$ generates scale transformations. Recalling that
the variable $z$ in conformal field theory has its origin in
$\exp{t+ix}$ (cf.\ \ct{Gins89}) one finds that $\Lp0$ corresponds to
time translations. Thus it may be identified with the energy which
should be bounded below in any sensible quantum field theory. In
fact, vertex operator algebras can be regarded as a rigorous
mathematical definition of chiral algebras in physics \ct{MooSei89}.
Then the formal variable $z$ can be thought of as a local complex
coordinate and the above relations \Ref{res0} and \Ref{res0} can be
realized by contour integrals. The vertex operators
$\Vp$ correspond to holomorphic chiral fields i.e.\ they can be
viewed as operator-valued distributions on a local coordinate chart
of a Riemannn surface. In this context the three terms of the Jacobi
identity are geometrically interpreted as the three ways of cutting
the Riemann sphere with four punctures into two spheres
with three punctures
\ct{FreZhu92, Zhu90}.

Since vertex operators are operator valued formal Laurent series we
can give an alternative formulation (see \ct{Borc92}, e.g.) of the
axioms of a vertex algebra using the mode expansion
\[ \Vp=\sum_{n\in\Z}\p_nz^{-n-1}. \lb{mode} \] One has \ben
\item (Regularity)
\[ \p_n\f=0\FOR{ $n$ sufficiently large}, \lb{mode1} \]
\item (Vacuum)
\[ \1_n\p=\delta_{n+1,0}\p, \lb{mode2} \]
\item (Injectivity)
\[ \p_n=0\quad\forall n\in\Z\quad\iff\quad\p=0, \lb{mode3} \]
\item (Conformal vector)
\[ \bw_{n+1}=\Ln, \lb{mode4} \]
\item (Jacobi identity)
\[ \sum_{i\ge0}(-1)^i{l \choose i}\Big(\p_{l+m-i}(\f_{n+i}\x)-
(-1)^l\f_{l+n-i}(\p_{m+i}\x)\Big)=\sum_{i\ge0}{m \choose i}(\p_{l+i}
\f)_{m+n-i}\x, \lb{mode5} \]
for all $\p,\f,\x\in\cF,\ l,m,n\in\Z$.
\een

In what follows we will frequently make use of two important formulas
which are the special cases $m=0$ and $l=0$, respectively, of eqn.\
\Ref{mode5}:\\
{\bf (Associativity formula)}
\[ (\p_l\f)_n=\sum_{i\ge0}(-1)^i{l \choose i}\Big(\p_{l-i}\f_{n+i}-
                (-1)^l\f_{l+n-i}\p_i\Big), \lb{assform} \]
{\bf (Commutator formula)}
\[ [\p_m,\f_n]=\sum_{i\ge0}{m \choose i}(\p_i\f)_{m+n-i},
                 \lb{comform} \]
for all $\p,\f\in\cF,\ l,m,n\in\Z$.

To get a feeling of the formalism and the axioms it is instructive
to derive some important properties of vertex algebras.
Iterating \Ref{vert4b} and using translation \Ref{trans} we find that
$\Lm1$ indeed generates translations,
\[ \cV\left(\exp{z_0\Lm1}\p,z\right)=\cV(\p,z+z_0). \lb{prop1} \]
Moreover, the vacuum is translation invariant because \Ref{vert4b} for
$\p=\1$ together with the vacuum axiom \Ref{vert2} and injectivity
\Ref{vert3} gives
\[ \Lm1\1=0. \lb{prop2} \]

Taking $\Res{z_0}{\Res{z_1}{z_1^n({\rm Jacobi\ identity})}}$
in the special case $\p=\bw$ we obtain
\[ [\Ln,\Vf]=\sum_{i\ge-1}{n+1 \choose i+1}\cV(\Lp{i}\f,z)z^{n-i}.
                                                           \lb{prop3} \]
In particular,
\[ [\Lm1,\Vf] &=& \dz\Vf, \lb{prop4} \\
   \   [\Lp0,\Vf] &=& \left(z\dz+\D_\f\right)\Vf
\IF{}\f\in\Fz{\D_\f}. \lb{prop5} \]
Using the well-known formula
$\exp AB\exp{-A}=\sum_{n=0}^\infty\frac1{n!}({\rm ad}_A)^nB
           \equiv\sum_{n=0}^\infty\frac1{n!}[A,[A,\ldots[A,B]]\ldots]$
and eqns. \Ref{trans} and \Ref{scale} above equations give,
respectively, \\[2mm]
{\bf (Translation property)}
\[ \exp{y\Lm1}\Vf\exp{-y\Lm1}=\cV(\f,z+y), \lb{prop6} \]
{\bf (Scaling property)}
\[ \exp{y\Lp0}\Vf\exp{-y\Lp0}=\exp{y\D_\f}\cV(\f,\exp yz)
\IF{}\f\in\Fz{\D_\f}, \lb{prop7} \]
for every $y\in z_0\C[\![z_0]\!]$. Thus $\Lp0$ generates scale
transformations. Note that \Ref{prop5} also implies
\[ \f_n\Fz{m}\subset\Fz{\D_\f+m-n-1}
\IF{}\f\in\Fz{\D_\f}, \lb{prop8} \]
which means that the operator $\f_n$ shifts the grading by $\D_\f-
n- 1$, i.e.\ it can be assigned ``degree'' $\D_\f- n- 1$. In view
of this relation the reader might wonder again why we use subscripts
in round brackets for the grading of $\cF$ and for the Virasoro
generators in contrast to the naked subscripts occurring in the mode
expansion \Ref{mode} of a vertex operator. This possibly causes some
confusion but stems from the fact that we employ two different mode
expansions. In conformal field theory we are familiar with the
expansion
\[ \p(z)\equiv\Vp=\sum_{n\in\Z}\p_{(n)}z^{-n-\D_\p}, \lb{prop9} \]
which depends on the conformal weight of the field $\p(z)$. To exhibit
explicitly the Virasoro algebra in the definition of a vertex algebra
we used this expansion for the vertex operator associated with the
conformal vector (stress-energy tensor!) in \Ref{vert4}. It is quite
easy to convert results obtained in one expansion into the other
formalism, namely, simply by shifting the grading:
\[ \p_n&\equiv&\p_{(n+1-\D_\p)}, \\
      \p_{(n)}&\equiv&\p_{n-1+\D_\p}, \]
for any homogeneous element $\p\in\cF$. For example we can rewrite
\Ref{prop8} as
\[ \f_{(n)}\Fz{m}\subset\Fz{m-n}, \]
so that $\f_{(n)}$ always has ``degree'' $-n$ irrespective of $\f$.
The mode expansion \Ref{prop9} is therefore the more natural one
because it respects the grading of $\cF$. On the other hand for formal
calculus it is more useful to stick to an expansion which does not refer
to the conformal weight of a state. Hence we shall almost everywhere
in the formulas assume the mode expansion \Ref{mode}.

Note that the Jacobi identity is obviously invariant under
$(\p,z_1,z_0)\leftrightarrow(\f,z_2,-z_0)$. This symmetry property
together with \Ref{del2}, \Ref{del4}, \Ref{prop1} and \Ref{vert3}
finally yields \\[2mm]
{\bf (Skew-symmetry)}
\[ \VP{0}\f=\exp{z_0\Lm1}\cV(\f,-z_0)\p, \lb{prop10} \]
or, in components,
\[ \p_n\f=-(-1)^n\f_n\p+\sum_{i\ge1}\frc1{i!}(-1)^{i+n+1}
     \Lm1^i(\f_{n+i}\p). \lb{prop11} \]

In particular, we observe that the vertex operator $\Vp$ ``creates''
the state $\p\in\cF$ when applied to the vacuum:
\[ \Vp\1=\exp{z\Lm1}\p, \lb{prop12} \]
by \Ref{vert2}. In components,
\[ \p_n\1=\cases{0 &for $n\ge0$ \cr
                   \p &for $n=-1$ \cr
 \frac1{(-n-1)!}\Lm1^{-n-1}\p &for $n\le-2$.} \lb{prop13} \]
Hence the vacuum satisfies
\[ \Ln\1=0\qquad\forall n\ge{-1}. \lb{prop14} \]

We shall denote by $\Pz{\D}$ the space of {\bf(conformal) highest
weight vectors {\rm or} primary states} of weight $\D$ satisfying
\[ (\Ln-\d_{n0}\D)\p=0\qquad\forall n\ge0. \lb{prop15} \]
Thus in any vertex algebra the vacuum is a primary state of weight zero.
We immediately infer from \Ref{prop3} that, for $\p\in\Pz{\D}$,
\[ [\Ln,\Vp] = z^n\left\{z\dz+(n+1)\D\right\}\Vp \qquad\forall
n\in\Z, \lb{prop16} \] or
\[ [\Ln,\p_m]=\left\{(\D-1)(n+1)-m\right\}\p_{m+n} \qquad\forall
m,n\in\Z, \lb{prop17} \]
i.e.\ $\Vp$ is a so called {\bf(conformal) primary field} of weight
$\D$.
We can rewrite \Ref{prop16} as
\[ [\Ln,z^{\D(n+1)}\Vp]=z^{n+1}\dz\left\{z^{\D(n+1)}\Vp\right\}, \]
so that, by \Ref{proj},
\[ \exp{y\Ln}\Vp\exp{-y\Ln}=
                 \left(\frac{\partial z_1}{\partial z}\right)^\D\VP{1}
   \qquad\forall n\neq0, \lb{prop18} \]
for every $y\in z_0\C[\![z_0]\!]$ where
$z_1=(z^{-n}-ny)^{-1/n}=z(1-nyz^n)^{-1/n}$.

We shall provide a certain subspace of the Fock space $\cF$ with the
structure of a Lie algebra (cf.\ \ct{Borc86},\ct{Borc92},\ct{FLM88}).
We define a bilinear product on $\cF$ by
\[ [\p,\f]:=\p_0\f, \lb{Lie1} \]
which turns out to be antisymmetric on the quotient space \FLie
due to skew-symmetry \Ref{prop11}. Putting $l=m=n=0$ in the Jacobi
identity \Ref{mode5} we get
$\p_0(\f_0\x)-\f_0(\p_0\x)=(\p_0\f)_0\x$.
But this equation translates precisely into the classical Jacobi
identity for Lie algebras,
\[ [[\p,\f],\x]+[[\f,\x],\p]+[[\x,\p],\f]=0, \]
on \FLie. Note that we may identify the Lie algebra \FLie with the
Lie algebra of commutators of operators $\p_0$, $\p\in\cF$. Indeed,
the commutator formula \Ref{comform} for $m=n=0$,
\[ [\p_0,\f_0]=([\p,\f])_0, \lb{Lie2} \]
together with definition \Ref{Lie1} tells us that in the adjoint
representation $\p$ acts on \FLie as the operator $\p_0$. Moreover, if
\( \p= \Lm1\f\in \Lm1\Fz{0} \) for some $\f\in\Fz{0}$ then
\( \p_0= \Res{z}{\cV(\Lm1\f,z)}= \Res{z}{\dz\Vf}= 0 \) by \Ref{vert4b}
and \Ref{res1}. In other words, dividing out the subspace $\Lm1\cF$
reflects the fact that the zero mode $\p_0$ of a vertex operator
$\Vp$ remains unchanged when a total derivative is added to $\Vp$.

The Lie algebra \FLie is too large for further investigations.
In physical applications such as string theory a distinguished role
is played by the primary states of weight $\D=1$ which we shall call
{\bf physical states} from now on. In fact, we learn from eqn.\
\Ref{prop17} that for a physical state $\p$ the corresponding zero
mode operator $\p_0$ commutes with the Virasoro algebra thereby
preserving all subspaces $\Pz{n}$ of primary states of weight $n$.
In particular, it maps physical states into physical states, i.e.
$[\Pz1,\Pz1]\subset\Pz1\mod\Lm1\Pz0$.
Hence it is natural to look in detail at the {\bf Lie algebra of
physical states},
\[ \ggF:=\PLie, \lb{Lie2a} \]
where we used the fact that
\[ \Lm1\Fz0\cap\Pz1=\Lm1\Pz0 \lb{Lie3} \]
in any vertex algebra. To see this we start from the following identity:
\[ \Lp{n}\Lm1\p=(n+1)\Lp{n-1}\p+\Lm1\Lp{n}\p
                \qquad\forall\p\in\cF,n\in\Z. \]
Then the inclusion ``$\supseteq$'' in \Ref{Lie3} obviously holds. On
the other hand, let $\p\in\Fz0$ and demand \(\Lp{n}\Lm1\p\stackrel!=
0\ \forall n\ge1\). Hence
\[ \Lp{n-1}\p=-\frc1{n+1}\Lm1\Lp{n}\p\qquad\forall n\ge1, \]
which by induction yields the inclusion ``$\subseteq$'' in \Ref{Lie3}
when the regularity axiom \Ref{mode1} is applied to the right-hand
side.

When defining the Lie algebra \FLie we had to divide out the
space $\Lm1\cF$ for mathematical reasons. Surprisingly, this reasoning
is motivated by physical considerations (cf.\ \ct{GSW88}).
Suppose that $\cF$ is equipped with an inner product (\_,\_) such
that the operator $\Lm{n}$ is the adjoint of $\Ln$ . Then
\( (\Lm1\f,\p)= (\f,\Lp1\p)= 0\ \forall \f\in\cF, \p\in\Pz{n},
   n\in\Z \),
i.e.\ the space $\Lm1\cF$ is orthogonal to all primary states. In
particular, $\Lm1\Pz{0}$ consists of {\bf null physical
states}, physical states orthogonal to all physical states including
themselves. Hence the Lie algebra $\ggF$ is obtained from $\Pz1$ by
dividing out (unwanted) null physical states.

It is well known that there are additional null physical states in
$\Pz1$ if and only if the central charge takes the critical value
$c=26$, namely the space $(\Lm2+\frac32\Lm1^2)\Pz{-1}$ (see
\ct{GSW88} for the calculations). The existence of these additional
null physical states is used in the proof of the no-ghost-theorem
\ct{GodTho72}.

\subsection{Toroidal compactification of the bosonic string}
It is by no means obvious that nontrivial examples of vertex (operator)
algebras exist. However, a class of vertex algebras is provided by the
following result (see \ct{Borc86},\ct{FLM88},\ct{Dong1}): \\
Associated with each nondegenerate even lattice $\L$ is a vertex
algebra \vertex. If in addition $\L$ is positive definite then
\vertex\ has the structure of a vertex operator algebra.

The rest of this section will be concerned with the explicit
construction of the vertex algebra stated above.
For physical motivations of the construction below the reader may
consult the articles \ct{GodOli85}, \ct{Godd86} and \ct{GodOli86} or the
comprehensive review \ct{LeScheWa89}.

Let $\L$ be an even lattice of rank $d<\infty$ with a symmetric
nondegenerate $\Z$-valued $\Z$-bilinear form $\_\X\_$ and
corresponding metric tensor $\y^{\m\n}$, $1\le\m,\n\le d$ ($\L$ even
means that $\vr^2 \in2\Z$ for all $\vr\in\L$). The vertex algebra
\vertex\ which we shall construct can be thought of as a bosonic
string theory with $d$ spacetime dimensions compactified on a torus.
Thus $\L$ represents the allowed momentum vectors of the theory.

Introduce ``zero mode states'' $\Pr$, $\vr\in\L$, which are by
definition orthonormal,
\[ (\Pr,\Ps)=\delta_{\vr,\vs}, \]
and oscillators $\amm$, $m\in\Z$, $1\le\m\le d$, satisfying the
commutation relations
\[ [\amm,\ann]=m\y^{\m\n}\delta_{m+n,0}, \]
and the hermiticity conditions
\[ (\amm)^\dagger=\alpha^\m_{-m}, \]
and acting on zero mode states by
\[ \amm\Pr&=&0\IF{}m>0, \\ p^\m\Pr&=&r^\m\Pr, \]
where $p^\m\equiv\a^\m_0$ and $r^\m$ are the components of
$\vr\in\L$. While the operators $\amm$ for $m>0$ by definition act as
annihilation operators, the creation operators $\amm$, $m<0$, generate
the Fock space from the zero mode states. For convenience let us define
\[ \vr(m):=\sum_{\m=1}^dr_\m\amm\equiv\vr\X\val_m \]
for $\vr\in\L$, $m\in\Z$, such that
\[ [\vr(m),\vs(n)]=m(\vr\X\vs)\delta_{m+n,0}. \lb{osccom} \]
We denote the $d$-fold Heisenberg algebra spanned by the oscillators
by \[ \hat{\bf h}:=\{\vr(m)\|\vr\in\L,m\in\Z\}, \]
and for the vector space of finite products of creation operators
($\equiv$ algebra of polynomials on the oscillators) we write
\[ S(\hat{\bf h}^-):=
   \bigoplus_{N\in\N}\bigg\{\prod_{i=1}^N\vr_i(-m_i)\|
                    \vr_i\in\L,m_i>0\ {\rm for}\ 1\le i\le N\bigg\}, \]
where ``$S$'' stands for ``symmetric'' because of the fact that the
creation operators commute with each other.

If we introduce formally position operators $q^\m$, $1\le\m\le d$,
commuting with $\amm$ for $m\neq0$ and satisfying
\[ [q^\n,p^\m]=i\y^{\m\n}, \]
then we find that
\[ \exp{i\vr\.\vq}\Ps=\P_{\vr+\vs}, \]
i.e.\ the zero mode states can be generated from the vacuum $\P_\vo$:
\[ \Pr=\exp{i\vr\.\vq}\P_\vo. \]
Thus the operators $\exp{i\vr\.\vq}$, $\vr\in\L$, may be identified
with the zero mode states and form an abelian group which is called
the group algebra of the lattice $\L$ and is denoted by $\C[\L]$. One
might expect the full Fock space $\cF$ of the vertex algebra to be
$S(\hat{\bf h}^-)\XO\C[\L]$. However, it turns out that we shall
need to replace the group algebra $\C[\L]$ by something more delicate
in order to adjust the signs in the Jacobi identity for the vertex
algebra. We will multiply $\exp{i\vr\.\vq}$ by a so called cocycle
factor $c_\vr$ which is a function of momentum $\vp$. This means that
it commutes with all oscillators $\amm$ and satisfies the eigenvalue
equations
\[ c_\vr\Ps=\e(\vr,\vs)\Ps. \]
More specifically, we define operators
$\exp{\vr}:=\exp{i\vr\.\vq}c_\vr$ and impose the conditions
\[ \exp\vr\exp\vs&=&\e(\vr,\vs)\exp{\vr+\vs}, \lb{coc1} \\
      \exp\vr\exp\vs&=&(-1)^{\vr\.\vs}\exp\vs\exp\vr, \lb{coc2} \\
      \exp\vr\exp{-\vr}&=&(-1)^{\frac12{\vr}^2}, \lb{coc3} \\
      \exp{\vo}&=&1, \lb{coc4} \]
which are equivalent to requiring, respectively,
\[ \e(\vr,\vs)\e(\vr+\vs,\vt)&=&
      \e(\vr,\vs+\vt)\e(\vs,\vt), \lb{ecoc1} \\
      \e(\vr,\vs)&=&(-1)^{\vr\.\vs}\e(\vs,\vr), \lb{ecoc2} \\
      \e(\vr,-\vr)&=&(-1)^{\frac12{\vr}^2}, \lb{ecoc3} \\
      \e(\vo,\vo)&=&1. \lb{ecoc4} \]
For example, associativity of the product $\exp\vr\exp\vs\exp\vt$
and \Ref{coc1} yield \Ref{ecoc1}. Note that the coycle condition
\Ref{ecoc1} implies $\e(\vo,\vo)=\e(\vo,\vr)=\e(\vr,\vo)\ \forall\vr$.
It is not difficult to show that it is always possible to construct
cocycles with these properties (see \ct{GodOli86}, e.g.).\footnote{
Without loss of generality we can assume that the function $\e$ is
bimultiplicative, i.e.\ $\e(\vr+\vs,\vt)=\e(\vr,\vt)\e(\vs,\vt)$ and
$\e(\vr,\vs+\vt)=\e(\vr,\vs)\e(\vr,\vt)$, $\forall\vr,\vs,\vt$.
Together with \Ref{ecoc3} and the normalization condition \Ref{ecoc4}
this then implies that
$\e(m\vr,n\vr)=[\e(\vr,\vr)]^{mn}=(-1)^{\frac{1}{2} mn{\vr}^2}$,
$\forall\vr,\ m,n\in\Z$.} Also note that
every 2-cocycle $\e:\L\times\L\to\{\pm1\}$ corresponds to a central
extension $\hat{\L}$ of $\L$ by $\{\pm1\}$:
\[ 1\to\{\pm1\}\to\hat{\L}\to\L\to1, \]
where we put $\hat{\L}=\{\pm1\}\times\L$ as a set and define a
multiplication in $\hat{\L}$ by
\[ (\rho,\vr)*(\sigma,\vs):=(\e(\vr,\vs)\rho\sigma,\vr+\vs)
   \FOR{}\r,\s\in\{\pm1\},\ \vr,\vs\in\L. \]
We will take the twisted group algebra $\C\{\L\}$ consisting of the
operators $\exp\vr$, $\vr\in\L$, instead of $\C[\L]$. This means
nothing but working with the section in the double cover $\hat{\L}$
of the lattice $\L$.

To summarize: The Fock space associated with the lattice $\L$ is defined
to be
\[ \cF:=S(\hat{\bf h}^-)\XO\C\{\L\}. \]
Note that the oscillators $\vr(m)$, $m\ne0$, act only on the first
tensor factor, namely, creation operators as multiplication operators
and annihilation operators via the adjoint representation, i.e.\ by
\Ref{osccom}. The zero mode operators $\a^\m_0$, however, are only
sensible for the twisted group algebra,
\[ \vr(0)\exp\vs=(\vr\X\val_0)\exp\vs=(\vr\X\vs)\exp\vs
     \qquad\forall\vr,\vs\in\L,  \lb{zeroact} \]
while the action of $\exp\vr$ on $\C\{\L\}$ is given by \Ref{coc1}.

We shall define next the (untwisted) vertex operators $\Vp$ for $\p\in
\cF$. For $\vr\in\L$ we introduce the formal sum
\[ \vr(z):=\sum_{m\in\Z}\vr(m)z^{-m-1}, \lb{current} \]
which is an element of $\hat{\bf h}[\![z,z^{-1}]\!]$ and may be regarded
as a generating function for the operators $\vr(m)$, $m\in\Z$, or as a
``current'' in contrast to the ``states'' in $\cF$. It is convenient to
split the current $\vr(z)$ into three parts:
\[ \vr(z)=\vr_<(z)+\vr(0)+\vr_>(z), \] where
\[ \vr_<(z):=\sum_{m>0}\vr(-m)z^{m-1},\qquad
   \vr_>(z):=\sum_{m>0}\vr(m)z^{-m-1}. \]
We will employ the usual normal ordering procedure, i.e.\ colons
indicate that in the enclosed expressions, $q^\n$ is written to the
left of $p^\m$, as well as the creation operators are to be placed to
the left of the annihilation operators.

For $\exp\vr\in\C\{\L\}$, we set
\[ \cV(\exp\vr,z):=\exp{\int\vr_<(z)dz}\exp\vr z^{\vr(0)}
     \exp{\int\vr_>(z)dz}, \lb{vertexop1} \]
using an obvious formal integration notation, i.e.\
\[ \int\vr_<(z)dz&:=&\sum_{m>0}\frac1m\vr(-m)z^m, \\
      \int\vr_>(z)dz&:=&-\sum_{m>0}\frac1m\vr(m)z^{-m}. \]
This can be written in a way more familiar to physicists by introducing
the Fubini-Veneziano coordinate field,
\[ Q^\m(z)\equiv q^\m-ip^\m\ln z+i\sum_{m\in\Z}\frac1m\amm z^{-m}, \]
which really only has a meaning when exponentiated. We find that the
vertex operator in \Ref{vertexop1} takes the familiar form
\[ \cV(\exp\vr,z)=\:\exp{i\vr\.\vQ(z)}\:c_\vr, \]
and the current $\vr(z)$ becomes
\[ \vr(z)=i\dz[\vr\X\vQ(z)]. \]
This shows that the vertex operators in \Ref{vertexop1} are indeed
already normal ordered and carry a cocycle factor hidden in the
elements of the twisted group algebra $\C\{\L\}$.

Let
$\p=\big[\prod_{j=1}^N\vs_j(-n_j)\big]\XO\exp\vr$ be
a typical homogeneous element of $\cF$ and define
\[ \Vp&:=&\ord\;\cV(\exp\vr,z)\prod_{j=1}^N\frac1{(n_j-1)!}
       \left(\dz\right)^{n_j-1}\vs_j(z)\;\ord \lb{vertexop2}\\
    &\equiv&i\;\ord\;\exp{i\vr\.\vQ(z)}\prod_{j=1}^N
         \frac1{(n_j-1)!)}\left(\dz\right)^{n_j}(\vs_j\X\vQ(z))
         \;\ord\;c_\vr. \nn \]

Extending this definition by linearity we finally obtain a
well-defined map
\[ \cV:\cF&\to&(\End\cF)[\![z,z^{-1}]\!], \non
      \p&\mapsto&\sum_{n\in\Z}\p_nz^{-n-1}. \]
We shall prove the first four axioms in the definition of a vertex
algebra.
\ben
\item {\bf (Regularity)}
Note that $\cF$ contains only states with a finite occupation number
of creation operators and the vertex operators are normal ordered
expressions. Having this in mind it is clear that $\p_n\f=0$ for $n$
large enough (depending on $\p,\f\in\cF$) since annihilation
operators are always attached to negative powers of the formal
variables.
\item {\bf (Vacuum)}
We choose the vacuum to be the zero mode state with no momentum and
without any creation operators, i.e.\
\[ \1:=1\XO\exp{\vo}, \]
so that
\[ \cV(\1,z)=\:\exp{i{\vo}\.\vQ(z)}\:c_{\vo}
            ={\rm id}_\cF \]
by the normalization condition \Ref{coc4}.
\item {\bf (Injectivity)}
Observe that, when acting on the vacuum, only terms involving creation
operators survive in the expression for a vertex operator. Then it is
obvious that
\[ \p_{-1}\1=\Res{z}{z^{-1}\Vp\1}=\p\qquad\forall \p\in\cF. \]
In particular, $\Vp=0$ implies $\p=0$.
\item {\bf (Conformal vector)}
We claim that the element
\[ \bw:=\2\sum_{\m=1}^d\ve^{(\m)}(-1)\ve_{(\m)}(-1)(\XO\exp{\vo}) \]
provides a conformal vector of dimension $d$ which is independent
of the choice of the basis $\{\ve_{(\m)}\}$ of $\L$ with dual basis
$\{\ve^{(\m)}\}$ (w.r.t. $\y^{\m\n}$). By \Ref{vertexop2} and
\Ref{current}, we have
\[ \cV(\bw,z)&=&\2\sum_{\m=1}^d\:\ve^{(\m)}(z)
                 \ve_{(\m)}(z)\: \non
              &=&\2\sum_{m,n\in\Z}\:\val_m\X\val_n\:\,
                 z^{-m-n-2}. \]
(Note that in the last step we had to rely on nondegeneracy of the
lattice, i.e.\ we used the completeness relation
 $\sum_{\m=1}^d(\ve^{(\m)})_\r(\ve_{(\m)})_\s=\y_{\r\s}$.)
Thus
\[ \Ln\equiv\bw_{n+1}=\2\sum_{m\in\Z}\:\val_m\X\val_{n-m}\:,
                                                  \lb{Virosc} \]
in agreement with the well-known expression from string theory. Using
the oscillator commutation relations one indeed finds that the $\Ln$'s
obey \Ref{vert4a} with central charge $c=d$ (see \ct{GSW88} for the
calculation). To establish the translation property of $\Lm1$ we find
that
\[ \Lm1\exp\vr&=&\vr(-1)\exp\vr, \\
   \Lm1\vr(-m)&=&m\vr(-m-1)\quad\mbox{for }m>0, \]
by \Ref{zeroact} and \Ref{osccom}; but, on the other hand,
\[ \dz\cV(\exp\vr,z)&=&\:\vr(z)\cV(\exp\vr,z)\:
                     =\cV(\vr(-1)\exp\vr,z), \\
     \dz\cV(\vr(-m),z)&=&\frac1{(m-1)!}\left(\dz\right)^m\vr(z)
                       =\cV(m\vr(-m-1),z), \]
by \Ref{vertexop1}, \Ref{vertexop2} and \Ref{current}.
Together with the derivation property of $\Lm1$ and $\dz$ this proves
\Ref{vert4b}.
Finally, let $\p=\big[\prod_{j=1}^N\vs_j(-n_j)\big]\XO\exp\vr$ be
a typical homogeneous element of $\cF$. Then
\[ \Lp0\p&=&\bigg(\2\vp^2+\sum_{m\ge1}\val_{-m}\X\val_m\bigg)
   \left\{\bigg[\prod_{j=1}^N\vs_j(-n_j)\bigg]\XO\exp\vr\right\} \non
 &=&\bigg(\2\vr^2+\sum_{j=1}^Nn_j\bigg)\p \lb{numberop} \]
yields the desired grading of $\cF$. Furthermore we observe that the
spectrum of $\Lp0$ is nonnegative and the eigenspaces of $\Lp0$ are
finite-dimensional provided that $\L$ is a positive definite lattice;
while if $\L$ is Lorentzian then $\vr^2$
can be arbitrarily negative
so that the spectrum of $\Lp0$ is unbounded from above as well as
from below.
\een
It is not surprising that by far the hardest axiom to prove is the
Jacobi identity because it contains most information about a vertex
algebra. We will skip the proof and refer the interested reader to
Ref.\ \ct{FLM88}.

We turn now to the analysis of the Lie algebra of physical states,
$\ggL$, and work out some of its commutators. Let us first list the
simplest physical states:
\ben
\item {\bf Tachyonic states:}
\[ \ggL^{[0]}:=\{\exp\vr\|\vr\in\L_2\}; \lb{phys0} \]
\item {\bf Photonic states:}
\[ \ggL^{[1]}:=\{\vs(-1)\XO\exp\vr\|
                 \vr\X\vs=0,\vs\in\L,\vr\in\L_0\}; \lb{phys1} \]
\item {\bf Massive spin 2 states:}
\[ \ggL^{[2]}&:=&\Bigm\{\Big[
                        (\vs\X\vr)\vt(-2)+(\vt\X\vr)\vs(-2)
                        -2\vs(-1)\vt(-1)\Big]\XO\exp\vr\Bigm| \non
              & &\hspace{1cm}\vs\X\vt=2(\vs\X\vr)(\vt\X\vr),
                     \vs,\vt\in\L,\vr\in\L_{-2}\Bigm\}; \lb{phys2} \]
\een
where $\L_n:=\{\vr\in\L\|\vr^2=n\ (\in2\Z)\}$
denotes the set of lattice
vectors of squared length $n$ and the superscript of $\ggL$ counts the
oscillator excitations.
The relevant physical state conditions for above polarization vectors
$\vs,\vt\in\L$ follow immediately from \Ref{numberop} and
\[ \Lp{m}\p &=& \sum_{k=1 \atop n_k>m}^Nn_k\bigg[\vs_k(m-n_k)
         \prod_{j=1 \atop j\neq k}^N\vs_j(-n_j)\bigg]\XO\exp\vr
                +m\sum_{k=1}^N\d_{m,n_k}(\vs_k\X\vr)
   \bigg[\prod_{j=1 \atop j\neq k}^N\vs_j(-n_j)\bigg]\XO\exp\vr \non
            & & {}+\sum_{k<k'}^Nn_kn_{k'}\d_{m,n_k+n_{k'}}
                   (\vs_k\X\vs_{k'})
   \bigg[\prod_{j=1 \atop j\neq k,k'}^N\vs_j(-n_j)\bigg]\XO\exp\vr
   \lb{virstate} \]
for $\p=\big[\prod_{j=1}^N\vs_j(-n_j)\big]\XO\exp\vr\in\cF$.
Above formula also exhibits an explicit example for the regularity
axiom \Ref{mode1}, namely, that
$\Lp{m}\p=0$ for $m>\max_{j\neq k}(n_j+n_k)$. We want to stress again
that the physical states in $\ggL$ are only defined modulo $\Lm1\Pz0$
which means for example that $\vr(-1)\XO\exp\vr= \Lm1(\exp\vr)\equiv0$
in $\ggL$ for $\vr\in\L_0$.

For the antisymmetric product \Ref{Lie1} on \PLie we obtain
\[ [\exp\vr,\exp\vs]&:=&\exp\vr_0\exp\vs
    = \Res{z}{\exp{\int\vr_<(z)dz}\exp\vr z^{\vr(0)}
              \exp{\int\vr_>(z)dz}(1\XO\exp\vs)} \non
   &=&\res_z\bigg[\sum_{m\ge0}\sS_m(\vr)z^{m+\vr\.\vs}\exp\vr\exp\vs
                 \bigg] \non
   &=&\cases{0  & if $\vr\X\vs\ge0$, \cr
             \e(\vr,\vs)\sS_{-1-\vr\.\vs}(\vr)\XO\exp{\vr+\vs}
                & if $\vr\X\vs<0$,} \lb{comm1} \]
where we used the Schur polynomials
$\sS_m(\vr)\equiv\sS_m(\vr(-1),\vr(-2),\ldots,\vr(-m))$
which are defined via the generating function
\[ \exp{\sum_{n>0}\frac1n\vr(-n)z^n}=\sum_{m\ge0}
                 \sS_m(\vr(-1),\vr(-2),\ldots,\vr(-m))z^m, \]
e.g.,
\[ \sS_0(\vr)&=&1 \non
   \sS_1(\vr)&=&\vr(-1), \non
   \sS_2(\vr)&=&\frc1{2!}\Big(\vr(-1)^2+\vr(-2)\Big), \non
   \sS_3(\vr)&=&\frc1{3!}\Big(\vr(-1)^3+3\vr(-2)\vr(-1)+2\vr(-3)\Big).
   \]
For notational convenience we put
$\sS_m(\vr):=0\ \forall m<0,\vr\in\L$. We also find that
\leqn0{[\vs(-1)\XO\exp\vr,\exp\vt]}\vspace{-5ex}
\[ &=&\Res{z}{\ord\exp{\int\vr_<(z)dz}\exp\vr z^{\vr(0)}
              \exp{\int\vr_>(z)dz}\vs(z)\ord(1\XO\exp\vt)} \non
   &=&\res_z\bigg[\sum_{m\ge0}\sS_m(\vr)z^{m+\vr\.\vt}
                \bigg((\vs\X\vt)z^{-1}+
                      \sum_{n>0}\vs(-n)z^{n-1}\bigg)
                \XO\exp\vr\exp\vt\bigg] \non
   &=&\cases{0  & if $\vr\X\vt\ge1$, \cr
             \e(\vr,\vt)\bigg[(\vs\X\vt)\sS_{-\vr\.\vt}(\vr)+
             \sum_{m=0}^{-1-\vr\.\vt}\sS_m(\vr)\vs(m+\vr\X\vt)\bigg]
             \XO\exp{\vr+\vt}
                & if $\vr\X\vt\le0$;} \lb{comm2} \]
\leqn0{[\vs(-1)\XO\exp\vr,\vu(-1)\XO\exp\vt]}\vspace{-5ex}
\[ &=&\Res{z}{\ord\exp{\int\vr_<(z)dz}\exp\vr z^{\vr(0)}
              \exp{\int\vr_>(z)dz}\vs(z)\ord(\vu(-1)\XO\exp\vt)} \non
   &=&\res_z\bigg[\sum_{m\ge0}\sS_m(\vr)z^{m+\vr\.\vt}
              \bigg(\Big(\vs\X\vu-(\vr\X\vu)(\vs\X\vt)\Big)z^{-2}
              +\Big((\vs\X\vt)\vu-(\vr\X\vu)\vs\Big)(-1)z^{-1} \non
   & &\hspace{1cm}
              +\sum_{n>0}\Big(\vs(-n)\vu(-1)-(\vr\X\vu)\vs(-n-1)\Big)
                   z^{n-1}\bigg)\XO\exp\vr\exp\vt\bigg] \non
   &=&\cases{0  & if $\vr\X\vt\ge2$, \cr
             \e(\vr,\vt)\bigg[
             \Big(\vs\X\vu-(\vr\X\vu)(\vs\X\vt)\Big)
             \sS_{1-\vr\.\vt}(\vr)+
             \Big((\vs\X\vt)\vu-(\vr\X\vu)\vs\Big)(-1)
             \sS_{-\vr\.\vt}(\vr) \cr
             \phantom{\e(\vr,\vt)\bigg[\ }
             +\sum_{m=0}^{-1-\vr\.\vt}
             \Big(\vs(-m-\vr\X\vt)\vu(-1)
                  -(\vr\X\vu)\vs(-m-1-\vr\X\vt)\Big)
             \sS_m(\vr)\bigg]\XO\exp{\vr+\vt}
                & if $\vr\X\vt\le1$.} \lb{comm3} \]
These formulas simplify drastically in the special case where
$\L$ is a positive definite even lattice. Obviously,
$\Fz{0}=\C\1$ and the spectrum of $\Lp0$ is nonnegative
so that $\ggL=\Pz1=\Fz1$. Its elements are easy to describe,
\[ \ggL=\{\C\exp\vr\|\vr\in\L_2\}\oplus\{\vs(-1)\|\vs\in\L\}. \]
The commutators become
\[ [\vr(-1),\vs(-1)]&=&0, \lb{pos1} \\[.5em]
   [\vr(-1),\exp{\vs}]&=&(\vr\X\vs)\exp{\vs}, \lb{pos2} \\[.5em]
   [\exp{\vr},\exp{\vs}]
   &=&\cases{0  & if $\vr\X\vs\ge0$, \cr
             \e(\vr,\vs)\exp{\vr+\vs}  & if $\vr\X\vs=-1$, \cr
             -\vr(-1)  & if $\vr\X\vs=-2$.} \lb{pos3} \]
Note that in this special case the Schwarz inequality yields
$|\vr\X\vs|\le2$. Moreover, $\vr\X\vs=-1\iff\vr+\vs\in\L_2$ and
$\vr\X\vs=-2\iff\vr+\vs=0$ for $\vr,\vs\in\L_2$.

We are not interested here in this well-understood case of positive
definite $\L$ which leads to a finite-dimensional Lie algebra $\ggL$,
but rather the case of Lorentzian lattices, which is of course
far more complicated.

We have seen that a special role is played by the norm 2 vectors of
$\L$ which we call {\bf real roots} of the lattice. The {\bf
reflection} $\ew_\vr$ associated with a real root $\vr$ is defined
as $\ew_\vr(\vx)=\vx-(\vx\X\vr)\vr$ for $\vx\in\L$. It is easy to see
that a reflection in a real root is an automorphism of the lattice.
The hyperplanes perpendicular to these real roots divide the real
vector space $\R\XO\L$ into regions called {\bf Weyl chambers}. The
reflections in the real roots of $\L$ generate a group called the
{\bf Weyl group} $\eW$ of $\L$, which acts simply transitively on the
Weyl chambers of $\L$. This means that if we fix one Weyl chamber
$\eC$ then any real root from the interior of another Weyl chamber
can be transported via Weyl reflection to a unique real root in $\eC$.
The real roots $\vr_i$ that are perpendicular to the faces of $\eC$
and have inner product at most 0 with the elements of $\eC$ are
called the {\bf simple roots} of $\eC$. The {\bf Coxeter-Dynkin
diagram} $\eG$ of $\eC$ is the set of simple roots of $\eC$, drawn as
a graph with one vertex for each simple root of $\eC$ and two vertices
corresponding to the distinct roots $\vr_i, \vr_j$ are joined by
$-\vr_i\X\vr_j$ lines.\footnote{Let us denote the group of graph
automorphisms of the Coxeter-Dynkin diagram by $\Aut(\eG)$. Note that
an automorphism $\s\in\Aut(\eG)$ induces an automorphism of $\L$ by
$\s(\vr_i):=\vri{\s(i)}\ \forall i$. Hence $\Aut(\eG)$ may be
identified with the group of automorphisms of $\L$ fixing $\eC$.
Furthermore, one can show that $\s\ew_{\vr_i}\s^{-1}=
\ew_{\vri{\s(i)}}$ and that $\eW\cap\Aut(\eG)=1$. Then the group
of all autochronous automorphisms of the lattice $\L$ is a split
extension of its Weyl group by $\Aut(\eG)$,
\[ 0\longrightarrow\eW\stackrel\iota\longrightarrow
                   \Aut(\L)^+\stackrel\pi\longrightarrow
                   \Aut(\eG)\longrightarrow0,\qquad\img\iota=\ker\pi,
                                                       \nn \]
i.e.\ it is equivalent to a semidirect product of the Weyl group and
the group of graph automorphisms:
\[ \Aut(\L)^+=\eW{\bbl o}\Aut(\eG).  \nn \]
The full automorphism group of $\L$ is just the autochronous subgroup
extended by the negative of the identity operation (which interchanges
the forward and backward light cones). }

Returning to the vertex algebra \vertex\ associated with the even
Lorentzian lattice $\L$, we immediately infer from \Ref{phys0} and
\Ref{phys1} that, for any simple root $\vr_i$, the elements
$\exp{\vr_i}$, $\exp{-\vr_i}$, and $\vr_i(-1)$ describe physical
states, i.e.\ they lie in $\Pz1$. Define generators for a Lie algebra
$\ggA$ by
\[ e_i&\mapsto&\exp{\vr_i}, \\
   f_i&\mapsto&-\exp{-\vr_i}, \\
   h_i&\mapsto&\vr_i(-1). \]
Then, by \Ref{pos1} -- \Ref{pos3}, we find the following relations to
hold:
\[ [h_i,h_j]&=&0, \\ {}
   [h_i,e_j]&=&a_{ij}e_j,\quad [h_i,f_j]=-a_{ij}f_j, \\  {}
   [e_i,f_j]&=&\d_{ij}h_i, \]
where we defined the {\bf Cartan matrix} $A=(a_{ij})$ associated with
$\eC$ by $a_{ij}:=\vr_i\X\vr_j$. The elements $h_i$ obviously
form a basis for an abelian subalgebra of $\ggA$ called the {\bf
Cartan subalgebra} $\ghA$. In technical terms, from above commutators
we learn that the elements $\{e_i,f_i,h_i\|i\}$ generate the so
called free Lie algebra associated with $A$. But even more is true;
for we can show that the {\bf Serre relations}
\[ \ad{e_i}{1-a_{ij}}e_j=0, \quad
   \ad{f_i}{1-a_{ij}}f_j=0, \lb{Serre} \]
are fulfilled for all $i,j$. To see this we recall that $\cF$ is
$\L$-graded by construction,
\[ \cF=\bigoplus_{\vx\in\L}S(\hat{\bf h}^-)\XO\exp{\vx}
      \equiv\bigoplus_{\vx\in\L}\cF^{(\vx)}. \]
Then the Lie algebra of physical states inherits a natural
$\L$-gradation from $\cF$ by defining
\[ \ggL^{(\vx)}:=\ggL\cap\Big[S(\hat{\bf h}^-)\XO\exp{\vx}\Big], \]
so that
\[ [\ggL^{(\vx)},\ggL^{(\vy)}]\subset\ggL^{(\vx+\vy)} \lb{Liegrad} \]
for $\vx,\vy\in\L$.\footnote{Some of the subspaces $\ggL^{(\vx)}$ may
be empty, e.g.\ for $\vx^2>2$.}In particular,
\[ \ad{\exp\vr}j\exp\vs\in\ggL^{(j\vr+\vs)}\quad\forall j\ge0,\
                                          \vr,\vs\in\L_2. \]
{}From \Ref{numberop} we infer that the element $\ad{\exp\vr}j\exp\vs$
has an $\Lp0$ eigenvalue of at least $\frac12(j\vr+\vs)^2=
1+j(j+\vr\X\vs)$. Comparing this with the physical state condition
$\Lp0\p=\p$ we conclude that
\[ \ad{\exp\vr}j\exp\vs=0\FOR{}j\ge1-\vr\X\vs. \]
Having established the Serre relations, the Gabber-Kac theorem
\ctz{Thm.9.11}{Kac90} tells us that the Lie algebra $\ggA$ generated
by the elements $\{e_i,f_i,h_i\|i\}$ is just the {\bf Kac Moody algebra}
associated with the Cartan matrix $A$. Namely, the latter is defined
as above free Lie algebra divided by the maximal ideal intersecting
$\ghA$ trivially, and the theorem states that this maximal ideal is
generated by the elements
$\{\ad{e_i}{1-a_{ij}}e_j,\ \ad{f_i}{1-a_{ij}}f_j\|i\ne j\}$.

We would like to emphasize the remarkable fact that the
physical state condition $\Lp0\p=\p$ accounts for all Serre relations
which are usually very difficult to deal with in the theory of
Kac Moody algebras; or, in string theory language, the absence of
particles with squared mass below the tachyon reflects the validity
of the Serre relations for the Lie algebra $\ggA$.

To summarize (cf.\ \ct{Borc86}): The physical states
$\{\exp{\vr_i},\exp{-\vr_i},\vr_i(-1)\| i\}$ generate via
multiple commutators the Kac Moody algebra $\ggA$ associated with the
Cartan matrix $A=(\vr_i\X\vr_j)$ which is a subalgebra of the Lie
algebra of physical states, $\ggL$. Only in the Euclidean case
these two Lie algebras coincide. In general, we have a {\it proper}
inclusion
\[ \ggA\subset\ggL, \lb{inclus} \]
and the characterization of the
elements of $\ggL$ not contained in the Lie algebra $\ggA$, is the
key problem for the vertex operator construction of hyperbolic
Kac Moody algebras. The special feature of
\Ref{inclus} is that the root system of the Kac Moody algebra $\ggA$
is well understood though its root multiplicities are not completely
known for a single example; whereas the root system of $\ggL$ is
certainly not compatible with that of a Kac Moody algebra although
the root multiplicities are always known. Thus a complete
understanding of \Ref{inclus} requires a ``mechanism'' which tells us
how $\ggA$ has to be filled up with physical states to reach the
complete Lie algebra of physical states.
For the special case of the unique self-dual
Lorentzian lattice $\III$, this was accomplished
in \ct{Borc90} by the addition of imaginary simple roots,
or, equivalently, by adjoining new generators to the Kac Moody
algebra $L_\infty$($=\ggA$, where the infinite matrix $A$
corresponds to the Coxeter-Dynkin diagram built up from the Leech
roots \ct{BoCoQuSl84}), thereby furnishing the transition to
the ``fake monster'' Lie algebra $\gfake$ \ct{Borc90}.
See also \ct{NeSa90} for an attempt to determine the structure
constants of this algebra.

\section{Discrete DDF construction}
As can be seen from eqns.\ \Ref{phys0} -- \Ref{phys2} and \Ref{virstate}
the Virasoro conditions $(\Ln-\d_{n0})\p=0,\ n\ge0,$ which should be
obeyed by physical states $\p$, become increasingly complicated at
higher excitations. In fact, we cannot hope to arrive at a general
description of the physical states by this method of calculating
polarization vectors. However, there is an elegant resolution of this
problem by Del Giudice, Di Vecchia and Fubini \ct{DeDiFu72} which
allows an explicit construction of all the physical excited states.
The idea is to find a set of operators that commute with the Virasoro
operators, and which when applied successively to the tachyonic ground
states give all possible physical states. These operators form a closed
algebra called the {\bf spectrum generating algebra}. It turns out that
the latter consists of transversal DDF operators $A^i_n$,
$1\le i\le d-2$, $n\in\Z$, describing the transversal modes of the
string, and of longitudinal DDF operators $\sL_n$, $n\in\Z$
for the longitudinal excitations.
We shall now introduce the discrete version of these operators taking
into account that the momenta lie on the even lattice $\L$ so that we
are not allowed to use Lorentz transformations to rotate them
into convenient frames. Apparently, the longitudinal DDF operators
have so far not been considered in this discrete context.
\subsection{Realization of DDF operators}
Let $\vk$ be a primitive lightlike lattice vector, i.e., $\vk\in\L_0$
and $\frac1n\vk\notin\L_0\ \forall n>1$. Using \Ref{comm3} we can
immediately write down the commutator of physical states
$\vxi(-1)\exp{m\vk}$ and $\vet(-1)\exp{n\vk}$, $m,n\in\Z$:
\[ [\vxi(-1)\exp{m\vk},\vet(-1)\exp{n\vk}]
    &=& \e(m\vk,n\vk)(\vxi\X\vet)m\vk(-1)\exp{(m+n)\vk} \non
    &=& m(\vxi\X\vet)\d_{m+n,0}\vk(-1), \]
since $\vxi\X\vk=\vet\X\vk=\vk\X\vk=0$ and
$\vk(-1)\exp{n\vk}=\frac1n\Lm1(\exp{n\vk})\equiv0$ for $n\ne0$.
Recall that we assumed the cocycle $\e$ to be bimultiplicative so
that $\e(m\vk,n\vk)=(-1)^{\frac12mn\vk^2}=1$.
We define the {\bf transversal DDF operator} $A^\vxi_m=A_m(\vxi,\vk)$
as the zero mode operator corresponding to the physical state
$\vxi(-1)\exp{m\vk}$,
\[ A^\vxi_m&:=&\left(\vxi(-1)\exp{m\vk}\right)_0 \lb{DDFdef} \\
            &=&\Res{z}{\cV(\vxi(-1)\exp{m\vk},z)} \nn \\
            &=&\Res{z}{\vxi(z)\cV(\exp{m\vk},z)}, \]
where normal ordering in the last line is unnecessary due to
$\vxi\X\vk=0$. According to \Ref{Lie2}
the above commutator then translates into
\[ [A^\vxi_m,A^\vet_n]&=& m(\vxi\X\vet)\d_{m+n,0}(\vk(-1))_0 \non
                      &=& m(\vxi\X\vet)\d_{m+n,0}\vk(0). \]
We observe that apart from the operator $\vk(0)=\vk\X\val_0$, this
is just an oscillator commutation relation like \Ref{osccom} but now
for $d-2$ oscillators since the space
\(\L^\bot(\vk):= \{\vxi\| \vxi\X\vk=0, \,\vxi\equiv\vxi\mod\Z\vk\}\)
has indeed dimension $d-2$. Moreover, it is clear from \Ref{prop17}
that these operators commute with the Virasoro algebra,
\[ [\Ln,A^\vxi_m]=0\qquad\forall n,m\in\Z. \]
Since we shall encounter the DDF operators only when acting on
physical states with certain momentum $\vr$, say, the
operator $\vk(0)$ can be thought of as an integer $\vk\X\vr$.
The crucial feature of the DDF construction is then that for given
momentum $\vr$, one has to find a lightlike vector $\vk=\vk(\vr)$ such
that $\vk\X\vr=1$.
In this case the transversal DDF operators $A^\vxi_m(\vk)$ realize
precisely the algebra of $d-2$ transversal oscillators on the ground
state $\exp\vr$. Indeed, we learn from \Ref{comm2} that the DDF
operators $A^\vxi_m(\vk)$ for positive $m$ annihilate the tachyonic
ground state $\exp\vr$,
\[ A^\vxi_m(\vk)|\vr\>=0\qquad\forall m>0, \]
the operator $A^\vxi_0(\vk)=\vxi(0)$ acts diagonally with eigenvalue
$\vxi\X\vr$, while the operators $A^\vxi_m(\vk)$ for negative $m$
when applied to the ground state generate new physical states called
{\bf transversal DDF states},
\[ A^{\vxi_1}_{-m_1}\ldots A^{\vxi_N}_{-m_N}|\vr\>
   \equiv\Big[\vxi_1(-1)\exp{-m_1\vk},\Big[\ldots,\Big[
          \vxi_N(-1)\exp{-m_N\vk},\exp\vr\Big]\ldots\Big]\Big], \]
where we wrote $\exp\vr\equiv|\vr\>$ to make contact with the
standard physics notation. For later purposes we denote the $d-2$-fold
Heisenberg algebra spanned by the transversal DDF operators by
\[ \hat{\bf t}:=\{A^{\vxi}_m\|\vxi\in\L^\bot(\vk),\ m\in\Z\}, \]
and the vector space of finite products of creation operators
($\equiv$ algebra of polynomials on the transversal oscillators) is
written as
\[ S(\hat{\bf t}^-):=
   \bigoplus_{N\in\N}\bigg\{\prod_{i=1}^N A^{\vxi_i}_{-m_i}\|
     \vxi_i\in\L^\bot(\vk),\ m_i>0\ \forall i \bigg\}, \]
where ``$S$'' stands for ``symmetric'' because of the fact that the
creation operators commute with each other.

The above identification of DDF physical states with multiple
commutators in the Lie algebra $\ggL$ will be our main guide in the
analysis of hyperbolic Lie algebras; for the
DDF construction allows us to write down elements
of the Kac Moody algebra $\ggA$ {\it explicitly} and to
introduce the notion of {\it polarization} into the framework of
these algebras.

Recall that the photonic physical states in \Ref{phys1} deserve the
attribute ``transversal'' in the sense that the polarization vector
$\vs$ in $\vs(-1)\exp{\vr}$ has to be orthogonal to the momentum
vector $\vr$. Thus, we cannot expect to obtain ``longitudinal''
physical states in a straightforward way. Nevertheless, there is
a ``dirty trick'' \ct{Brow72}.
Let $\vr\in\L$, $\vk\in\L_0$ and suppose that $\vk\X\vr\ne0$. Then
eq.\ \Ref{prop3} yields
\[ [\Ln,\cV(\vr(-1)\exp{\vk},z)]
     &=&z^n\left\{z\dz+n+1\right\}\cV(\vr(-1)\exp{\vk},z) \non
     & &{}+\2n(n+1)(\vk\X\vr)\cV(\exp{\vk},z)z^{n-1}. \]
The unwanted term on the right-hand side which destroys the conformal
transformation properties \Ref{prop18} can be removed by the following
trick: introduce the formal series
\[ \vkx:=z\frac{\vk(z)}{\vr\X\vk}-1
        =\frac1{\vr\X\vk}\sum_{n\neq 0}\vk(n)z^{-n}
         +\bigg(\frac{\vk(0)}{\vr\X\vk}-1\bigg), \lb{kstar}  \]
and define
\[ \log\Big(z\frac{\vk(z)}{\vr\X\vk}\Big)
   &=&\log\Big(1+\vkx\Big) \non
   &:=&\sum_{i\ge1}\frac{(-1)^{i+1}}{i}(\vkx)^i, \lb{logkstar} \]
{\it which is only defined on states with momentum $\vs$ such that
$\vs\X\vk=\vr\X\vk$}: if the second term on the right hand side
of \Ref{kstar} does not vanish on a given state, an infinite number
of terms will contribute when \Ref{logkstar} is applied to it.
This means that the above series is not
(algebraically) summable on the whole space $\cF$. In particular,
it is not summable on the vacuum state $\1\equiv|\vo\>$
which, in view of \Ref{prop13}, makes it impossible to recover {\it
the} state corresponding to the $\log$ series: there does not exist
a universal state whose vertex operator is given by $\log(1+\vkx)$.
Luckily, however, we shall only need the action of this $\log$
series on states with momentum $\vr - n \vk$ (with $n\in \N$),
so that the resulting series
will be well-defined. And if this is the case then we may indeed find
a state whose vertex operator has the same action as the $\log$ series.
Thus, the $\log$ series should be interpreted as some sort of generating
series for a class of genuine vertex operators which can be revealed
by acting on states. Keeping in mind this subtlety let us perform some
calculations in connection with the $\log$ series.
\[ [\Ln,\vkx]
     &=&z^n\left\{z\dz+n\right\}\vkx+nz^n, \]
since the current $\vk(z)$ is a primary field of weight 1. For the
formal series $\log\Big(1+\vkx\Big)$ we therefore obtain
\[ \bigg[\Ln,\log\Big(1+\vkx\Big)\bigg]
     &=&\sum_{i\ge1}(-1)^{i+1}(\vkx)^{i-1}[\Ln,\vkx]
              \qquad\mbox{since $\vk\in\L_0$} \non
     &=&z^{n+1}\dz\log\Big(1+\vkx\Big)+nz^n, \]
so that
\[ \bigg[\Ln,\dz\log\Big(1+\vkx\Big)\bigg]
    =z^n\left\{z\dz+n+1\right\}\dz\log\Big(1+\vkx\Big)+n^2z^{n-1}. \]
Using this formula and the fact that
\[ \dz\log\frac{\vk(z)}{\vr\X\vk}=\dz\log\Big(1+\vkx\Big)-z^{-1}, \]
we deduce that
\[ \bigg[\Ln,\dz\log\frac{\vk(z)}{\vr\X\vk}\bigg]
      =z^n\left\{z\dz+n+1\right\}
       \bigg(\dz\log\frac{\vk(z)}{\vr\X\vk}\bigg)+n(n+1)z^{n-1}. \]
Putting everything together we conclude that the {\bf DDF vertex
operators}
\[ \cY_{\vk}(\vr,z):=
        \cV(\vr(-1)\exp{\vk},z)-\2(\vr\X\vk)
     \dz\log\bigg(\frac{\vk(z)}{\vr\X\vk}\bigg)\cV(\exp{\vk},z),
                                                  \lb{longvert}  \]
enjoy the correct conformal transformation properties for primary
fields of weight 1:\footnote{ This is in perfect agreement with
\ct{Brow72} since we employ a different normal ordering prescription
for $p^\m$ and $q^\n$; we use $\:q^\n p^\m\:=q^\n p^\m$ in contrast
to the ``standard'' symmetric normal ordering $\Ord q^\n{p^\m}\Ord=
\frac12(q^\n p^\m+p^\m q^\n)= \:q^\n p^\m\:- \frac i2\y^{\m\n}$ which
leads to
\[ \cV_{\rm symm.}(\vr(-1)\exp{\vk},z)=
   \cV(\vr(-1)\exp{\vk},z)+\frc12(\vk\X\vr)\cV(\exp{\vk},z)z^{-1},
   \nn \]
so that indeed
\[ \cV_{\rm symm.}(\vr(-1)\exp{\vk},z)-\2(\vk\X\vr)
        \dz\log\Big(1+\vkx\Big)\cV(\exp{\vk},z)=
        \cV(\vr(-1)\exp{\vk},z)-\2(\vk\X\vr)
        \dz\log\bigg(\frac{\vk(z)}{\vr\X\vk}\bigg)\cV(\exp{\vk},z).
   \nn \]}
\[ [\Ln,\cY_{\vk}(\vr,z)]
   =z^n\left\{z\dz+n+1\right\}\cY_{\vk}(\vr,z)\qquad\forall n\in\Z. \]
For $\vr\X\vk\ne0$ we call $\cY_{\vk}(\vr,z)$ {\bf longitudinal vertex
operator} since otherwise we recover the transversal vertex operator
$\cV(\vr(-1)\exp{\vk},z)$.\footnote{Apparently, the essential
$\log$ term was missed in \ct{Fren85}.} Also
note that the $\log$ term in \Ref{longvert} does not require normal
ordering because of $\vk\in\L_0$. We define the {\bf longitudinal
Virasoro operator} $\sL_m$ as the zero mode operator of the
longitudinal vertex operator $\cY_{m\vk}(\vr,z)$,
\[ \sL_m&:=&-\Res{z}{\cY_{m\vk}(\vr,z)} \non
        &\equiv&\Res{z}{-\cV(\vr(-1)\exp{m\vk},z)+\frac{m}{2}(\vr\X\vk)
     \dz\log\bigg(\frac{\vk(z)}{\vr\X\vk}\bigg)\cV(\exp{m\vk},z)}. \]
These operators satisfy the commutation relations of a
Virasoro algebra with central charge $c=24$. To see this, we first
note that
\[ [\vr(-1)\exp{m\vk},\vr(-1)\exp{n\vk}]
   &=&\e(m\vk,n\vk)
\bigg[m(\vr^2-mn(\vr\X\vk))\vk(-1)+(n-m)(\vr\X\vk)\vr(-1)\bigg]
      \exp{(m+n)\vk} \non
   &=&(n-m)(\vr\X\vk)\vr(-1)\exp{(m+n)\vk}
 +m(\vr^2 +m^2(\vr\X\vk))\d_{m+n,0}\vk(-1) \]
by \Ref{comm3} so that
\leqn0{\left[\Res{z_1}{\cV(\vr(-1)\exp{m\vk},z_1)}\,,\,
         \Res{z_2}{\cV(\vr(-1)\exp{n\vk},z_2)}\right]}\vspace{-3ex}
\[  =(m-n)(\vr\X\vk)\Res{z}{-\cV(\vr(-1)\exp{(m+n)\vk},z)}
 +m(\vr^2+m^2(\vr\X\vk))\d_{m+n,0}\vk(0). \]
It is also clear that the commutator of two $\log$ terms vanishes due
to lightlikeness of $\vk$. Finally, we have to calculate the cross
commutator:
\leqn0{\left[\Res{z_1}{\cV(\vr(-1)\exp{m\vk},z_1)}\,,\,
         \res_{z_2}\bigg[\frac{d}{dz_2}
                         \log\bigg(\frac{\vk(z_2)}{\vr\X\vk}\bigg)
                         \cV(\exp{n\vk},z_2)\bigg]\right]}\vspace{-3ex}
\[ &=&\res_{z_2}\left\{
      \frac{d}{dz_2}\log\bigg(\frac{\vk(z_2)}{\vr\X\vk}\bigg)
      \bigg[\Res{z_1}{\cV(\vr(-1)\exp{m\vk},z_1)}\,,\,
                      \cV(\exp{n\vk},z_2)\bigg]+\right. \non
   & &\phantom{\res_{z_2}\Bigg\}} \left.+
      \bigg[\Res{z_1}{\cV(\vr(-1)\exp{m\vk},z_1)}\,,\,
      \frac{d}{dz_2}\log\bigg(\frac{\vk(z_2)}{\vr\X\vk}\bigg)\bigg]
                      \cV(\exp{n\vk},z_2)\right\}. \]
To calculate these two commutators we first recall the following version
of the commutator formula \Ref{comform}:
\[ \Big[\Res{z_1}{\VP1},\VF2\Big]&=&\cV(\p_0\f,z_2) \non
                         &\equiv&\cV([\p,\f],z_2). \lb{comform2} \]
{}From \Ref{comm2} and \Ref{comm3} we therefore deduce that
\[ \bigg[\Res{z_1}{\cV(\vr(-1)\exp{m\vk},z_1)}\,,\,
         \cV(\exp{n\vk},z_2)\bigg]
   =n(\vr\X\vk)\cV(\exp{(m+n)\vk},z_2), \]
and
\[ \bigg[\Res{z_1}{\cV(\vr(-1)\exp{m\vk},z_1)}\,,\,
         \cV(\vk(-1),z_2)\bigg]
   =m(\vr\X\vk)\cV(\vk(-1)\exp{m\vk},z_2), \]
respectively. The last formula then yields
\leqn0{\bigg[\Res{z_1}{\cV(\vr(-1)\exp{m\vk},z_1)}\,,\,
       \frac{d}{dz_2}\log\bigg(\frac{\vk(z_2)}{\vr\X\vk}\bigg)\bigg]}
\vspace{-3ex}
\[ &=&\frac{d}{dz_2}\left[\sum_{i\ge1}(-1)^{i+1}
                     \bigg(\frac{\vk(z_2)}{\vr\X\vk}-1\bigg)^{i-1}
                     m(\vr\X\vk)\cV(\vk(-1)\exp{m\vk},z_2)\right] \non
   &=&\frac{d}{dz_2}\left[m(\vr\X\vk)\cV(\exp{m\vk},z_2)\right] \non
   &=&m^2(\vr\X\vk)\cV(\vk(-1)\exp{m\vk},z_2). \]
Collecting above commutators we finally get
\[ [\sL_m,\sL_n]
   &=&(m-n)(\vr\X\vk)\Res{z}{-\cV(\vr(-1)\exp{(m+n)\vk},z)}
      +m(\vr^2+m^2(\vr\X\vk))\d_{m+n,0}\vk(0) \non
   & &{}-\frac{n^2}{2}(\vr\X\vk)^2\Res{z}
         {\frac{d}{dz}\log\bigg(\frac{\vk(z)}{\vr\X\vk}\bigg)
         \cV(\exp{(m+n)\vk},z)} \non
   & &{}-\frac{nm^2}{2}(\vr\X\vk)^2\Res{z}
         {\cV(\vk(-1)\exp{(m+n)\vk},z)} \non
   & &{}+\frac{m^2}{2}(\vr\X\vk)^2\Res{z}
         {\frac{d}{dz}\log\bigg(\frac{\vk(z)}{\vr\X\vk}\bigg)
         \cV(\exp{(m+n)\vk},z)} \non
   & &{}+\frac{mn^2}{2}(\vr\X\vk)^2\Res{z}
         {\cV(\vk(-1)\exp{(m+n)\vk},z)} \non
   &=&(m-n)\sL_{m+n}
+[((\vr\X\vk)^2+(\vr\X\vk))m^3+\vr^2m]\d_{m+n,0}\vk(0). \]
As for the central term, we shall assume from now on that
$\vr\X\vk=1$ so that the factor in the central term reads
$2m^3+\vr^2m$.
The standard form $\frac{c}{12}(m^3-m)$ can be obtained by redefining
\(\sL_0\to \sL_0+ (1+\frac12 \vr^2)
\vk(0)\) so that we end up with
\[ [\sL_m,\sL_n]=(m-n)\sL_{m+n}+2(m^3-m)\d_{m+n,0}\vk(0). \]
We conclude that the longitudinal Virasoro operators $\sL_m$, when
applied to physical states with momentum $\vr$, realize
a Virasoro algebra, $\Vir_\sL$, with central charge $c_\sL=24$.
Remarkably, this Virasoro algebra is universal in the sense that its
central charge does {\it not} depend on the dimension of the lattice.

Let us proceed by determining the commutator of the transversal DDF
operators and the longitudinal Virasoro operators.
\[ [\sL_m,A^\vxi_n]
   &=&\bigg[\Res{z_1}{\cV(\vxi(-1)\exp{n\vk},z_1)}\,,\,
            \Res{z_2}{\cV(\vr(-1)\exp{m\vk},z_2)}\bigg] \non
   &=&\Res{z}{\cV([\vxi(-1)\exp{n\vk},\vr(-1)\exp{m\vk}],z)} \non
   &=&\Res{z}{n(\vxi\X\vr)\cV(\vk(-1)\exp{(m+n)\vk},z)
              -n(\vr\X\vk)\cV(\vxi(-1)\exp{(m+n)\vk},z)} \non
   &=&-n(\vr\X\vk)A^\vxi_{n+m}+n(\vr\X\vxi)\d_{m+n,0}\vk(0) \]
by \Ref{comm3}. Obviously, we can remove the second term by choosing
$\vxi$ orthogonal to $\vr$; and if we make our standard assumption that
$\vr\X\vk=1$ we arrive at the important formula
\[ [\sL_m,A^\vxi_n]=-nA^\vxi_{n+m}. \lb{semiprod} \]

We claim that the tachyonic ground state $\exp\vr$ is annihilated by
the longitudinal Virasoro operators $\sL_m$ for nonnegative $m$,
\[ \sL_m|\vr\>=0\qquad\forall m\ge0. \]
First note that the operator
$\sL_0=-\vr(0)+(1+\frac12\vr^2)\vk(0)$ acts diagonally
with eigenvalue $(1-\frac12\vr^2)$ which indeed vanishes
because $\vr\in\L_2$. Next, using the $\L$-gradation \Ref{Liegrad}
of $\ggL$ we observe that the state $\sL_m|\vr\>$ carries momentum
$\vr+m\vk$. But $\frac12(\vr+m\vk)^2=1+m$ contradicts the physical
state condition $\Lp0\p=\p$ for positive $m$ in view of
\Ref{numberop} unless the state itself vanishes. We conclude that
only the operators $\sL_m$ for negative $m$ generate new physical
states when applied to the ground state
\[ \sL_{-n_1}\ldots \sL_{-n_P}|\vr\>\quad\in\Pz1 \lb{longstate} \]
for $n_1,\ldots,n_P\ge1$. Further, we can
verify that the state $\sL_{-1}|\vr\>$ is a null physical state, i.e.
the action of the operator $\sL_{-1}$ is essentially the same as
the action of $\Lm1$:
\[ \sL_{-1}|\vr\>=\e(\vr,\vk)\Lm1|\vr-\vk\>, \lb{longzero} \]
which vanishes as an element of $\ggL$! To prove this equation we first
deduce from \Ref{comm2} that
\[ \Res{z}{-\cV(\vr(-1)\exp{-\vk},z)}(\exp\vr)
   &=&-[\vr(-1)\exp{-\vk},\exp\vr] \non
   &=&\e(\vr,\vk)(\vr-2\vk)(-1)\XO\exp{\vr-\vk}. \]
The calculations for the $\log$ term have to be performed explicitly:
\leqn0{\Res{z}{-\2\dz\log\vk(z)\cV(\exp{-\vk},z)}(\exp\vr)}\vspace{-3ex}
\[ &=&\Res{z}{-\2\dz\log\vk(z)\cV(\exp{-\vk},z)}(\exp\vr) \non
   &=&-\2\Res{z}{\Big[\dz\log\Big(1+\vkx\Big)-z^{-1}\Big]
                 \cV(\exp{-\vk},z)}(\exp\vr) \non
   &=&-\2\Res{z}{\exp{-\vk}\sum_{i\ge1}\frac{(-1)^{i+1}}{i}(\vkx)^i
                 \vk(z)\sum_{m\ge0}\sS_m(-\vk)z^{m-1}}(\exp\vr) \non
   & &{}+\2\Res{z}{z^{-1}\sum_{m\ge0}\sS_m(-\vk)z^{m-1}\exp{-\vk}
                   \exp{\vr}} \non
   &=&-\2\Res{z}{\sum_{i\ge1}\frac{(-1)^{i+1}}{i}
                    \bigg(\sum_{n>0}\vk(-n)z^n\bigg)^i
                 \bigg(z^{-1}+\sum_{n>0}\vk(-n)z^{n-1}\bigg)
                 \sum_{m\ge0}\sS_m(-\vk)z^{m-1}\exp{-\vk}\exp{\vr}} \non
   & &{}-\2\e(-\vk,\vr)\vk(-1)\exp{\vr-\vk} \non
   &=&-\e(-\vk,\vr)\vk(-1)\exp{\vr-\vk}. \]
Putting together above two results we obtain
$\e(\vr,\vk)(\vr-\vk)(-1)|\vr-\vk\>$ as desired. Thus, using the
commutation relations and \Ref{longzero}, we can rewrite any
state of the form \Ref{longstate} as a linear combination of states
not containing $\sL_{-1}$. As a basis for states of the form
\Ref{longstate} in $\ggL$ we may therefore choose
\[ \sL_{-n_1}\ldots \sL_{-n_P}|\vr\>, \lb{longspan} \]
with fixed ordering $n_1\ge\ldots\ge n_P\ge2$.

We turn now to the no-ghost theorem applied to our discrete
construction.
We fix a tachyonic groundstate $\exp\vr\equiv|\vr\>$, $\vr\in\L_2$,
and suppose that there exists a lightlike vector $\vk=\vk(\vr)\in\L_0$
such that $\vr\X\vk=1$. Then we can always find $d-2$ orthonormal
lattice vectors $\vxi_i$, $1\le i\le d-2$, orthogonal to both $\vr$
and $\vk$. If we put $A^i_m\equiv A^{\vxi_i}_m$ then the no-ghost
theorem \ct{GodTho72} tells us that the states
\[ A^{i_1}_{-m_1}\ldots A^{i_N}_{-m_N}
   \sL_{-n_1}\ldots \sL_{-n_P}|\vr\> \lb{physspan} \]
for $i_1,\ldots,i_N\in\{1,\ldots,d-2\}$, $m_1,\ldots,m_N\ge1$ and
$n_1\ge\ldots\ge n_P\ge1$,
account for {\it all} physical states (including null physical states!)
with momentum
\[ \vr-\bigg(\sum_{a=1}^Nm_a+\sum_{b=1}^Pn_b\bigg)\vk. \]
Reformulated in the language of the Lie algebra $\ggL$, the subspace
\[ \ggL(\vr):=\bigoplus_{n\in\N}\ggL^{(\vr-n\vk)},\qquad\vr\in\L_2, \]
is spanned by elements of the form
\[ A^{i_1}_{-m_1}\ldots A^{i_N}_{-m_N}
   \sL_{-n_1}\ldots \sL_{-n_P}|\vr\>, \]
where $i_1,\ldots,i_N\in\{1,\ldots,d-2\}$, $m_1,\ldots,m_N\ge1$ and
$n_1\ge\ldots\ge n_P\ge2$.

Note that due to \Ref{semiprod} we had to fix some ordering of the
operators in \Ref{physspan}. Historically, this was the reason for
replacing the longitudinal Virasoro operators $\sL_m$ by {\bf
longitudinal DDF operators} $A^-_m$ which commute with the transversal
DDF operators. To see this, we define the standard normal
ordering of the transversal DDF operators by
\[ \:A^i_mA^j_n\::=\cases{A^i_mA^j_n & if $m\le n$, \cr
                          A^j_nA^i_m & if $m>n$, \cr} \lb{order} \]
and define
\[ \eL_n:=\2\sum_{i=1}^{d-2}\sum_{m\in\Z}\:A^i_mA^i_{n-m}\:. \]
Comparing this with \Ref{Virosc} we immediately infer that the
$\eL_n$'s obey a Virasoro algebra, $\Vir_\eL$, with central charge
$c_\eL=d-2$.
Furthermore, it is straightforward to show that
\[ [\eL_m,A^i_n]=-nA^i_{n+m}. \]
Hence, if we define
\[ A^-_n:=\sL_n-\eL_n=
          \sL_n-\2\sum_{i=1}^{d-2}\sum_{m\in\Z}\:A^i_mA^i_{n-m}\:, \]
we get
\[ [A^i_m,A^-_n]=0\qquad\forall m,n\in\Z,\ 1\le i\le d-2. \]
The last equation can be used to show that longitudinal DDF operators
form a ``coset'' Virasoro algebra, $\Vir_{A^-}$, with central charge
$c_{A^-}=c_\sL-c_\eL=26-d$:
\[ [A^-_m,A^-_n]&=&[\sL_m-\eL_m,\sL_n-\eL_n] \non
                &=&[\sL_m,\sL_n]-[\eL_m,\eL_n] \non
                &=&(m-n)A^-_{m+n}+\frc{26-d}{12}(m^3-m)\d_{m+n,0}. \]
Thus we may rewrite the basis of all physical states (including null
states!) as
\[ A^{i_1}_{-m_1}\ldots A^{i_N}_{-m_N}
   A^-_{-n_1}\ldots A^-_{-n_P}|\vr\> \lb{DDFspan} \]
where $i_1,\ldots,i_N\in\{1,\ldots,d-2\}$, $m_1,\ldots,m_N\ge1$
and $n_1\ge\ldots\ge n_P\ge1$, which exhibits explicitly how the
space of physical states with momentum $\vr-n\vk$, $n\ge0$, splits
into a tensor product of the algebra of polynomials in the transversal
oscillators with a Virasoro Verma module:
\[ \Pz1(\vr):=\bigoplus_{n\ge0}\Pz1^{(\vr-n\vk)}
            =S(\hat{\bf t}^-)\XO V(26-d,0), \]
where $V(c,h)$ denotes the irreducible highest weight $\Vir$-module.
In other words, we may regard the associative algebra
\[ S(\hat{\bf t}^-)\XO\Vir_{A^-}, \]
as the {\bf spectrum generating algebra} associated with $\vr$,
since it generates all physical states with momentum $\vr-n\vk$,
$n\in\N$, by acting on the fixed tachyonic groundstate $|\vr\>$.
In particular, we observe how the critical dimension $d=26$ arises:
in $26$ dimensions the longitudinal and the transversal modes decouple
because the coset Virasoro module $V(26-d,0)$ becomes trivial.
Moreover, \Ref{DDFspan} enables us to write down a formula for
the dimension of the physical subspaces with momentum $\vr-n\vk$,
$\vr\in\L_2$:
\[ \dim\ggL^{(\vr-n\vk)}=p_{d-1}(n)-p_{d-1}(n-1), \lb{physdim} \]
where $p_{d-1}(n)$ counts the partitions of $n\in\N$ into ``parts'' of
$d-1$ ``colours'' , i.e.\
\[ \phi(q)^{1-d}&:=&\prod_{l=1}^\infty(1-q^l)^{1-d} \non
            &=&\sum_{n\in\N}p_{d-1}(n)q^n \non
            &=&1+(d-1)q+\2(d-1)(d+2)q^2+\frc16(d-1)d(d+7)q^3+\ldots,
                                                          \lb{Euler} \]
in terms of the generating Euler function $\phi(q)$. Hence
\[ p_{d-1}(n)-p_{d-1}(n-1)&=&\sum_{l=0}^np_{d-2}(l)p_1(n-l)
                            -\sum_{l=0}^{n-1}p_{d-2}(l)p_1(n-l-1) \non
                          &=&p_{d-2}(n)+\D_{d-2}(n), \]
with $\D_{d-2}(n):=\sum_{l=0}^{n-1}p_{d-2}(l)[p_1(n-l)-p_1(n-l-1)]$;
for example:
\[ \D_{d-2}(1)&=&0, \non
   \D_{d-2}(2)&=&1, \non
   \D_{d-2}(3)&=&d-1, \non
   \D_{d-2}(4)&=&\2(d-1)(d+2). \nn \]
Explicitly,
\[ \sum_{n=0}^\infty\dim\ggL^{(\vr-n\vk)}q^n
    =1+(d-2)q+\2(d-1)dq^2+\frc16(d-1)(d^2+4d-6)q^3+\ldots \]
Note that the second term in \Ref{physdim} is due to the null physical
states.

Since we will mainly focus on a deeper understanding
of the Kac Moody algebra $\ggA$ the question arises how to make
contact between the elegant DDF formulation of $\ggL$ and the
construction of $\ggA$ in terms of generators and relations. In other
words, we have to face the problem how to separate the DDF states
contained in $\ggA$ from those which cannot be generated by the set
$\{e_i,f_i,h_i\|i\}$. Note that a special case of \Ref{comm1} gives
us a recipe for writing physical states $\vxi(-1)\exp\vr$ as Lie algebra
commutators:
\[ [\exp\vs,\exp\vt]&=&\e(\vs,\vt)\vs(-1)\exp{\vs+\vt} \non
                    &=&\2\e(\vs,\vt)(\vs-\vt)(-1)\exp{\vs+\vt}
                                                     \lb{tachcomm} \]
for $\vs,\vt\in\L_2$ such that $\vs\X\vt=-2$. The last equality
is obtained by adding the null physical state (``total derivative'')
$\frac12\e(\vs,\vt)\Lm1\exp{\vs+\vt}$. Hence we may put $\vxi=\vs-\vt$
and $\vr=\vs+\vt$. This observation will be useful later.

We conclude with a comment that will be crucial for
the discrete DDF construction of $\0$. So far, we have tacitly
assumed the DDF vectors $\vk$ and $\vr$ to be elements of the root
lattice. However, inspection of the computations presented above
shows that all arguments remain valid if only $\vk^2=0,
\vr^2=2, \vr\X\vk=1$ and $\vxi\X\vr= \vxi\X\vk= 0$. Thus, there
is actually no need to assume the
vectors $\vk$ and $\vr$ to be on the root lattice
as long as these conditions are satisfied. In particular,
under these circumstances we may choose $\vk$ and $\vr$ on
the rational extension $\Q\XO\L$, and the discrete
DDF construction still works. All our formulas will continue to
make sense, whereas the interpretation of physical states and the
identification of Lie algebra elements need some care. This subtlety
arises because, rigorously speaking, we are dealing with a
{\bf generalized vertex algebra} associated with $\Q\XO\L$,
into which the original vertex algebra (associated with $\L$) can be
embedded. The generalized vertex operators are then defined as
in \Ref{vertexop1} and \Ref{vertexop2}, but are no longer elements
of $(\End\cF)[\![z,z^{-1}]\!]$; rather, the generalized vertex
operator associated with a typical homogeneous element
$\p=\big[\prod_{j=1}^N\vs_j(-n_j)\big]\XO\exp\vr$ (where now
$\vr\in\Q\XO\L$) is an element of
of $(\End\cD_\vr)[\![z,z^{-1}]\!]$ with
\[ \cD_\vr:=\bigoplus_{\vs\in\Q\XO\L \atop \vr\.\vs\in\Z}\cF^{(\vs)}. \]
This means that the modes of the generalized vertex operators
are {\it not} well defined operators on the whole Fock space $\cF$,
but only on certain of its subspaces.

\section{The Hyperbolic Algebra $\0$ and the DDF Construction}
We now want to apply the concepts developed in the foregoing
chapters to the study of Kac Moody algebras $\ggA$ whose Cartan
matrix $A$ is of hyperbolic type, choosing the hyperbolic Kac Moody
algebra $\0$ as our principal example. We remind the reader
that hyperbolic algebras are distinguished from the more general
algebras based on arbitrary indefinite Cartan matrices by the
additional requirement that the removal of any point from the
Dynkin diagram leaves a Kac Moody algebra which is either of affine
or finite type (for a review of hyperbolic root systems,
see \ct{Mood79}). As shown in
\ct{Kac90}, the rank can then be 10 at most, and the root lattice
must be Minkowskian, i.e.\ with metric signature $(+\ldots+|-)$. There
are altogether three hyperbolic algebras of maximal rank. Of these,
$\0$ is not only the most interesting, containing $E_8$ and its affine
extension $\9$ as subalgebras, but also distinguished by the
fact that it has only {\it one} affine subalgebra that can be obtained
by removing a point from the $\0$ Dynkin diagram, while the other
two rank 10 algebras contain at least two regular affine subalgebras,
(see e.g. \ct{Sacl89}). Furthermore, the root lattice $Q(\0 )$
coincides with the (unique) 10-dimensional even unimodular Lorentzian
lattice $\II$ \ct{Con83}, whereas the root lattices of the
other two maximal rank hyperbolic algebras are not self-dual.

Overall, our knowledge about Kac Moody algebras of
hyperbolic type is rather limited. As already explained in
section 2.2, they are generally defined in terms of multiple
commutators of the basic generators $e_i, f_i, h_i$ and the
multilinear Serre relations \Ref{Serre}. In contradistinction to the
finite and affine cases, a manageable representation of all the Lie
algebra elements obtained in this way has not yet been found.
In principle, the string vertex operator construction
provides a more concrete realization with the additional advantage
that the Serre relations \Ref{Serre} are built in
from the outset (see the discussion at the end of section 2.2),
but the problem of characterizing the missing elements
belonging to $\ggL$ and not to $\ggA$ in \Ref{inclus} remains.
We emphasize that we face essentially the same problem
if instead we want to define a Borcherds-type algebra \ct{Borc92}
based on $\II$, because we then would have to supply the
missing generators ``by hand'' by adding extra imaginary simple
roots, which again presupposes knowledge of what
the missing Lie algebra elements are (not to mention the
potential arbitrariness as to the number of ways in
which this can be consistently done).

As already mentioned, our analysis makes use of a discretized version
of the DDF construction and relies in a crucial way on the
identification of Lie algebra elements with physical Fock space states.
In the previous section we have seen that a central role is played
the tachyon momentum $\va$ of the ground state (so
${\va}^2 =2$)  and the null vector $\vk$, subject to the condition
$\vk \X \va = 1$. For continous momenta $\va$, we can always
find suitable $\vk = \vk (\va )$; moreover, we can rotate
these vectors into a convenient frame by means of a Lorentz
transformation \ct{Scher75}. On the lattice, however, the
full Lorentz invariance is broken to a discrete subgroup (containing
the Weyl group generated by the fundamental Weyl reflections), and
for generic roots $\vL$, the associated DDF vectors $\va$ and
$\vk$ will {\it not} be elements of the root lattice $\II$ in
general\footnote{To make this explicit in the notation, we designate
the tachyon momentum by $\va$ rather than $\vr$ as in the previous
sections.}.
Nevertheless, we employ these vectors in our analysis
because we can still use the associated (transversal and
longitudinal) DDF operators to construct a complete basis for any
root space of the Lie algebra of physical states $\ggI$.
The corresponding root space of the Kac Moody algebra $\ggA$ is then
a proper subspace thereof.
As we will see, longitudinal states are absent only for
level zero and level one; this accounts for the
comparative simplicity of the corresponding multiplicity formulas.

Although it is possible in principle (with some effort) to extend
our discussion to other hyperbolic Kac Moody algebras, the following
points must be kept in mind. Our method may not apply to
strictly hyperbolic algebras, which by definition have no affine, but
only finite subalgebras, because their associated Weyl chambers contain
no null vectors (i.e. they lie entirely within the light-cone),
so the DDF operators cannot be defined.
On the other hand, the Weyl chambers of arbitrary Kac Moody
algebras of indefinite type generically contain several linearly
independent null directions, a feature that will
greatly complicate (if not vitiate) the application of our method,
because one must then deal with at least two different sets of photon
momenta for the DDF operators. Moreover, if the algebra contains
more than one regular affine subalgebra, the level of a root is
no longer uniquely defined; for indefinite algebras, which are
not hyperbolic (such as the fake monster), it is not even clear
whether this notion can be sensibly defined at all. We thus begin to
understand the possible importance of the fact that the fundamental
Weyl chamber of $\0$ touches the light-cone at precisely one edge.
In view of the limitations of the method, we will make no attempt
to state the results in the most general way.

In subsection 4.1 we will summarize the pertinent results about
$\0$. In section 4.2, we apply the discrete DDF construction
to level-zero and level-one elements of $\ggA$,
thereby recovering some known results. In section 4.3, we turn
to the level-two states, analyzing one example in complete detail.

\subsection{Basic results about $\0$ }
The hyperbolic Kac Moody algebra $\0$ is defined via its
Coxeter-Dynkin diagram and the Serre relations following from it.
As already mentioned,
the root lattice $Q(\0)$ coincides with the
unique 10-dimensional even unimodular Lorentzian lattice $\II$.
The latter can be defined as the lattice of all points
$\vx=(x_1,\ldots,x_{9}|x_0)$ for which the $x_\m$'s are all in $\Z$ or
all in $\Z+\frac12$ and which have integer inner product with the
vector $\vl=(\frac12,\ldots,\frac12\|\frac12)$, all norms and inner
products being evaluated in the Minkowskian metric
$\vx^2=x_1^2+\ldots+x_9^2-x_0^2$ (cf.\ \ct{Serr73}).
In more physical parlance, we are dealing with a subcritical
open bosonic string moving in 10-dimensional space-time fully
compactified on a torus (hence ``finite in all directions'' \ct{Moor93}
), so that the momenta lie on $\II$.
According to \ct{Con83}, a set of positive norm simple
roots for $\II$ is given by the ten vectors $\vri{-1}, \vri0, \vri1,
\ldots, \vri8$ in $\II$ for which $r_i^2=2$ and $\vr_i\X\vro=-1$
where the {\bf Weyl vector} is $\vro=(0,1,2,\ldots,8|38)$ with
$\vr^2=-1240$. Explicitly,
$$ \begin{array}{l@{\quad=\quad(}r@,r@,r@,r@,r@,r@,r@,r@,r@{\|}l}
   \vri{-1} & 0 & 0 & 0 & 0 & 0 & 0 & 0 & 1 & -1 & 0),  \non
   \vri{ 0} & 0 & 0 & 0 & 0 & 0 & 0 & 1 & -1 & 0 & 0),  \non
   \vri{ 1} & 0 & 0 & 0 & 0 & 0 & 1 & -1 & 0 & 0 & 0),  \non
   \vri{ 2} & 0 & 0 & 0 & 0 & 1 & -1 & 0 & 0 & 0 & 0),  \non
   \vri{ 3} & 0 & 0 & 0 & 1 & -1 & 0 & 0 & 0 & 0 & 0),  \non
   \vri{ 4} & 0 & 0 & 1 & -1 & 0 & 0 & 0 & 0 & 0 & 0),  \non
   \vri{ 5} & 0 & 1 & -1 & 0 & 0 & 0 & 0 & 0 & 0 & 0),  \non
   \vri{ 6} & -1 & -1 & 0 & 0 & 0 & 0 & 0 & 0 & 0 & 0),  \non
   \vri{ 7} & \2 & \2 & \2 & \2 & \2 & \2 & \2 & \2 & \2 & \2),  \non
   \vri{ 8} & 1 & -1 & 0 & 0 & 0 & 0 & 0 & 0 & 0 & 0). \nn
\end{array} $$
These simple roots indeed generate the reflection group of $\II$.
The corresponding Coxeter-Dynkin diagram looks as follows
\[ \unitlength1mm
   \begin{picture}(66,10)
   \multiput(1,0)(8,0){9}{\circle*{2}}
   \put(49,8){\circle*{2}}
   \multiput(2,0)(8,0){8}{\line(1,0){6}}
   \put(49,1){\line(0,1){6}}
   \end{picture}   \]
and is associated with the Cartan matrix:
$$ A\equiv (a_{ij})=(\vr_i\X\vr_j)=\left(
   \begin{array}{rrrrrrrrrr}
              2 &-1 & 0 & 0 & 0 & 0 & 0 & 0 & 0 & 0  \non
             -1 & 2 &-1 & 0 & 0 & 0 & 0 & 0 & 0 & 0  \non
              0 &-1 & 2 &-1 & 0 & 0 & 0 & 0 & 0 & 0  \non
              0 & 0 &-1 & 2 &-1 & 0 & 0 & 0 & 0 & 0  \non
              0 & 0 & 0 &-1 & 2 &-1 & 0 & 0 & 0 & 0  \non
              0 & 0 & 0 & 0 &-1 & 2 &-1 & 0 & 0 & 0  \non
              0 & 0 & 0 & 0 & 0 &-1 & 2 &-1 & 0 &-1  \non
              0 & 0 & 0 & 0 & 0 & 0 &-1 & 2 &-1 & 0  \non
              0 & 0 & 0 & 0 & 0 & 0 & 0 &-1 & 2 & 0  \non
              0 & 0 & 0 & 0 & 0 & 0 &-1 & 0 & 0 & 2  \nn
   \end{array} \right) $$
whose inverse
$$ A^{-1}=-\left(
   \begin{array}{rrrrrrrrrr}
              0 & 1 & 2 & 3 & 4 & 5 & 6 & 4 & 2 & 3  \non
              1 & 2 & 4 & 6 & 8 &10 &12 & 8 & 4 & 6  \non
              2 & 4 & 6 & 9 &12 &15 &18 &12 & 6 & 9  \non
              3 & 6 & 9 &12 &16 &20 &24 &16 & 8 &12  \non
              4 & 8 &12 &16 &20 &25 &30 &20 &10 &15  \non
              5 &10 &15 &20 &25 &30 &36 &24 &12 &18  \non
              6 &12 &18 &24 &30 &36 &42 &28 &14 &21  \non
              4 & 8 &12 &16 &20 &24 &28 &18 & 9 &14  \non
              2 & 4 & 6 & 8 &10 &12 &14 & 9 & 4 & 7  \non
              3 & 6 & 9 &12 &15 &18 &21 &14 & 7 &10  \nn
   \end{array} \right) $$
we shall need below. The $\9$ {\bf null root} is
\[  \vd=\sum_{i=0}^8n_i\vr_i
       = (0,\ 0,\ 0,\ 0,\ 0,\ 0,\ 0,\ 0,\ 1\|1), \]
where the coefficients $n_i$ (called {\bf marks} of $\9$)
can be read off from
\[       \left[\begin{array}{*{9}{c}}
                &   &   &   &    &    &  3 &   &   \\
               0& 1 & 2 & 3 &  4 &  5 &  6 & 4 & 2
               \end{array} \right]. \]
The fundamental Weyl chamber $\eC$ of $\0$ is the convex cone
generated by the
{\bf fundamental weights} $\vL_i$ \footnote{Notice
that our convention is opposite to the one adopted in \ct{KaMoWa88}.
The fundamental weights here are {\it positive} and satisfy
$\vL_i \X \vr_j = - \d_{ij}$
Thus, we will be dealing with ``lowest weight'' rather than ``highest
weight'' representations in accordance with physics usage.},
\[ \vL_i = - \sum_{j=-1}^8 (A^{-1})_{ij} \vr_j
 \FOR{$i=-1,0,1,\ldots 8$}                           \]
with the inverse Cartan matrix from above. Thus,
\[ \vL \in \eC    \Longleftrightarrow
     \vL =     \sum_{i=-1}^8 k_i \vL_i  \]
for $  k_i \in \Z_+$.
A special feature of $\0$ is that we need
not distinguish between root and weight lattice,
since these are the same for self-dual root lattices\footnote{In the
remainder, we will consequently denote arbitrary roots by
$\vL$ and reserve the letter $\vr$ for
real roots (i.e.\ $\vr^2=2$).}.
Since Weyl transformations preserve multiplicities and since every
root can be brought into $\eC$ by means of a Weyl transformation,
the structure of the algebra is completely understood once the
root spaces for roots belonging to $\eC$ are under control.
Note also that the null root plays a special role: the first
fundamental weight is just $\vL_{-1} = \vd$, and all null-vectors
in $\eC$ must be multiples of $\vL_{-1}$ since ${\vL}_i^2 <0$
for all other fundamental weights.

As described in section 2.2, the algebra $\ggA = \0$ consists of
all multiple commutators of the Chevalley-Serre generators $e_i, f_i,
h_i$ with $i=-1,0,1,\ldots,8$. It is a standard result \ct{Kac90}
that this algebra can be written as a direct sum
\[ \ggA = \gn_+ \oplus \gh \oplus \gn_-            \]
where the subalgebras $\gn_-$ and $\gn_+$ are defined to consist
of all linear combinations of multiple commutators of
the form $[f_{i_1},[f_{i_2},\ldots[f_{i_{n-1}},f_{i_n}]\ldots]]$ and
$[e_{i_1},[e_{i_2},\ldots[e_{i_{n-1}},e_{i_n}]\ldots]]$, respectively,
modulo the multilinear Serre relations \Ref{Serre}.
Since $\gn_+$ and $\gn_-$ are conjugate
and thus enjoy analogous properties, it is enough in practice to
consider only multiple commutators made out of $e_i$ generators
(corresponding to positive roots). To classify such commutators one
introduces the {\bf level} $\cl \in \Z$ of a root, such that positive
$\cl$ counts the number of $e_{-1}$ generators in
$[e_{i_1},[e_{i_2},\ldots[e_{i_{n-1}},e_{i_n}]\ldots]]$ (similarly, if
$\cl$ is negative, $-\cl$ counts the number of $f_{-1}$ generators
in $[f_{i_1},[f_{i_2},\ldots[f_{i_{n-1}},f_{i_n}]\ldots]]$). In terms
of the corresponding root $\vL = \vr_{i_1} +\ldots+ \vr_{i_n}$,
$\cl$ is defined by
\[   \cl := - \vL \X \vd  \]
Observe that $\cl$ is not preserved under arbitrary $\0$ Weyl
transformations, but only under the subgroup $\eW (\9 )$ corresponding
to the $\9$ subalgebra. Therefore, we can freely use this notion
also for roots $\vL$ which are not in $\eC$, but can be brought
into $\eC$ by an $\9$ Weyl transformation.

The level derives its importance from the fact that it grades the
algebra $\0$ with respect to its affine subalgebra $\9$ \ct{FeiFre83}.
The subspaces belonging to a fixed level can be decomposed into
irreducible representations of $\9$, the level being equal
to the eigenvalue of the central term of the $\9$ algebra on this
representation (the full $\0$ algebra contains $\9$ representations
of {\it all} integer levels!). Let us emphasize that for general
hyperbolic algebras there would be a separate grading associated
with every regular affine subalgebra, and therefore the graded
structure would no longer be unique.
An important result is the following
\ct{FeFrRi93}.
\begin{theo}
Suppose that $x$ is an element of $\0$ at level $n$.
Then it can be represented as a linear combination of $n$-fold
commutators of level-one elements, viz.
\[ x = [x_1,[x_2,\ldots[x_{n-1},x_n]\ldots]]    \]
where each $x_i$ contains exactly one generator $e_{-1}$
in the right-most position\footnote{All level-one elements can be
cast into this form by use of the Jacobi identity and by taking
appropriate linear combinations.}, i.e.
\[ x_1 = [e_{i_1},[e_{i_2},\ldots[e_{i_k},e_{-1}]\ldots]]     \]
with $i_\nu \in \{0,1,\ldots,8 \}$, and similarly for the other $x_i$.
\lb{thm1} \end{theo}
We are going to make use of this result in the next section
in order to effectively construct higher level elements.

As already mentioned, little is known about the general structure
of this algebra. Partial progress has been made in determining
the multiplicity of certain roots, i.e.\ the number of
linearly independent Lie algebra elements in the corresponding
root space. Although the general form of the multiplicity
formulas for arbitrary levels appears to be beyond reach for the
moment, the following results for levels $\cl \leq 2$ have been
established. For $\cl =0$ and $\cl =1$, we have \ct{Kac90}
\[  \mult(\vL) = p_8 (1- \2 {\vL}^2 )   \]
i.e.\ the multiplicities are just given by the number of transversal
states; we will see in the next section that the corresponding states
are indeed transversal. For $\cl =2$, it was shown in \ct{KaMoWa88}
that\footnote{The derivation of this result in \ct{KaMoWa88} is based
on the $\9$ decomposition \label{foot1}
$$
L(\vL_0) \wedge L(\vL_0) \cong L(\vL_1)\XO V\Big(\frc12 ,
       \frc1{16} \Big)
$$
where $L(\vL_i)$ denotes the irreducible $\9$ module with lowest
weight $\vL_i$ and
$V\Big( \frc12 ,\frc1{16} \Big)$ the irreducible Virasoro module
with $c=\frc12$ and $h= \frc1{16}$ (by abuse of notation, we use
the same labels for the $\9$ weights as for the $\0$ weights).
Observe that the module $L(\vL_1)$ precisely corresponds to the
ideal generated by the double commutator $[[e_0, e_{-1}], e_{-1}]$.
For higher levels, analogous decompositions contain more than one term
on the right hand side, and it seems no longer possible to divide out
the Serre relations by this method.}
\[  \mult(\vL) = \xi (3- \2 {\vL}^2)  \lb{mult2}  \]
where
\[  \sum_{n \geq 0} \xi (n) q^n =
\frac{1}{\phi (q)^8}
\Big[ 1- \frac{\phi (q^2)}{\phi (q^4]} \Big]  \]
and the Euler function $\phi (q)$ is defined in \Ref{Euler}.
For sufficiently large (negative) ${\vL}^2$, one can check
from this formula that there are roots $\vL$ such that
$\mult(\vL) > p_8 (1-\2 {\vL}^2)$; this clearly implies
the presence of longitudinal states.
Beyond $\cl = 2$, no general formula seems to be known although
the multiplicities can be determined recursively from the
Peterson formula (see e.g. \ct{KMPS90}).

In the physical interpretation, the multiplicity of a root $\vL$ is
nothing but the number of linearly
independent polarization states of the associated
vertex operator of momentum $\vL$. Given a root $\vL \in \eC$, we
call a polarization vector $\vxi$ {\bf transversal} if
$\vxi \X \vL = \vxi \X \vd = 0$, and {\bf longitudinal}
otherwise. This terminology is, of course, physically motivated.
We also define the {\bf little group} $\eW (\vL ,\vd )$ to be
that subgroup of the full Weyl group of $\0$ which
leaves the vectors $\vL$ and $\vd$ invariant.
Unless $\vL$ is collinear with $\vd$ (corresponding to
$\cl = 0$), $\eW(\vL,\vd)$ is a {\it finite}
subgroup of $\eW (\0 )$, as well as a discrete subgroup of
$SO(8)$. As an example consider $\cl =1$;
then $\vL = \vL_0 = \vri{-1} +2\vd$
and $\eW (\vL,\vd)$ is isomorphic to the Weyl group of $E_8$.
In fact, for $\vL \in \eC$, it is known (\ctz{Prop.3.12}{Kac90})
that $\eW (\vL,\vd)$ is generated by the
reflections $\ew_i$ corresponding to those simple roots $\vr_i$
for which $\vL\X\vr_i= \vd\X\vr_i= 0$. This indicates
that the little group will not be quite as useful in this context as
it is in conventional quantum field theory, because it becomes
trivial for sufficiently high levels. However, at
low levels, this problem does not yet arise, and the polarization
states and hence the elements belonging to the root space
$\ggI^{(\vL)}$ can be classified as
representations of $\eW (\vL,\vd)$.

Any root $\vL \in \eC$ can be represented in the form
\[ \vL = \cl \vri{-1} + M \vd + \vb   \]
where $\cl$ is the level of $\vL$ and
$\vb$ an element of the $E_8$-root lattice $Q(E_8)$
($\vb$ need not be positive by itself as only $M \vd + \vb$
must be positive). We now define the {\bf DDF decomposition}
of $\vL$ by
\[ \vL = \va - n \vk(\va), \lb{DDFdec1} \]
where
\[ \vk (\va ) := - \frc1{\cl} \vd  \lb{DDFdec2} \]
and
\[ n = 1 - \2 \vL^2 = 1 + (M-\cl ) \cl - \2 \vb^2. \]
By construction, $\va$ obeys $\va^2 =2$ and is therefore associated
with a tachyon state, and $n$ is the number of steps required to
reach the root $\vL$ by starting from $\va$ and decreasing the
momentum by $\vk$ at each step ($n$ is non-negative because
${\vL}^2 \leq 2$; note also that $\vk$ is always a {\it negative} root,
so $\vL$ is positive for all $n$). Obviously, for
$\cl >1$, neither $\vk$ nor $\va$ belong to the lattice in general.
As a consequence, the intermediate DDF states associated
with momenta $\va -m \vk$ not on the lattice will not correspond
to elements of the algebra. On the other hand, states
associated with the root $\vL$ do belong to the algebra of
physical states, and the DDF decomposition enables
us to write down all possible polarization states associated with the
root $\vL \in \eC$ in terms of transversal and longitudinal DDF
states; the totality of these states constitutes the complete set of
elements in the root space $\ggI^{(\vL)}$.

Of course, we could also try to apply the DDF decomposition to
roots $\vL$ not $\eW (\9)$-equivalent to roots in $\eC$. Whenever
we succeed in finding a suitable null vector $\vk$ on the lattice
obeying $\vL \X \vk= 1$, we can also find a Weyl
transformation $\ew$ such that $\ew (\vk) = -\vd$ since $\vd$ is the
only primitive null vector in $\eC$. Since, $\vL \X \vk =
- \ew (\vL) \X \vd$ is just the level, it follows that
$\ew (\vL )$ is a level-one root with tachyon momentum
\[ \va=\vri{-1} +
   \big( \2 \vb^2 \big) \vd + \vb  \lb{tachmom} \]
Therefore, nothing is gained by
searching for DDF vectors outside the $\eW (\9)$ transforms
of the fundamental Weyl chamber.

\subsection{The DDF states at levels zero and one}
Although the multiplicity formulas for levels $\cl =0$ and $\cl =1$
are understood \ct{Kac90},
we here derive them once more, because our
explicit DDF representation of the level-one elements has apparently
not been exhibited in the literature so far. The level-zero
elements make up the $\9$ subalgebra of $\0$. The allowed (positive
and negative) roots are all $\vr\in\II$ obeying $\vr^2=2$ and
$\vr\X\vd=0$ (hence having no $\vri{-1}$ component), and $m\vd$ for
$m \in \Z^\times$. These correspond to the tachyonic and photonic
states with multiplicities $1$ and $8$, respectively:
\[ |\vr\>&\equiv&\exp\vr \FOR{}\ \vr^2=2, \lb{E9-1} \\
   \vxi_i(-1)|m\vd\>&\equiv&\vxi_i(-1)\exp{m\vd}, \lb{E9-2} \]
where $\vxi_i\X\vd=0$ and $\vxi_i$ has no component along
$\vd$ (i.e.\ $\vri{-1}\X\vxi_i=0$). The Cartan subalgebra of
$\9$ is spanned by the states
\[ \vd(-1)|\vo\>&=:&K, \lb{E9-3} \\
   (\vr_{-1}+\vd)(-1)|\vo\>&=:&d, \lb{E9-4} \\
   \vxi_i(-1)|\vo\>& &\FOR{}\ i=1,\ldots,8\ , \lb{E9-5} \]
where $K$ represents the central element, $d$ is the derivation of $\9$,
and \(\{\vxi_i(-1)|\vo\>\| i=1,\ldots,8\}$ span the Cartan subalgebra of
$E_8$. This is the standard ``light-cone'' basis of $\gh(\9)$ in the
sense that $K$ and $d$ are lightlike.
As for the commutators we rewrite \Ref{comm3} and \Ref{pos1} --
\Ref{pos3} as
\[ \Big[\vet(-1)|\vo\>,\vze(-1)|\vo\>\Big]&=&0, \\[.5em]
   \Big[\vet(-1)|\vo\>,\vxi_i(-1)|m\vd\>\Big]
       &=&m(\vet\X\vd)\vxi_i(-1)|m\vd\>, \\[.5em]
   \Big[\vet(-1)|\vo\>,|\vr\>\Big]&=&(\vet\X\vr)|\vr\>, \\[.5em]
   \Big[\vxi_i(-1)|m\vd\>,\vxi_j(-1)|n\vd\>\Big]
       &=&m\d_{m+n,0}(\vxi_i\X\vxi_j)\vd(-1)|\vo\>, \\[.5em]
   \Big[\vxi_i(-1)|m\vd\>,|\vr\>\Big]
       &=&(\vxi_i\X\vr)|\vr+m\vd\>, \\[.5em]
   \Big[|\vr\>,|\vs\>\Big]
       &=&\cases{0  & if $\vr\X\vs\ge0$, \cr
                 \e(\vr,\vs)|\vr+\vs\>  & if $\vr\X\vs=-1$, \cr
                 -\vr(-1)|m\vd\>  & if $\vr+\vs=m\vd$,} \]
for $\vet,\vze\in\gh(\9)$ and $\9$ roots $\vr,\vs$.
To see that photonic states with all required
transversal polarizations can be generated by commuting tachyonic
states, we recall \Ref{tachcomm} (a special case of \Ref{comm1}):
choosing $\vs = \vr_i$ and $\vt = m\vd - \vr_i$ (where $\vr_i$ is any
simple root of $\9$), we obtain all transversal polarizations.
There is obviously no way to generate longitudinal states, because the
polarization vectors $\vxi_i$ would then have to have components
along $\vri{-1}$, which we cannot generate by commuting
tachyonic states belonging to $\9$ roots only.
Since we can ignore null physical states (for which
$\vxi \propto \vd$), we can in addition impose the requirement
$\vxi \X \vr_{-1} = 0$, so $\vxi \in \{\vr_1,\ldots,\vr_8\}$, so that
by taking appropriate linear combinations we can arrange that
$\vxi_i \X \vxi_j = \d_{ij}$ with $\vxi_i \X \vd =
\vxi_i \X \vr_{-1} = 0$ for $i,j=1,\ldots,8$.
It is clear that an infinity of conjugate $\9$ subalgebras in $\0$
can be obtained by Weyl conjugation of these states with
elements of $\eW( \0)$ not in $\eW (\9)$.

Let us now turn to the level-one roots. Inspection of the inverse
Cartan matrix shows that the only such roots
in $\eC$ are of the form
\[ \vL = k_{-1} \vL_{-1} + \vL_0 = \vr_{-1} + (2+ k_{-1}) \vd   \]
corresponding to the DDF decomposition \Ref{DDFdec1} with
$\va = \vr_{-1}$, $\vk = - \vd$ and $n=2+ k_{-1}$. Since all these
vectors are elements of the lattice, we can straightforwardly apply
the DDF construction to obtain the physical states
\[ A_{-m_1}^{i_1}\cdots A_{-m_N}^{i_N}|\vr_{-1}\>, \lb{E10-lev1}  \]
where $m_1+\ldots+m_N=2+k_{-1}$ and with the polarization vectors
chosen as above. Recall that
$A^i_{-m}\equiv\Big(\vxi_i(-1)\exp{m\vd}\Big)_0$. These states
correspond to the multiple commutators
\[   \Big[\vxi_{i_1}(-1) | m_1 \vd \> ,
     \Big[\ldots,
     \Big[\vxi_{i_N}(-1) | m_N \vd \> ,
          | \vr_{-1} \> \Big]\ldots\Big]\Big], \]
as we have already shown. Moreover, we can explicitly verify that
they form the basic representation of $\9$ with lowest weight vector
$|\vr_{-1}\>$. To see this we have to work out the commutators of the
$\9$ elements \Ref{E9-1} -- \Ref{E9-5} with the level-one states
\Ref{E10-lev1}:
\[  \Big[\vet(-1)|\vo\>,
            A_{-m_1}^{i_1}\cdots A_{-m_N}^{i_N}|\vr_{-1}\>\Big]
=   \Big((m_{i_1}+\ldots+m_{i_N})
    \vd \X \vet +\vr_{-1} \X \vet \Big)
      A_{-m_1}^{i_1}\cdots A_{-m_N}^{i_N}|\vr_{-1}\>,   \]
\[  \Big[\vxi_j(-1)|n\vd\>,
            A_{-m_1}^{i_1}\cdots A_{-m_N}^{i_N}|\vr_{-1}\>\Big]
   &=&A_{-n}^j\Big(A_{-m_1}^{i_1}\cdots
                   A_{-m_N}^{i_N}|\vr_{-1}\>\Big) \non
   &=&\cases{-\sum_{k=1}^Nn\d_{j,i_k}\d_{n,m_k}
                  \prod_{l\ne k}A_{-m_l}^{i_l}|\vr_{-1}\>
             & if $n<0$, \cr \cr
             A_{-n}^jA_{-m_1}^{i_1}\cdots A_{-m_N}^{i_N}|\vr_{-1}\>
             & if $n>0$,} \]
\[  \Big[|\vs\>,
            A_{-m_1}^{i_1}\cdots A_{-m_N}^{i_N}|\vr_{-1}\>\Big]
   &=&-\sum_{k=1}^N(\vs\X\vxi_{i_k})
      A_{-m_1}^{i_1}\cdots A_{-m_{k-1}}^{i_{k-1}}
      \Big[|\vs+m_k\vd\>,A_{-m_{k+1}}^{i_{k+1}}\cdots A_{-m_N}^{i_N}
      |\vr_{-1}\>\Big]  \non
   & &{}+A_{-m_1}^{i_1}\cdots A_{-m_N}^{i_N}
         \Big[|\vs\>,|\vr_{-1}\>\Big]. \]
The first commutator tells us that the Cartan subalgebra of $\9$
acts diagonally on the DDF states, giving the components
of the lowest weight $\vet \X \vL$ of the representation.
The second commutator which directly
follows from the definitions \Ref{Lie1} and \Ref{DDFdef},
reveals that the $\9$ elements corresponding to multiples of the
null root $\vd$ act by multiplication with a DDF operator. The last
commutator is obtained by rewriting the DDF states \Ref{E10-lev1}
as multiple commutators and repeated application of the following
version of the Jacobi identity:
\[ \Big[|\vs\>,A^i_{-m}\p\Big]
   &\equiv&\Big[|\vs\>,\Big[\vxi_i(-1)|m\vd\>,\p\Big]\Big] \non
   &=&\Big[\Big[|\vs\>,\vxi_i(-1)|m\vd\>\Big],\p\Big]
      +\Big[\vxi_i(-1)|m\vd\>,\Big[|\vs\>,\p\Big]\Big] \non
   &=&-m(\vs\X\vxi_i)\Big[|\vs+m\vd\>,\p\Big]
      +A^i_{-m}\Big[|\vs\>,\p\Big]  \]
for any state $\p$.
Note that the commutator $[|\vs\>,|\vr_{-1}\>]$ above can be
evaluated using \Ref{comm1}. For example it vanishes whenever $\vs$
is a negative root of $\9$ as it should be since $\vr_{-1}$ is
a lowest weight vector; furthermore, we always get a level-one
state since $\vs$ is a $\9$ root.
Weyl-equivalent level-one states can be generated by Weyl conjugation
with elements $\ew \in \eW$ leaving the level fixed, i.e.\
$\ew \in \eW (\9 )$. The tachyonic momentum $\vri{-1}$ is then
mapped to a vector of the form \Ref{tachmom}
with $\va = \ew (\vri{-1})$, while $\vd$ is left invariant.
The polarizations used above must be replaced
by the rotated polarization vectors $\vxi_{\ew(i)}:=\ew(\vxi_i)$
with corresponding changes in the DDF vectors. Denoting the
rotated DDF operators by
$A^{\ew(i)}_{-m}\equiv A^{\ew(\vxi_i)}_{-m}$, we obtain the new states
\[ A_{-m_1}^{\ew(i_1)}\cdots A_{-m_N}^{\ew(i_N)}|\ew(\vr_{-1})\>. \]
Notice that although we are using transversal indices these now
transform under different little groups (which are all conjugate
to $\eW (E_8)$). The multiplicity formula for the level-zero and
level-one roots \ct{Kac90}
\[ \mult(\vL) = p_8 (n) = p_8 (1- \2{\vL}^2 )     \]
can be read off immediately from \Ref{E9-1},
\Ref{E9-2} and \Ref{E10-lev1}.
This multiplicity formula holds likewise for roots related by an
arbitrary Weyl rotation to a level-one root.

As already mentioned before, the states \Ref{E10-lev1} transform
covariantly under the corresponding little group $\eW (\vr_{-1}, \vd )$,
which is just the Weyl group of $E_8$. Now it is known that
$\eW (E_8) = D_4(2) \times (\Z_2)^2 $, where
$D_4 (2)$ is the Chevalley group of order $2^{12} 3^5 5^2 7$,
or, equivalently, the set of $SO(8)$ matrices with entries in the
field $\Z_2$ (see e.g. \ct{Car72}). Since it is the maximal
discrete subgroup of $SO(8)$ of this type in the sense that the
little groups of all higher level roots will be much smaller,
this also explains why the states \Ref{E10-lev1} look
``$SO(8)$ covariant'' (although the polarization indices $i,j,\dots$
should by no means be regarded as $SO(8)$ indices!). As we will see
the higher level root spaces will exhibit much less symmetry.

\subsection{Higher Level: Generalities}
Before turning to the discussion of an explicit example of a
level-two root space, we would like to explain the general ideas
underlying the description of higher level elements in terms of the
DDF construction. As we have already mentioned, the DDF states
constitute a complete basis of physical states for any allowed
momentum on the root lattice. Consequently, the root space
$E_{10}^{(\vL)}$ is a (proper for $\ell >1$) subspace of $\ggI^{(\vL)}$
for any root $\vL$ (this inclusion is a special case of \Ref{inclus}).
The physical states are explicitly given by \Ref{DDFspan} or,
equivalently, by \Ref{physspan}. Anticipating that the final results
are somewhat simpler in terms of \Ref{physspan}, we will use
the basis
\[ A^{i_1}_{-m_1}(\va) \ldots A^{i_M}_{-m_M}(\va)
   \sL_{-n_1}(\va)\ldots \sL_{-n_N}(\va)|\va\> , \lb{DDFbasis} \]
explicitly indicating the dependence of the DDF operators
and their polarizations on the tachyon momentum $\va$ and the
associated lightlike vector $\vk (\va) = -\frc1{\ell} \vd$,
and assuming $n_i \geq 2$ from now on to exclude null states.
Since $\ell \neq 1$, we have
\[ A^i_{-m} (\va) \equiv \Big( \big(\vxi_i (\va) \big)(-1)
   \exp{\frc{m}{\ell}  \vd} \Big)_0  \]
with an extra factor of $\frc1{\ell}$ in the exponent,
as appropriate for level $\ell$ by \Ref{DDFdec2}.
In accordance with the DDF decomposition $\vL = \va - n \vk (\va)$,
the indices obey the sum rule $m_1+\ldots+m_M+ n_1+\ldots+n_N= n$.
We emphasize once more that neither $\va$ nor $\vk (\va)$ need
be on the root lattice for $\ell >1$ any more.
The problem of characterizing the root
spaces of the hyperbolic Kac Moody algebra is now no longer one of
dividing out the Serre relations \Ref{Serre} (these are
automatically taken care of by the vertex operator formalism
as we pointed out already), but rather one of identifying the
missing states which cannot be generated by multiple
commutators of the Chevalley generators $e_i$ or $f_i$.
The above representation immediately yields the following upper
bound on the root multiplicities \ct{Borc86}
\[ {\rm mult} \, (\vL ) \leq
 p_9 (1-\2{\vL}^2) - p_9 (-\2{\vL}^2).  \]

To effectively construct higher level elements we invoke Theorem 1
of section 4.1. For instance, given a level-two root
$\vL$ in the fundamental Weyl chamber $\eC$, we write
\[  \vL = \vr + \vs + m\vd , \lb{Lrs}  \]
where $\vr$ and $\vs$ are real positive level-one roots (i.e.
containing the simple root $\vri{-1}$ exactly once and obeying
$\vr^2 = \vs^2 = 2$). In general, there will be
many different ways to split $\vL$ in this manner, as well as
different integers $m$. For fixed value of $m$, these decompositions
are related by the little group, which leaves $\vL$ and $\vd$ fixed,
but varies $\vr$ and $\vs$. Thus, we work with a fixed
decomposition and then act on the resulting commutator states
with the little group so as to obtain all possible states with
the same value of $m$. The commutator to be computed is
\[ \Big[ A^{i_1}_{-m_1}(\vs)\ldots A^{i_M}_{-m_M}(\vs)|\vs\> \, ,\,
A^{j_1}_{-n_1}(\vr)\ldots A^{j_N}_{-n_N}(\vr)|\vr\> \Big]
\lb{comm-lev1}      \]
where $m_1 + \ldots m_M+n_1+ \ldots + n_N = m$. For the special
example to be discussed below, this expression can be evaluated
with the help of the formulas given in the appendices.
Expanding it in terms of the basis \Ref{DDFbasis}, we arrive at
\[ \Ref{comm-lev1} = \sum_{p_1+\ldots+q_Q=n \atop k_1,\ldots,k_P}
 c_{k_1\ldots k_P}^{i_1\ldots i_M j_1\ldots j_N}
   A^{k_1}_{-p_1}(\va) \ldots A^{k_P}_{-p_P}(\va)
   \sL_{-q_1}(\va)\ldots \sL_{-q_Q}(\va) |\va\>  \lb{lev2-dec}\]
with the ``Clebsch Gordan coefficients''
$c_{k_1\ldots k_P}^{i_1\ldots i_M j_1\ldots j_N}$, into which
all the information about the missing states is encoded.
For the Fock space states, this equality holds of course
only modulo terms $\Lm1 (\dots)$, which can however be ignored for
the Lie algebra, as they are factored out by \Ref{Lie2a}.
\Ref{lev2-dec} is the crucial formula, containing both transversal
and longitudinal excitations\footnote{This formula also shows why
the fake monster Lie algebra of \ct{Borc86} is, in a certain sense,
much simpler (though bigger) than $\0$.
The longitudinal components generated by commuting
two transversal DDF states decouple in 26 dimensions, and
therefore only the terms {\it without} longitudinal states
survive in the expansion \Ref{lev2-dec}. To be sure, one must still
prove that indeed {\it all} transversal states can be generated in this
way if one takes into account the imaginary
simple roots.}.
For the calculations, we note that the polarization vectors
$\vxi_i (\vr)$ and $\vxi_i (\vs)$ can always be  chosen
orthonormal and such that they agree for $i=1,\ldots,7$;
from \Ref{Lrs} we then see that $\vxi_i (\va ) = \vxi_i (\vr)$
as well for these values of the indices. As for the remaining
components $\vxi_8 (\vr)$, $\vxi_8 (\vs)$ and $\vxi_8 (\va)$,
one can convince oneself that their differences are proportional
to the null vector $\vd$. Since such contributions drop out
in the non-zero mode parts of the DDF operators (cf. the
discussion after \Ref{Lie2}), the respective DDF operators
are really the same except for their zero mode parts and
the crucial fact that their photon momenta depend on the level.
We stress that this would not be true if the Weyl
chamber contained more than one null direction.

Just as for the level-one states, one can determine how the states
\Ref{lev2-dec} transform under $\9$. Suppressing the label $(\va)$
on the DDF operators to make the formulas less cumbersome,
this calculation requires the commutators (for $\ell =2$)
\leqn0{\Big[\vet(-1)|\vo\>,
            A^{i_1}_{-m_1}\ldots A^{i_M}_{-m_M}
            \sL_{-n_1}\ldots \sL_{-n_N}|\va\>\Big]}\vspace{-3ex}
\[ =\Big(\frc1{\ell}(m_1+\ldots+n_N)\vd\X\vet+\va\X\vet\Big)
    A^{i_1}_{-m_1}\ldots A^{i_M}_{-m_M}
    \sL_{-n_1}\ldots \sL_{-n_N}|\va\>. \]
The scalar product in parantheses is easily seen to reduce to
$\vet\X\vL$, giving the components of the lowest weight of the
representation. Furthermore,
\leqn0{\Big[\vxi_j(-1)|n\vd\>,
            A^{i_1}_{-m_1}\ldots A^{i_M}_{-m_M}
            \sL_{-n_1}\ldots \sL_{-n_N}|\va\>\Big]}\vspace{-3ex}
\[ =A^j_{-\ell n}\Big(
   A^{i_1}_{-m_1}\ldots A^{i_M}_{-m_M}
   \sL_{-n_1}\ldots \sL_{-n_N}|\va\>\Big), \lb{comml2E9a} \]
(notice that the index on the first operator is $(-\ell n)$ rather
than $(-n)$!) and
\leqn0{\Big[|\vs\>,
   A^{i_1}_{-m_1}\ldots A^{i_M}_{-m_M}
   \sL_{-n_1}\ldots \sL_{-n_N}|\va\>\Big]}\vspace{-3ex}
\[ &=&-\sum_{k=1}^M(\vs\X\vxi_{i_k})
                   A^{i_1}_{-m_1}\ldots A^{i_{k-1}}_{m_{k-1}}
                   \Big[|\vs+\frc1{\ell}m_k\vd\>,
                   A^{i_{k+1}}_{m_{k+1}}\ldots A^{i_M}_{m_M}
                   \sL_{-n_1}\ldots \sL_{-n_N}|\va\>\Big] \non
   & &{}+\sum_{l=1}^N(\vs\X\va)
                   A^{i_1}_{-m_1}\ldots A^{i_M}_{-m_M}
                   \sL_{-n_1}\ldots \sL_{n_{l-1}}
                   \Big[|\vs+\frc1{\ell}n_l\vd\>,
                   \sL_{n_{l+1}}\ldots \sL_{n_N}|\va\>\Big] \non
   & &{}+A^{i_1}_{-m_1}\ldots A^{i_M}_{-m_M}
         \sL_{-n_1}\ldots\sL_{-n_N}\Big[|\vs\>,|\va\>\Big].
                  \lb{comml2E9} \]
Observe that there are no contributions from the logarithmic terms
in $\sL_{-m}$ to the last commutator because $\vs\X\vd = 0$
for any $\9$ root $\vs$. The proof of these formulas is
analogous to the proof of the corresponding formulas for the
level-one states in the previous section, save for the following
important caveat. When building
up the states from the tachyonic groundstate $|\va\>$
by successive application of the DDF operators, the intermediate
states, whose momenta are not on the root lattice, do {\it not}
belong to the Kac Moody algebra, because the Lie bracket
with arbitrary elements is in general not defined due to branch cuts
in the relevant operator product expansions\footnote{We note that
the cocycle conditions \Ref{coc1}--\Ref{ecoc4} can be solved on a
rational extension of the root lattice \ct{FLM88}.}.
Therefore, the ``commutators'' in \Ref{comml2E9} are neither
commutators in $\0$ nor even in the Lie algebra of physical
states $\ggL$; nevertheless,
the above calculation does make sense because all relevant
products of momenta are integer, and therefore the
generic branch cuts are absent. So we must keep
in mind that only the final result including summation according
to \Ref{lev2-dec} is an element of $\0$ again.
The fact that the direct construction of the DDF states has no Lie
algebra analog beyond level one explains the emergence of
longitudinal states as well as the disappearance of certain
transversal states.

In the next section, we will work out one non-trivial
level-two root, arriving at a complete description of its
root space in terms of DDF states, which decompose into
irreducible representations of the little group $\eW (\vL ,\vd )$;
as a by-product, we verify the multiplicity formula
of \ct{KaMoWa88} for a concrete example. The comparative simplicity
of the representation obtained in this
manner is perhaps best appreciated by noting that the number
of Lie brackets needed to represent any of its
elements in terms of Chevalley generators is equal to
$(-\vro\X\vL-1)$, where $\vro$ is the Weyl vector.

\subsection{A Level-Two Example: $\vL = \vL_7$}
Any level-two root in $\eC$ must be of the form
$\vL_1+n\vd$ or $\vL_7+n\vd$  or $2\vL_0+n\vd$ for some $n\in\N$.
We will here only discuss the root $\vL = \vL_7$, dual to the
simple root $\vri7$. Explicitly, $\vL_7$ is given by
\[ \vL_7=\left[\begin{array}{*{9}{c}}
                &   &   &   &    &    &  7 &   &   \\
               2& 4 & 6 & 8 & 10 & 12 & 14 & 9 & 4
               \end{array} \right]
        =(0,\ 0,\ 0,\ 0,\ 0,\ 0,\ 0,\ 0,\ 0\|2), \]
so $\vL_7^2 = -4$. Its decomposition into two level-one
tachyonic roots is $\vL_7=\vr+\vs+2\vd$, where
\[ \vr:=\vr_{-1}
      &=&\left[\begin{array}{*{9}{c}}
                &   &   &   &    &    &  0 &   &   \\
               1& 0 & 0 & 0 &  0 &  0 &  0 & 0 & 0
               \end{array} \right]
        =(0,\ 0,\ 0,\ 0,\ 0,\ 0,\ 0,\ 1, -1\|0), \non
   \vs&:=&
         \left[\begin{array}{*{9}{c}}
                &   &   &   &    &    &  1 &   &   \\
               1& 2 & 2 & 2 &  2 &  2 &  2 & 1 & 0
               \end{array} \right]
       = (0,\ 0,\ 0,\ 0,\ 0,\ 0,\ 0, -1, -1\|0). \nn \]
Since $n=1-\2\vL_7^2=3$ we have the DDF decomposition $\vL_7=\va-3\vk$
where $\vk:=-\2\vd$ and
\[ \va:=\vr+\vs-\vk
       =(0,\ 0,\ 0,\ 0,\ 0,\ 0,\ 0,\ 0, -\frc32\|\2). \nn \]
As expected, neither $\vk$ nor $\va$ are elements
of $\II$. Nevertheless, since $\va\X\vk=1$, the action of the
DDF operators $A^i_{-n}(\vk)$ on the tachyonic ground-state $|\va\>$
is perfectly well-defined as we already pointed out.
As for the three sets of polarization vectors associated with
the tachyon momenta $|\vr\>$, $|\vs\>$ and $|\va\>$, respectively,
a convenient choice is
\[ \vxa&\equiv&\vxa(\vr)=\vxa(\vs)=\vxa(\va)
                 \FOR{}\ \a=1,\ldots,7\ , \non
   \vxi_1&:=&(1,\ 0,\ 0,\ 0,\ 0,\ 0,\ 0,\ 0,\ 0\|0), \non
         &\vdots& \non
   \vxi_7&:=&(0,\ 0,\ 0,\ 0,\ 0,\ 0,\ 1,\ 0,\ 0\|0); \nn \\[.5em]
   \8(\vr)&:=&(0,\ 0,\ 0,\ 0,\ 0,\ 0,\ 0,\ 1,\ 1\|1), \non
   \8(\vs)&:=&(0,\ 0,\ 0,\ 0,\ 0,\ 0,\ 0, -1,\ 1\|1), \non
   \8\equiv\8(\va)
              &:=&(0,\ 0,\ 0,\ 0,\ 0,\ 0,\ 0,\ 1,\ 0\|0).  \]
The little group is $\eW (\vL_7 , \vd ) = \eW (D_8 ) =
S_8 {\bbl o} (\Z_2 )^7$ of order $2^{14}3^1 5^1 7^1$. This group is
generated by the fundamental reflections $ \{ \ew_0 ,\ew_1, \ew_2 ,
\ew_3 ,\ew_4 ,\ew_5 ,\ew_6 ,\ew_8 \}$. On the polarization vectors
$\vxi_i (\va )$ it acts as follows:
\[  \ew_0 (\vxi_7) &=& \vxi_8 \;\; , \;\;
    \ew_0 (\vxi_8) = \vxi_7, \non
    \ew_1 (\vxi_6) &=& \vxi_7 \;\; , \;\;
    \ew_1 (\vxi_7) = \vxi_6, \non
    \ew_2 (\vxi_5) &=& \vxi_6 \;\; , \;\;
    \ew_2 (\vxi_6) = \vxi_5, \non
    \ew_3 (\vxi_4) &=& \vxi_5 \;\; , \;\;
    \ew_3 (\vxi_5) = \vxi_4, \non
    \ew_4 (\vxi_3) &=& \vxi_4 \;\; , \;\;
    \ew_4 (\vxi_4) = \vxi_3, \non
    \ew_5 (\vxi_2) &=& \vxi_3 \;\;, \;\;
    \ew_5 (\vxi_3) = \vxi_2, \non
    \ew_6 (\vxi_1) &=& - \vxi_2 \;\; , \;\;
    \ew_6 (\vxi_2) = - \vxi_1, \non
    \ew_8 (\vxi_1) &=& \vxi_2  \;\; , \;\;
    \ew_8 (\vxi_2) = \vxi_1,  \lb{WeylPol}    \]
and as the identity on all those that have not been listed.
Furthermore, $\{ \ew_1, \dots ,\ew_6, \ew_8 \}$ leave $\vr$ and $\vs$
invariant, whereas
\[  \ew_0 (\vr ) = \vr + \vr_0 \;\;\; , \;\;\,
    \ew_0 (\vs ) = \vs - \vr_0.    \nn     \]
The Weyl group element
$\ew = \ew_0 \ew_1 \ew_2 \ew_3 \ew_4 \ew_5 \ew_8 \ew_6 \ew_5
\ew_4 \ew_3 \ew_2 \ew_1 \ew_0$ interchanges $\vr$ and $\vs$.

There are three sets of DDF operators
acting on different tachyonic ground states $|\vr\>$, $|\vs\>$, and
$|\va\>$, respectively. Now, since $\ggI^{(\vL_7)}$ is spanned by the
$192$ transversal and the $9$ longitudinal DDF states
\[ A^i_{-1}A^j_{-1}A^k_{-1}|\va\>&,& \non
           A^i_{-2}A^j_{-1}|\va\>&,& \non
                   A^i_{-3}|\va\>&,& \non
         A^i_{-1}\sL_{-2}|\va\>&,& \non
                 \sL_{-3}|\va\>&,& \nn  \]
we can express any element of the root space $\0^{(\vL_7)}$ as
a linear combination of above elements modulo $\Lm1$ terms. This is
done by using the formulas from the appendix and solving the resulting
(overdetermined!) systems of linear equations for the coefficients.
One gets the following results:
\[ \LL \Big[|\vs\>,A^\a_{-1}A^\b_{-1}|\vr\>\Big]&=&
    \e\bigg\{-\2A^\a_{-2}A^\b_{-1}
             -\2A^\b_{-2}A^\a_{-1}
             -A^\a_{-1}A^\b_{-1}A^8_{-1} \RR \non
   & &\phantom{\bigg\{}
      +\frc1{24}\d^{\a\b}\Big[A^8_{-3}+3A^8_{-1}\sL_{-2}
      -4A^8_{-1}A^8_{-1}A^8_{-1}\Big]\bigg\}|\va\>, \]
\[ \LL \Big[|\vs\>,A^\a_{-1}A^8_{-1}|\vr\>\Big]&=&
    \e\bigg\{\frc14A^\a_{-3}+\2A^\a_{-2}A^8_{-1}
      -\2A^8_{-2}A^\a_{-1}
      -\frc14A^\a_{-1}\sL_{-2}\bigg\}|\va\>, \RR \HH \]
\[ \LL \Big[|\vs\>,A^8_{-1}A^8_{-1}|\vr\>\Big]&=&
    \e\bigg\{\frc{17}{24}A^8_{-3}+A^8_{-2}A^8_{-1}
      +\frc18A^8_{-1}\sL_{-2}
      +\frc16A^8_{-1}A^8_{-1}A^8_{-1}\bigg\}|\va\>, \RR \HH \]
\[ \LL \Big[|\vs\>,A^\a_{-2}|\vr\>\Big]&=&
    \e\bigg\{-\frc34A^\a_{-3}-\frc14A^\a_{-1}\sL_{-2}
      +A^\a_{-1}A^8_{-1}A^8_{-1}\bigg\}|\va\>, \RR \HH \]
\[ \LL \Big[|\vs\>,A^8_{-2}|\vr\>\Big]&=&
    \e\bigg\{-\2A^8_{-3}+\2A^8_{-1}\sL_{-2}\bigg\}|\va\>, \RR \HH \]
\[ \LL \Big[A^\a_{-1}|\vs\>,A^\b_{-1}|\vr\>\Big]&=&
    \e\bigg\{-\2A^\a_{-2}A^\b_{-1}
             +\2A^\b_{-2}A^\a_{-1}
             +A^\a_{-1}A^\b_{-1}A^8_{-1} \RR \non
   & &\phantom{\bigg\{}
      -\frc1{24}\d^{\a\b}\Big[A^8_{-3}+3A^8_{-1}\sL_{-2}
      -4A^8_{-1}A^8_{-1}A^8_{-1}\Big]\bigg\}|\va\>, \]
\[ \LL \Big[A^\a_{-1}|\vs\>,A^8_{-1}|\vr\>\Big]&=&
    \e\bigg\{-\frc14A^\a_{-3}+\2A^\a_{-2}A^8_{-1}
      +\2A^8_{-2}A^\a_{-1}
      +\frc14A^\a_{-1}\sL_{-2}\bigg\}|\va\>, \RR \HH \]
\[ \LL \Big[A^8_{-1}|\vs\>,A^8_{-1}|\vr\>\Big]&=&
    \e\bigg\{\frc{17}{24}A^8_{-3}
      +\frc18A^8_{-1}\sL_{-2}
      +\frc16A^8_{-1}A^8_{-1}A^8_{-1}\bigg\}|\va\>, \RR \HH \]
for $\a,\b=1,\ldots,7$ and with $\e\equiv\e(\vs,\vr)\e(\vk,\va)$;
contributions involving $\Lm{1}(\dots)$ have been
neglected in accordance with \Ref{Lie2a} (these extra terms are
listed in Appendix C). Let us also record the
following simple formula, which is an immediate consequence:
\[ -(-1)^{\d_{j8}+\d_{i8}}\Big[A^i_{-1}|\vs\>,A^j_{-1}|\vr\>\Big]
   -(-1)^{\d_{i8}}\Big[|\vs\>,A^i_{-1}A^j_{-1}|\vr\>\Big]
   &=&A^i_{-2}A^j_{-1}|\va\>, \]
Further careful analysis of the above results and usage of the little
Weyl group action \Ref{WeylPol} finally reveals that the following
states form a complete basis of the root space
$E_{10}^{(\vL_7)}$:
\[ A^i_{-2}A^j_{-1}|\va\>&&
   \FOR{$i,j$ arbitrary,} \non
   A^i_{-1}A^j_{-1}A^k_{-1}|\va\>&&
   \FOR{$i\neq j\neq k$,} \non
   \Big( A_{-3}^i - A^i_{-1}A^j_{-1}A^j_{-1} \Big) |\va\>&&
   \FOR{$i \neq j$,} \non
   \Big( 5A_{-3}^i + A^i_{-1}A^i_{-1}A^i_{-1}\Big) |\va\>&&
   \FOR{$i$ arbitrary,}  \non
   \Big( A_{-3}^i - A^i_{-1} \sL_{-2}\Big) |\va\>&&
   \FOR{$i$ arbitrary.}  \lb{E10-lev2}  \]
Remarkably, this choice is consistent with above eight commutator
equations and their Weyl-rotated analogs thereby proving the viability
of our method.
Altogether, we get $64 + 2\X56 + 2\X8 = 192 $
states in agreement with the formula \Ref{mult2} predicting
$\xi (3) = 192$ \ct{KaMoWa88}. Despite the fact that this number
coincides with the number of transversal states, our result
explicitly shows the appearance of longitudinal as well as the
disappearance of some transversal states. The above states form
irreducible representations of the little group, whose action
on the polarizations can be determined from \Ref{WeylPol}
in a straightforward fashion; in particular, the longitudinal
DDF operator is inert under the little Weyl group. We note that the
states \Ref{E10-lev2} do not even look ``$SO(8)$ covariant'' any more,
unlike the level-one states \Ref{E10-lev1}.

Having a complete description of the root space $E_{10}^{(\vL_7)}$,
we can now in principle explore root spaces associated with
other level-two roots of the form $\vL = \vL_7 + n\vd$ (i.e.\ the
{\bf root string} associated with $\vL_7$) by commuting the states
\Ref{E10-lev2} with the $\9$ elements \Ref{E9-2}. From
\Ref{comml2E9a} it is evident that all states obtained by
acting with a product $A^{i_1}_{-2m_1} \ldots A^{i_M}_{-2m_M}$
on any of the states \Ref{E10-lev2} belong to the
root space of $\vL = \vL_7 + (m_1 + \ldots m_M) \vd$ (note that
each operator $A^i_{-2m}$ shifts the momentum by $m\vd$!).
However, it is also clear that we cannot obtain all root space
elements in this way. For this, it is necessary to calculate DDF
commutators of the form \Ref{comm-lev1}. An alternative, more
elucidating way might be to consider the action of the Sugawara
generators defined by
\[   \eL^{{\rm Sug.}}_m :=  \frac{1}{2(\ell+h^\vee)}
  \sum_{n\in\Z}\Big\{\sum_{i=1}^8 \: A^i_n A^i_{m-n} \:
  +\sum_{\vs\in\Delta^{\rm real}(\9)} \: {\rm ad}_{|\vs\>}\,
     {\rm ad}_{|-\vs-m\vd\>} \: \Big\}   \]
on the states \Ref{E10-lev2}; here, $h^\vee=30$ is the dual Coxeter
number of $E_8$, the level is $\ell =2$, and the normal ordering of
the operators ${\rm ad}_{|\vr\>}\equiv(\exp\vr)_0$ is chosen as
\[ \:{\rm ad}_{|\vs+m\vd\>}\,{\rm ad}_{|\vt+n\vd\>}\:
                 :=\cases{
     {\rm ad}_{|\vs+m\vd\>}\,{\rm ad}_{|\vt+n\vd\>} & if $m\le n$, \cr
     {\rm ad}_{|\vt+n\vd\>}\,{\rm ad}_{|\vs+m\vd\>} & if $m>n$, \cr},
                                     \]
for $E_8$ roots $\vs,\vt$ and $m,n\in\Z$.
It is now not difficult to check that
\[ \eL^{\rm Sug.}_m |\va\> = 0  \]
for $m\geq 1$. Furthermore, when evaluating $\eL^{\rm Sug.}_0$
on the ground state $|\va\>$, only the term with $A_0^8 A_0^8$
contributes in the sum with our choice of polarization vectors.
With $A_0^8|\va\> = -2|\va\>$, we thus obtain
\[  \eL^{\rm Sug.}_0 |\va\> = \frc{1}{16} |\va\> , \]
showing that the state $|\va\>$ is a highest weight vector
of weight $h =\frc1{16}$ for the level-two Sugawara generators.
In accordance with the remarks in the footnote on page \pageref{foot1},
we therefore expect these states to belong to the irreducible
Virasoro module with $c=\frc12$ and $h=\frc1{16}$.
The problem that remains is to relate the Sugawara generators
to the longitudinal DDF operators. If this can be done, a completely
explicit description of {\it all} level-two root spaces is within
reach.
\\[.8cm]
{\bf Acknowledgments:} We are grateful to M. Koca and P. Slodowy
for discussions related to this work.

\begin{appendix}
\section{DDF states}
\def\r#1{\vr(-#1)}
\def\s#1{\vs(-#1)}
\def\k#1{\vk(-#1)}
\def\I#1{\vxi_i(-#1)}
\def\J#1{\vxi_j(-#1)}
\def\K#1{\vxi_k(-#1)}
\def\Ll#1{\vxi_l(-#1)}
\def\A#1{\vet_{i'}(-#1)}
\def\B#1{\vet_{j'}(-#1)}
\def\C#1{\vet_{k'}(-#1)}
\def\D#1{\vet_{l'}(-#1)}
\noi We here list the transversal and longitudinal DDF states, required
in section 4.5, up to oscillator number four. For the special
example discussed there, we must only evaluate them for the following
scalar products:
\( \vr^2= \vs^2= 2,\ \vk^2= 0, \ \vr\X\vk= \vs\X\vk= 1,\
   \vxi_i\X\vxi_j= \d_{ij}, \ \vet_{i'}\X\vet_{j'}= \d_{i'j'}, \
   \vxi_i\X\vr= \vxi_i\X\vk= \vet_{i'}\X\vs= \vet_{i'}\X\vk= 0, \
   \vxi_i\X\vs=: \d_{i\vs}, \ \vet_{i'}\X\vr=: \d_{i'\vr}, \
   \vet_{i'}\X\vxi_j=: g_{i'j}. \) Also put $\e\equiv\e(\vk,\vr)$,
$\e'\equiv\e(\vs,\vr)$, $\e''\equiv\e(\vs-\vk,\vr)$.

\medskip
\noi The transversal states are:
\[ \LL A^i_{-1}|\vr\>&=&\e\I1|\vr-\vk\>, \RR \HH \]
\[ \LL A^i_{-1}A^j_{-1}|\vr\>&=&\bigg\{\I1\J1
      +\2\d_{ij}\Big[\k1^2-\k2\Big]\bigg\}|\vr-2\vk\>, \RR \HH \]
\[ \LL A^i_{-2}|\vr\>&=&
      \bigg\{\I2-2\I1\k1\bigg\}|\vr-2\vk\>, \RR \HH \]
\[ \LL A^i_{-1}A^j_{-1}A^k_{-1}|\vr\>&=&\e\bigg\{\I1\J1\K1 \RR \non
   & &\phantom{\bigg\{}{}
      +\2\Big[\d_{ij}\K1+\d_{jk}\I1+\d_{ki}\J1\Big]
       \Big[\k1^2-\k2\Big]\bigg\}|\vr-3\vk\>, \]
\[ \LL A^i_{-2}A^j_{-1}|\vr\>&=&\e\bigg\{\I2\J1-2\I1\J1\k1 \RR \non
   & &\phantom{\bigg\{}{}
      -\frc23\d_{ij}\Big[2\k1^3-3\k2\k1+\k3\Big]\bigg\}|\vr-3\vk\>, \]
\[ \LL A^i_{-3}|\vr\>&=&\e\bigg\{\I3-3\I2\k1
      +\frc32\I1\Big[3\k1^2-\k2\Big]\bigg\}|\vr-3\vk\>, \RR \HH \]
\[ \LL A^i_{-1}A^j_{-1}A^k_{-1}A^l_{-1}|\vr\>&=&
      \bigg\{\I1\J1\K1\Ll1  \RR \non
   & &\phantom{\bigg\{}{}
      +\2\Big[\d_{ij}\K1\Ll1+\d_{ik}\J1\Ll1 \non
   & &\phantom{\bigg\{+\2\Big[}{}
              +\d_{il}\J1\K1+\d_{jk}\I1\Ll1 \non
   & &\phantom{\bigg\{+\2\Big[}{}
           +\d_{jl}\I1\K1+\d_{kl}\I1\J1\Big]\Big[\k1^2-\k2\Big] \non
   & &\phantom{\bigg\{}{}
      +\frc14\Big[\d_{ij}\d_{kl}+\d_{ik}\d_{jl}+\d_{il}\d_{jk}\Big]
       \Big[\k1^2-\k2\Big]^2\bigg\}|\vr-4\vk\>, \]
\[ \LL A^i_{-2}A^j_{-1}A^k_{-1}|\vr\>&=&
      \bigg\{\I2\J1\K1-2\I1\J1\K1\k1 \RR \non
   & &\phantom{\bigg\{}{}
      +\2\d_{jk}\I2\Big[\k1^2-\k2\Big] \non
   & &\phantom{\bigg\{}{}
      -\frc23\Big[\d_{ij}\K1+\d_{ik}\J1\Big]
       \Big[2\k1^3-3\k1\k2+\k3\Big] \non
   & &\phantom{\bigg\{}{}
      -\d_{jk}\I1\Big[\k1^3-\k1\k2\Big]\bigg\}|\vr-4\vk\>, \]
\[ \LL A^i_{-3}A^j_{-1}|\vr\>&=&\bigg\{\I3\J1-3\I2\J1\k1
      +\frc32\I1\J1\Big[3\k1^2-\k2\Big] \RR \non
   & &\phantom{\bigg\{}{}
      +\frc38\d_{ij}\Big[9\k1^4-18\k1^2\k2+8\k1\k3 \non
   & &\phantom{\bigg\{+\frc38\d_{ij}\Big[}{}
      +3\k2^2-2\k4\Big]\bigg\}|\vr-4\vk\>, \]
\[ \LL A^i_{-2}A^j_{-2}|\vr\>&=&
      \bigg\{\I2\J2-2\Big[\I2\J1+\J2\I1\Big]\k1 \RR \non
   & &\phantom{\bigg\{}{}
      +4\I1\J1\k1^2
      +\d_{ij}\Big[4\k1^4-8\k1^2\k2+4\k1\k3 \non
   & &\phantom{\bigg\{+4\I1\J1\k1^2+\d_{ij}\Big[}{}
      +\k2^2-\k4\Big]\bigg\}|\vr-4\vk\>, \]
\[ \LL A^i_{-4}|\vr\>&=&
      \bigg\{\I4-4\I3\k1+2\I2\Big[4\k1^2-\k2\Big] \RR \non
   & &\phantom{\bigg\{}{}
      -\frc43\I1\Big[8\k1^3-6\k1\k2+\k3\Big]\bigg\}|\vr-4\vk\>. \]

\medskip
\noi The longitudinal states are:
\[ \LL A^-_{-1}|\vr\>&=&
      \e\bigg\{-\r1+\k1\bigg\}|\vr-\vk\>=-\e\Lm1|\vr-\vk\>, \RR \HH \]
\[ \LL A^-_{-2}|\vr\>&=&
      \bigg\{-\r2+\frc{d-6}4\k2+2\r1\k1-\2\sum_{i=1}^{d-2}\I1^2
      +\frc{6-d}4\k1^2\bigg\}|\vr-2\vk\>, \RR \HH \]
\[ \LL A^i_{-1}A^-_{-2}|\vr\>&=&
      \e\bigg\{-\I3-\r2\I1+3\I2\k1+\frc{d-2}4\k2\I1 \RR \non
   & &\phantom{\bigg\{}{}
      +2\r1\I1\k1-\2\sum_{j=1}^{d-2}\I1\J1^2 \non
   & &\phantom{\bigg\{}{}
      -\frc{d+6}4\I1\k1^2\bigg\}|\vr-3\vk\>, \]
\[ \LL A^-_{-3}|\vr\>&=&
      \e\bigg\{-\r3+\frc{2d-16}3\k3+3\r2\k1
      -\sum_{i=1}^{d-2}\I2\I1+\frc32\k2\r1 \RR \non
   & &\phantom{\bigg\{}{}
      +\frc{35-4d}2\k2\k1-\frc92\r1\k1^2
      +2\sum_{i=1}^{d-2}\I1^2\k1 \non
   & &\phantom{\bigg\{}{}
      +\frc{8d-79}6\k1^3\bigg\}|\vr-3\vk\>, \]
\[ \LL A^i_{-1}A^j_{-1}A^-_{-2}|\vr\>&=&
      \bigg\{-\I3\J1-\J3\I1+3\Big[\I2\J1+\J2\I1\Big]\k1 \RR \non
   & &\phantom{\bigg\{}{}
      -\r2\I1\J1+\frc{d+2}4\k2\I1\J1 \non
   & &\phantom{\bigg\{}{}
      +2\r1\I1\J1\k1-\frc{d+18}4\I1\J1\k1^2 \non
   & &\phantom{\bigg\{}{}
      -\2\sum_{k=1}^{d-2}\I1\J1\K1^2 \non
   & &\phantom{\bigg\{}{}
      +\d_{ij}\bigg[\frc14\sum_{k=1}^{d-2}\K1^2\Big[\k2-\k1^2\Big]
      +\2\r2\Big[\k2-\k1^2\Big] \non
   & &\phantom{\bigg\{+\d_{ij}\bigg[}{}
      +\r1\Big[\k1^3-\k1\k2\Big]-\frc{d+18}8\k1^4
      +\frc{d+22}4\k1^2\k2 \non
   & &\phantom{\bigg\{+\d_{ij}\bigg[}{}
      -4\k1\k3-\frc{d+2}8\k2^2+\k4\bigg]\bigg\}|\vr-4\vk\>, \]
\[ \LL A^i_{-2}A^-_{-2}|\vr\>&=&
      \bigg\{-2\I4+8\I3\k1+2\k3\I1-\r2\I2 \RR \non
   & &\phantom{\bigg\{}{}
      +2\r2\I1\k1+\frc{d+2}4\I2\k2+2\I2\r1\k1 \non
   & &\phantom{\bigg\{}{}
      -\2\sum_{j=1}^{d-2}\I2\J1^2
      -\frc{d+42}4\I2\k1^2-\frc{d+14}2\k2\I1\k1 \non
   & &\phantom{\bigg\{}{}
      -4\r1\I1\k1^2+\sum_{j=1}^{d-2}\I1\J1^2\k1 \non
   & &\phantom{\bigg\{}{}
      +\frc{d+18}2\I1\k1^3\bigg\}|\vr-4\vk\>, \]
\[ \LL A^i_{-1}A^-_{-3}|\vr\>&=&
      \bigg\{-\I4+4\I3\k1-\r3\I1+\frc{2d-13}3\k3\I1 \RR \non
   & &\phantom{\bigg\{}{}
      +\frc52\I2\k2+3\r2\I1\k1-\sum_{j=1}^{d-2}\J2\J1\I1 \non
   & &\phantom{\bigg\{}{}
      -\frc{17}2\I2\k1^2+\frc32\k2\r1\I1+(11-2d)\k2\I1\k1 \non
   & &\phantom{\bigg\{}{}
      -\frc92\r1\I1\k1^2+2\sum_{j=1}^{d-2}\I1\J1^2\k1 \non
   & &\phantom{\bigg\{}{}
      +\frc{4d-14}3\I1\k1^3\bigg\}|\vr-4\vk\>, \]
\[ \LL A^-_{-4}|\vr\>&=&
      \bigg\{-\r4+\frc{5d-42}4\k4+4\r3\k1
      -\sum_{i=1}^{d-2}\I3\I1+\frc43\k3\r1 \RR \non
   & &\phantom{\bigg\{}{}
      +\frc{134-15d}3\k3\k1+2\r2\k2-\2\sum_{i=1}^{d-2}\I2^2
      +\frc{122-13d}8\k2^2 \non
   & &\phantom{\bigg\{}{}
      -8\r2\k1^2+5\sum_{i=1}^{d-2}\I2\I1\k1-8\k2\r1\k1 \non
   & &\phantom{\bigg\{}{}
      +\frc32\sum_{i=1}^{d-2}\k2\I1^2
      +\frc{43d-422}4\k2\k1^2+\frc{32}3\r1\k1^3 \non
   & &\phantom{\bigg\{}{}
      -\frc{13}2\sum_{i=1}^{d-2}\I1^2\k1^2
      +\frc{1394-129d}{24}\k1^4\bigg\}|\vr-4\vk\>, \]
\[ \LL A^-_{-2}A^-_{-2}|\vr\>&=&
      \bigg\{2\r4+\frc{26-d}4\k4-8\r3\k1
      +\sum_{i=1}^{d-2}\I3\I1+\frc{3d-78}3\k3\k1 \RR \non
   & &\phantom{\bigg\{}{}
      +\r2^2+\frc{6-d}2\r2\k2+\frc{d^2-16d-20}{16}\k2^2
      -4\r2\r1\k1 \non
   & &\phantom{\bigg\{}{}
      +\sum_{i=1}^{d-2}\r2\I1^2+\frc{d+10}2\r2\k1^2
      -3\sum_{i=1}^{d-2}\I2\I1\k1 \non
   & &\phantom{\bigg\{}{}
      +(d-6)\k2\r1\k1
      +\frc{6-d}4\sum_{i=1}^{d-2}\k2\I1^2
      +\frc{-d^2+12d+284}8\k2\k1^2  \non
   & &\phantom{\bigg\{}{}
      +4\r1^2\k1^2
      -2\sum_{i=1}^{d-2}\r1\I1^2\k1+(6-d)\r1\k1^3 \non
   & &\phantom{\bigg\{}{}
      +\frc14\sum_{i,j=1}^{d-2}\I1^2\J1^2
      +\frc{d-2}4\sum_{i=1}^{d-2}\I1^2\k1^2 \non
   & &\phantom{\bigg\{}{}
      +\frc{d^2-20d-236}{16}\k1^4\bigg\}|\vr-4\vk\>. \]

\section{DDF commutators}
\noi Some commutators for $\vr\X\vs=0$:
\[ \LL \Big[|\vs\>,A^i_{-1}A^j_{-1}|\vr\>\Big]&=&
      \e'\bigg\{\I1\J1\s1-\2\Big[\d_{i\vs}\J1+\d_{j\vs}\I1\Big]
      \Big[\s1^2+\s2\Big] \RR \non
   & &\phantom{\bigg\{}{}
      +\frc16\d_{ij}\Big[\s1^3-3\s1^2\k1+3\s1\k1^2-3\s1\k2 \non
   & &\phantom{\bigg\{+\frc16\d_{ij}\Big[}{}
      +3\s1\s2-3\s2\k1+2\s3\Big] \non
   & &\phantom{\bigg\{}{}
      +\frc16\d_{i\vs}\d_{j\vs}\Big[\s1^3+3\s1\s2+2\s3\Big]
      \bigg\}|\vr-2\vk+\vs\>, \]
\[ \LL \Big[|\vs\>,A^i_{-2}|\vr\>\Big]&=&
      \e'\bigg\{\I2\s1+\I1\Big[\s1^2-2\s1\k1+\s2\Big] \RR \non
   & &\phantom{\bigg\{}{}
      -\2\d_{i\vs}\Big[\s1^3-2\s1^2\k1+3\s1\s2 \non
   & &\phantom{\bigg\{-\2\d_{i\vs}\Big[}{}
      -2\k1\s2+2\s3\Big]
      \bigg\}|\vr-2\vk+\vs\>, \]
\[ \LL \Big[A^{i'}_{-1}|\vs\>,A^j_{-1}|\vr\>\Big]&=&
      \e'\bigg\{\d_{j\vs}\A3-\A2\J1
      +\d_{j\vs}\A2\Big[\s1-\k1\Big] \RR \non
   & &\phantom{\bigg\{}{}
      -\A1\J1\Big[\s1-\k1\Big] \non
   & &\phantom{\bigg\{}{}
      -\2\Big[\d_{i'\vr}\J1-\d_{j\vs}\A1\Big]
      \Big[\s1^2-2\s1\k1+\k1^2 \non
   & &\phantom{\bigg\{-\2\Big[\d_{i'\vr}\J1-\d_{j\vs}\A1\Big]\Big[}{}
      +\s2-\k2\Big] \non
   & &\phantom{\bigg\{}{}
      +\frc16\Big[\d_{i'\vr}\d_{j\vs}-g_{i'j}\Big]
      \Big[\s1^3-3\s1^2\k1+3\s1\k1^2 \non
   & &\phantom{\bigg\{+\frc16\Big[\d_{i'\vr}\d_{j\vs}-g_{i'j}\Big]
               \Big[}{}
      -\k1^3+3\s2\s1-3\s2\k1 \non
   & &\phantom{\bigg\{+\frc16\Big[\d_{i'\vr}\d_{j\vs}-g_{i'j}\Big]
               \Big[}{}
      -3\k2\s1+3\k2\k1 \non
   & &\phantom{\bigg\{+\frc16\Big[\d_{i'\vr}\d_{j\vs}-g_{i'j}\Big]
               \Big[}{}
      +2\s3-2\k3\Big]\bigg\}|\vr-2\vk+\vs\>. \]
\noi Some commutators for $\vr\X\vs=1$:
\[ \LL \Big[|\vs\>,A^i_{-1}A^j_{-1}A^k_{-1}|\vr\>\Big]&=&
      \e''\bigg\{\I1\J1\K1\s1
      -\2\Big[\d_{i\vs}\J1\K1+\d_{j\vs}\K1\I1 \RR \non
   & &\phantom{\bigg\{\quad}{}
      +\d_{k\vs}\I1\J1\Big]\Big[\s1^2+\s2\Big] \non
   & &\phantom{\bigg\{}{}
      +\frc16\Big[\d_{ij}\K1+\d_{jk}\I1+\d_{ki}\J1\Big]
      \Big[\s1^3-3\s1^2\k1 \non
   & &\phantom{\bigg\{\quad}{}
      +3\s1\k1^2-3\s1\k2+3\s1\s2 \non
   & &\phantom{\bigg\{\quad}{}
      -3\s2\k1+2\s3\Big] \non
   & &\phantom{\bigg\{}{}
      +\frc16\Big[\d_{i\vs}\d_{j\vs}\K1+\d_{j\vs}\d_{k\vs}\I1 \non
   & &\phantom{\bigg\{+\frc16\Big[}{}
      +\d_{k\vs}\d_{i\vs}\J1\Big]\Big[\s1^3+3\s1\s2+2\s3\Big] \non
   & &\phantom{\bigg\{}{}
      -\frc1{24}\d_{i\vs}\d_{j\vs}\d_{k\vs}\Big[\s1^4+6\s1^2\s2
      +3\s2^2 \non
   & &\phantom{\bigg\{-\frc1{24}\d_{i\vs}\d_{j\vs}\d_{k\vs}\Big[}{}
      +8\s1\s3+6\s4\Big] \non
   & &\phantom{\bigg\{}{}
      -\frc1{24}\Big[\d_{i\vs}\d_{jk}+\d_{j\vs}\d_{ki}
      +\d_{k\vs}\d_{ij}\Big]\Big[\s1^4-4\s1^3\k1-6\s1^2\k2 \non
   & &\phantom{\bigg\{\quad}{}
      +6\s1^2\k1^2+6\s1^2\s2-12\s1\k1\s2 \non
   & &\phantom{\bigg\{\quad}{}
      +6\k1^2\s2-6\s2\k2+3\s2^2 \non
   & &\phantom{\bigg\{\quad}{}
      +8\s1\s3-8\k1\s3+6\s4\Big]
      \bigg\}|\vr-3\vk+\vs\>, \]
\[ \LL \Big[|\vs\>,A^i_{-2}A^j_{-1}|\vr\>\Big]&=&
      \e''\bigg\{\I2\J1\s1-\2\d_{j\vs}\I2\Big[\s1^2+\s2\Big] \RR \non
   & &\phantom{\bigg\{}{}
      +\I1\J1\Big[\s1^2-2\s1\k1+\s2\Big] \non
   & &\phantom{\bigg\{}{}
      -\frc13\d_{j\vs}\I1\Big[\s1^3-3\s1^2\k1+3\s1\s2 \non
   & &\phantom{\bigg\{-\frc13\d_{j\vs}\I1\Big[}{}
      -3\k1\s2+2\s3\Big] \non
   & &\phantom{\bigg\{}{}
      -\2\d_{i\vs}\J1\Big[\s1^3-2\s1^2\k1+3\s1\s2 \non
   & &\phantom{\bigg\{-\2\d_{i\vs}\J1\Big[}{}
      -2\k1\s2+2\s3\Big] \non
   & &\phantom{\bigg\{}{}
      +\frc1{24}\d_{i\vs}\d_{j\vs}\Big[3\s1^4-8\s1^3\k1
      +18\s1^2\s2 \non
   & &\phantom{\bigg\{+\frc1{24}\d_{i\vs}\d_{j\vs}\Big[}{}
      +9\s2^2-24\s1\k1\s2-16\k1\s3 \non
   & &\phantom{\bigg\{+\frc1{24}\d_{i\vs}\d_{j\vs}\Big[}{}
      +24\s1\s3+18\s4\Big] \non
   & &\phantom{\bigg\{}{}
      +\frc16\d_{ij}\Big[\s1^4-6\s1^3\k1+12\s1^2\k1^2-8\s1\k1^3 \non
   & &\phantom{\bigg\{+\frc16\d_{ij}\Big[}{}
      +12\s1\k1\k2-6\s1^2\k2+12\k1^2\s2 \non
   & &\phantom{\bigg\{+\frc16\d_{ij}\Big[}{}
      -18\s1\k1\s2+6\s1^2\s2-6\s2\k2 \non
   & &\phantom{\bigg\{+\frc16\d_{ij}\Big[}{}
      +3\s2^2-4\s1\k3+8\s1\s3 \non
   & &\phantom{\bigg\{+\frc16\d_{ij}\Big[}{}
      -12\k1\s3+6\s4\Big]\bigg\}|\vr-3\vk+\vs\>, \]
\[ \LL \Big[|\vs\>,A^i_{-3}|\vr\>\Big]&=&
      \e''\bigg\{\I3\s1+\frc32\I2\Big[\s1^2-2\s1\k1+\s2\Big] \RR \non
   & &\phantom{\bigg\{}{}
      +\2\I1\Big[2\s1^3-9\s1^2\k1+9\s1\k1^2+6\s1\s2 \non
   & &\phantom{\bigg\{+\2\I1\Big[}{}
      -3\s1\k2-9\k1\s2+4\s3\Big] \non
   & &\phantom{\bigg\{}{}
      -\frc1{12}\d_{i\vs}\Big[5\s1^4-24\s1^3\k1
      +27\s1^2\k1^2-9\s1^2\k2 \non
   & &\phantom{\bigg\{-\frc1{12}\d_{i\vs}\Big[}{}
      +30\s1^2\s2-72\s1\k1\s2+27\k1^2\s2 \non
   & &\phantom{\bigg\{-\frc1{12}\d_{i\vs}\Big[}{}
      +15\s2^2-9\s2\k2+40\s1\s3 \non
   & &\phantom{\bigg\{-\frc1{12}\d_{i\vs}\Big[}{}
      -48\k1\s3+30\s4\Big]\bigg\}|\vr-3\vk+\vs\>, \]
\[ \LL \Big[A^{i'}_{-1}|\vs\>,A^j_{-1}A^k_{-1}|\vr\>\Big]&=&
      \e''\bigg\{\Big[\d_{j\vs}\d_{k\vs}+\d_{jk}\Big]\A4
      +\d_{jk}\A3\Big[\s1-2\k1\Big] \RR \non
   & &\phantom{\bigg\{}{}
      +\d_{j\vs}\d_{k\vs}\A3\Big[\s1-\k1\Big]
      -\A3\Big[\d_{j\vs}\K1+\d_{k\vs}\J1\Big] \non
   & &\phantom{\bigg\{}{}
      +\A2\J1\K1
      +\A1\J1\K1\Big[\s1-\k1\Big] \non
   & &\phantom{\bigg\{}{}
      -\A2\Big[\d_{j\vs}\K1+\d_{k\vs}\J1\Big]\Big[\s1-\k1\Big] \non
   & &\phantom{\bigg\{}{}
      +\2\d_{jk}\A2\Big[
      \s1^2-4\s1\k1+\s2+4\k1^2-2\k2\Big] \non
   & &\phantom{\bigg\{}{}
      +\2\Big[\d_{j\vs}\d_{k\vs}\A2+\d_{i'\vr}\J1\K1 \non
   & &\phantom{\bigg\{+\2\Big[}{}
      -\d_{j\vs}\A1\K1-\d_{k\vs}\A1\J1\Big] \non
   & &\phantom{\bigg\{\quad}{}
      \Big[\s1^2-2\s1\k1+\s2+\k1^2-\k2\Big] \non
   & &\phantom{\bigg\{}{}
      +\frc16\Big[\d_{j\vs}\d_{k\vs}\A1
      +[\d_{i'j}-\d_{i'\vr}\d_{j\vs}]\K1
      +[\d_{i'k}-\d_{i'\vr}\d_{k\vs}]\J1\Big] \non
   & &\phantom{\bigg\{\quad}{}
      \Big[\s1^3-3\s1^2\k1+3\s1\k1^2-\k1^3 \non
   & &\phantom{\bigg\{\quad\Big[}{}
      +3\s2\s1-3\s2\k1-3\k2\s1 \non
   & &\phantom{\bigg\{\quad\Big[}{}
      +3\k2\k1+2\s3-2\k3\Big] \non
   & &\phantom{\bigg\{}{}
      +\frac16\d_{jk}\A1
      \Big[\s1^3-6\s1^2\k1+12\s1\k1^2-7\k1^3 \non
   & &\phantom{\bigg\{+\frac16\d_{jk}\A1\Big]}{}
      +3\s2\s1-6\s2\k1-6\k2\s1 \non
   & &\phantom{\bigg\{+\frac16\d_{jk}\A1\Big]}{}
      +9\k2\k1+2\s3-2\k3\Big] \non
   & &\phantom{\bigg\{}{}
      +\frc1{24}\Big[\d_{i'\vr}\d_{j\vs}\d_{k\vs}
      -\d_{i'j}\d_{k\vs}-\d_{i'k}\d_{j\vs}\Big] \non
   & &\phantom{\bigg\{\quad}{}
      \Big[\s1^4-4\s1^3\k1+6\s1^2\s2+6\s1^2\k1^2 \non
   & &\phantom{\bigg\{\quad}{}
      -6\s1^2\k2-12\s1\s2\k1+8\s1\s3 \non
   & &\phantom{\bigg\{\quad}{}
      -4\s1\k1^3+12\s1\k1\k2-8\s1\k3 \non
   & &\phantom{\bigg\{\quad}{}
      +3\s2^2+6\s2\k1^2-6\s2\k2-8\s3\k1 \non
   & &\phantom{\bigg\{\quad}{}
      +6\s4+\k1^4-6\k1^2\k2+8\k1\k3 \non
   & &\phantom{\bigg\{\quad}{}
      +3\k2^2-6\k4\Big] \non
   & &\phantom{\bigg\{}{}
      +\frc1{24}\d_{i'\vr}\d_{jk}\Big[
      \s1^4-8\s1^3\k1+6\s1^2\s2 \non
   & &\phantom{\bigg\{+\frc1{24}\d_{i'\vr}\d_{jk}\Big[}{}
      +24\s1^2\k1^2-12\s1^2\k2 \non
   & &\phantom{\bigg\{+\frc1{24}\d_{i'\vr}\d_{jk}\Big[}{}
      -24\s1\s2\k1+8\s1\s3 \non
   & &\phantom{\bigg\{+\frc1{24}\d_{i'\vr}\d_{jk}\Big[}{}
      -28\s1\k1^3+36\s1\k1\k2 \non
   & &\phantom{\bigg\{+\frc1{24}\d_{i'\vr}\d_{jk}\Big[}{}
      -8\s1\k3+3\s2^2+24\s2\k1^2 \non
   & &\phantom{\bigg\{+\frc1{24}\d_{i'\vr}\d_{jk}\Big[}{}
      -12\s2\k2-16\s3\k1+6\s4+11\k1^4 \non
   & &\phantom{\bigg\{+\frc1{24}\d_{i'\vr}\d_{jk}\Big[}{}
      -30\k1^2\k2+16\k1\k3 \non
   & &\phantom{\bigg\{+\frc1{24}\d_{i'\vr}\d_{jk}\Big[}{}
      +9\k2^2-6\k4\Big]\bigg\}|\vr-4\vk+\vs\>, \]

\[ \LL \Big[A^{i'}_{-1}|\vs\>,A^j_{-2}|\vr\>\Big]&=&
      \e''\bigg\{-3\d_{j\vs}\A4+2\A3\J1
      +\d_{j\vs}\A3\big[5k1-3\s1\Big] \RR \non
   & &\phantom{\bigg\{}{}
      +\A2\J2+2\A2\J1\Big[\s1-2\k1\Big] \non
   & &\phantom{\bigg\{}{}
      +\J2\A1\Big[\s1-\k1\Big] \non
   & &\phantom{\bigg\{}{}
      +\2\d_{j\vs}\A2\Big[3\k2-7\k1^2-3\s2+10\s1\k1-3\s1^2\Big] \non
   & &\phantom{\bigg\{}{}
      +\2\d_{i'\vr}\J2\Big[-\k2+\k1^2+\s2-2\s1\k1+\s1^2\Big] \non
   & &\phantom{\bigg\{}{}
      +\A1\J1\Big[-\k2+3\k1^2+\s2-4\s1\k1+\s1^2\Big] \non
   & &\phantom{\bigg\{}{}
      +\2\d_{j\vs}\A1\Big[2\k3-5\k1\k2+3\k1^3-2\s3 \non
   & &\phantom{\bigg\{+\2\d_{j\vs}\A1\Big[}{}
      +5\s2\k1+3\s1\k2-7\s1\k1^2 \non
   & &\phantom{\bigg\{+\2\d_{j\vs}\A1\Big[}{}
      -3\s1\s2+5\s1^2\k1-\s1^3\Big] \non
   & &\phantom{\bigg\{}{}
      +\frc13\d_{i'\vr}\J1\Big[
      -2\k3+6\k1\k2-4\k1^3+2\s3 \non
   & &\phantom{\bigg\{+\frc13\d_{i'\vr}\J1\Big[}{}
      -6\s2\k1-3\s1\k2+9\s1\k1^2 \non
   & &\phantom{\bigg\{+\frc13\d_{i'\vr}\J1\Big[}{}
      +3\s1\s2-6\s1^2\k1+\s1^3\Big] \non
   & &\phantom{\bigg\{}{}
      +\frc1{24}\d_{i'\vr}\d_{j\vs}\Big[
      -3\s1^4+20\s1^3\k1-18\s1^2\s2 \non
   & &\phantom{\bigg\{+\frc1{24}\d_{i'\vr}\d_{j\vs}\Big[}{}
      -42\s1^2\k1^2+18\s1^2\k2+60\s1\s2\k1 \non
   & &\phantom{\bigg\{+\frc1{24}\d_{i'\vr}\d_{j\vs}\Big[}{}
      -24\s1\s3+36\s1\k1^3-60\s1\k1\k2 \non
   & &\phantom{\bigg\{+\frc1{24}\d_{i'\vr}\d_{j\vs}\Big[}{}
      +24\s1\k3-9\s2^2-42\s2\k1^2 \non
   & &\phantom{\bigg\{+\frc1{24}\d_{i'\vr}\d_{j\vs}\Big[}{}
      +18\s2\k2+40\s3\k1-18\s4-11\k1^4 \non
   & &\phantom{\bigg\{+\frc1{24}\d_{i'\vr}\d_{j\vs}\Big[}{}
      +42\k1^2\k2-40\k1\k3-9\k2^2+18\k4\Big] \non
   & &\phantom{\bigg\{}{}
      +\frc16\d_{i'j}\Big[
      \s1^4-6\s1^3\k1+6\s1^2\s2 \non
   & &\phantom{\bigg\{+\frc16\d_{i'j}\Big[}{}
      +12\s1^2\k1^2-6\s1^2\k2-18\s1\s2\k1 \non
   & &\phantom{\bigg\{+\frc16\d_{i'j}\Big[}{}
      +8\s1\s3-10\s1\k1^3+18\s1\k1\k2 \non
   & &\phantom{\bigg\{+\frc16\d_{i'j}\Big[}{}
      -8\s1\k3+3\s2^2+12\s2\k1^2-6\s2\k2 \non
   & &\phantom{\bigg\{+\frc16\d_{i'j}\Big[}{}
      -12\s3\k1+6\s4+3\k1^4-12\k1^2\k2 \non
   & &\phantom{\bigg\{+\frc16\d_{i'j}\Big[}{}
      +12\k1\k3+3\k2^2-6\k4\Big]\bigg\}|\vr-3\vk+\vs\>. \]

\section{Commutators for $\vL_7$}
\noi The following commutators are written in terms of the basis
\Ref{DDFspan} rather than \Ref{physspan} as in the main text.
We also list the contributions $\Lm{1} (\dots )$ which have been
omitted in section 4.5. and put $\e\equiv\e(\vs,\vr)\e(\vk,\va)$.
\[ \LL \Big[|\vs\>,A^\a_{-1}A^\b_{-1}|\vr\>\Big]&=&
    \e\bigg\{-\2A^\a_{-2}A^\b_{-1}
             -\2A^\b_{-2}A^\a_{-1}
             -A^\a_{-1}A^\b_{-1}A^8_{-1} \RR \non
   & &\phantom{\bigg\{}
      +\d^{\a\b}\Big[\frc1{24}A^8_{-3}+\frc18A^8_{-1}A^-_{-2}
      +\frc1{16}\sum_{\c=1}^7A^8_{-1}A^\c_{-1}A^\c_{-1}
      -\frc5{48}A^8_{-1}A^8_{-1}A^8_{-1}\Big]\bigg\}|\va\> \non
   & &{}
      +\Lm1\bigg\{\2\vxa(-1)\vxb(-1)
      +\d^{\a\b}\bigg[-\frc18\8(-2)+\frc1{12}\vL(-2)
      +\frc14\8(-1)^2 \non
   & &\phantom{{}+\Lm1\bigg\{}
      +\frc1{48}\vL(-1)^2
      -\frc18\8(-1)\Big[\vL(-1)-\vd(-1)\Big]\bigg]
      \bigg\}|\vL_7\>, \]
\[ \LL \Big[|\vs\>,A^\a_{-1}A^8_{-1}|\vr\>\Big]&=&
    \e\bigg\{\frc14A^\a_{-3}+\2A^\a_{-2}A^8_{-1}
      -\2A^8_{-2}A^\a_{-1}
      -\frc14A^\a_{-1}A^-_{-2} \RR \non
   & &\phantom{\bigg\{}
      -\frc18\sum_{\c=1}^7A^\a_{-1}A^\c_{-1}A^\c_{-1}
      -\frc18A^\a_{-1}A^8_{-1}A^8_{-1}\bigg\}|\va\> \non
   & &{}
      +\Lm1\bigg\{-\frc14\vxa(-2)
      -\frc14\vxa(-1)\Big[2\8(-1)-\vL(-1)
      +3\vd(-1)\Big]\bigg\}|\vL_7\>, \]
\[ \LL \Big[|\vs\>,A^8_{-1}A^8_{-1}|\vr\>\Big]&=&
    \e\bigg\{\frc{17}{24}A^8_{-3}+A^8_{-2}A^8_{-1}
      +\frc18A^8_{-1}A^-_{-2} \RR \non
   & &\phantom{\bigg\{}
      +\frc1{16}\sum_{\c=1}^7A^8_{-1}A^\c_{-1}A^\c_{-1}
      +\frc{11}{48}A^8_{-1}A^8_{-1}A^8_{-1}\bigg\}|\va\> \non
   & &{}
      +\Lm1\bigg\{-\frc98\8(-2)+\frc5{12}\vL(-2)-\vd(-2)
      -\frc14\8(-1)^2+\frc5{48}\vL(-1)^2 \non
   & &\phantom{{}+\Lm1\bigg\{}
      -\2\vd(-1)^2
      -\frc18\8(-1)\Big[\vL(-1)+7\vd(-1)\Big]
      \bigg\}|\vL_7\>, \]
\[ \LL \Big[|\vs\>,A^\a_{-2}|\vr\>\Big]&=&
    \e\bigg\{-\frc34A^\a_{-3}-\frc14A^\a_{-1}A^-_{-2}
      -\frc18\sum_{\c=1}^7A^\a_{-1}A^\c_{-1}A^\c_{-1}
      +\frc78A^\a_{-1}A^8_{-1}A^8_{-1}\bigg\}|\va\> \RR \non
   & &{}
      +\Lm1\bigg\{\frc14\vxa(-2)
      -\frc14\vxa(-1)\Big[4\8(-1)-\vL(-1)
      +\vd(-1)\Big]\bigg\}|\vL_7\>, \]
\[ \LL \Big[|\vs\>,A^8_{-2}|\vr\>\Big]&=&
    \e\bigg\{-\2A^8_{-3}+\2A^8_{-1}A^-_{-2}
      +\frc14\sum_{\c=1}^7A^8_{-1}A^\c_{-1}A^\c_{-1}
      +\frc14A^8_{-1}A^8_{-1}A^8_{-1}\bigg\}|\va\> \RR \non
   & &{}
      +\Lm1\bigg\{-\2\8(-2)+\2\vL(-2)-\vd(-2)
      +\2\8(-1)^2+\frc18\vL(-1)^2 \non
   & &\phantom{{}+\Lm1\bigg\{}
      -\2\vd(-1)^2
      -\2\8(-1)\Big[\vL(-1)-\vd(-1)\Big]
      \bigg\}|\vL_7\>, \]
\[ \LL \Big[A^\a_{-1}|\vs\>,A^\b_{-1}|\vr\>\Big]&=&
    \e\bigg\{-\2A^\a_{-2}A^\b_{-1}
             +\2A^\b_{-2}A^\a_{-1}
             +A^\a_{-1}A^\b_{-1}A^8_{-1} \RR \non
   & &\phantom{\bigg\{}
      -\d^{\a\b}\Big[\frc1{24}A^8_{-3}+\frc18A^8_{-1}A^-_{-2}
      +\frc1{16}\sum_{\c=1}^7A^8_{-1}A^\c_{-1}A^\c_{-1}
      -\frc5{48}A^8_{-1}A^8_{-1}A^8_{-1}\Big]\bigg\}|\va\> \non
   & &{}
      +\Lm1\bigg\{-\2\vxa(-1)\vxb(-1)
      -\d^{\a\b}\bigg[-\frc18\8(-2)+\frc1{12}\vL(-2)
      +\frc14\8(-1)^2 \non
   & &\phantom{{}+\Lm1\bigg\{}
      +\frc1{48}\vL(-1)^2
      -\frc18\8(-1)\Big[\vL(-1)-\vd(-1)\Big]\bigg]
      \bigg\}|\vL_7\>, \]
\[ \LL \Big[A^\a_{-1}|\vs\>,A^8_{-1}|\vr\>\Big]&=&
    \e\bigg\{-\frc14A^\a_{-3}+\2A^\a_{-2}A^8_{-1}
      +\2A^8_{-2}A^\a_{-1}+\frc14A^\a_{-1}A^-_{-2} \RR \non
   & &\phantom{\bigg\{}
      +\frc18\sum_{\c=1}^7A^\a_{-1}A^\c_{-1}A^\c_{-1}
      +\frc18A^\a_{-1}A^8_{-1}A^8_{-1}\bigg\}|\va\> \non
   & &{}
      +\Lm1\bigg\{-\frc34\vxa(-2)
      -\frc14\vxa(-1)\Big[-2\8(-1)+\vL(-1)
      +\vd(-1)\Big]\bigg\}|\vL_7\>, \]
\[ \LL \Big[A^8_{-1}|\vs\>,A^8_{-1}|\vr\>\Big]&=&
    \e\bigg\{\frc{17}{24}A^8_{-3}
      +\frc18A^8_{-1}A^-_{-2}
      +\frc1{16}\sum_{\c=1}^7A^8_{-1}A^\c_{-1}A^\c_{-1}
      +\frc{11}{48}A^8_{-1}A^8_{-1}A^8_{-1}\bigg\}|\va\> \RR \non
   & &{}
      +\Lm1\bigg\{-\frc98\8(-2)+\frc5{12}\vL(-2)-\vd(-2)
      -\frc14\8(-1)^2+\frc5{48}\vL(-1)^2 \non
   & &\phantom{{}+\Lm1\bigg\{}
      -\2\vd(-1)^2
      -\frc18\8(-1)\Big[\vL(-1)-\vd(-1)\Big]
      \bigg\}|\vL_7\>. \]
\end{appendix}

\end{document}